\documentstyle[aps,pre]{revtex}

\begin{document}

\title{Spontaneous Symmetry Breaking and Exotic Quantum Order
in Integer Quantum Hall 
Systems under a Tilted Magnetic Field}

\author{Daw-Wei Wang$^{1,2}$, Eugene Demler$^2$, and S. Das Sarma$^1$}

\address{$^1$Condensed Matter Theory Center,
Department of Physics, University of Maryland,
College Park, MD 20842\\
$^2$Physics Department, Harvard University, Cambridge, MA 02138}

\date{\today}

\maketitle

\begin{abstract}
We use the microscopic Hartree-Fock approximation to investigate various
quantum phase transitions associated with possible spontaneous
symmetry breaking induced by a tilted magnetic field
in the integral quantum Hall regime of wide parabolic
wells and zero width double well (bilayer) systems.
Spin, isospin (associated with the layer index in the bilayer systems), and
orbital dynamics all play important roles in the quantum phase transitions
being studied. 
We propose a general class of variational wavefunctions that describe
several types of parity, spin, and translational symmetry breaking,
including spin and charge density wave phases.
In wide well systems at odd filling factors, we find
a many-body state of broken parity symmetry
for weak in-plane magnetic fields and an isospin
skyrmion stripe phase, which 
simultaneously has isospin and charge modulation,
for strong in-plane fields.
In wide well systems at even filling factors, we find direct first order
transitions between simple (un)polarized QH states with several
complex many-body states that are only slightly higher in energy
(within the Hartree-Fock theory)
than the ground state. We suggest that going beyond the approximations
used in this paper one may be able to stabilize such many-body
phases with broken symmetries (most likely the skyrmion stripe phase).
In a bilayer system 
at the filling factor $\nu=4N\pm 1$, 
where $N$ is an integer, we obtain
an isospin spiral stripe phase in addition to the known (in)commensurate 
phases and the stripe phase without isospin winding.
We do not find a charge or spin density wave
instability in the bilayer system at $\nu=4N+2$, except for 
the known commensurate canted antiferromagnetic phase.
Zero temperature quantum phase diagrams for these systems are calculated
in the parameter regime of experimental interest. 
We discuss the symmetry properties of our predicted quantum phase diagrams
and give a unified picture of these novel many-body phases.
We point out how quantum level crossing phenomena in many situations
(tuned by the applied tilted magnetic field) may lead to interesting
quantum phases and transitions among them. A conceptually new aspect of our
theory is the predicted possibility for the spontaneous breaking
of parity symmetry, 
which indicates a ``ferroelectric'' quantum order in
integer quantum Hall systems and has not 
been considered in the literature before.

\end{abstract}

\pacs{PACS numbers: 73.43.-f, 73.43.Nq, 73.43.Lp, 73.43.Cd}

\section{Introduction}
\label{introduction}

Unidirectional charge density wave order (also called stripe order) 
in quantum Hall (QH) systems has been 
extensively studied \cite{review} since the
first theoretical prediction in 1996 \cite{moessner96,fogler97} and the
first experimental observation in high Landau levels
via the magnetoresistance anisotropy measurement in 1999 
\cite{lilly99,du99}. Many related phenomena, e.g. transport
via internal edge state  
excitations \cite{macdonald00,lopatnikova01,eugene01}, liquid crystal
phases \cite{review,fradkin99}, reorientation
of stripe directions \cite{pan00,jungwirth99,zhu02}, 
and re-entrant integer quantum Hall effect
\cite{eisenstein01}, have been widely explored both theoretically
and experimentally in this context. 
Stripe formation at fractional filling factors, 
$\nu=(2N+1)/(4N+4)$, corresponding to the
composite fermion filling factor, $\nu^\ast=N+1/2$, 
was also proposed to have energy lower than the conventional Laughlin liquid,
composite-fermion Fermi sea or paired
composite-fermion state \cite{jain01}.
(Throughout this paper $N$ is zero or an integer, $N=0,1\cdots$, and
$\nu=1,2,3\cdots$ is the Landau level filling factor of the whole system.)
However, most of the stripe phases discussed in the literature so far
are formed by electrons
in the top (half-filled) level of the QH system at high 
half-odd-integer (HOI) filling
factors (e.g. $\nu=N+1/2$ for $N\ge 4$), while the  
stripe formation at integer filling factors
is seldom studied either theoretically or experimentally.

We develop a detailed theory in this paper for possible spontaneous
symmetry breaking and associated exotic quantum order in wide-well
and double-well integer quantum Hall systems by considering
all the symmetry properties of the realistic system Hamiltonians.
The great advantage of the quantum Hall systems in studying quantum 
critical phenomena is, however, the existence of various energy gaps 
(at the Fermi level) which enables us to carry out reasonable energetic 
calculations (within a mean-field Hartree-Fock theory) to quantitatively 
check whether the various possible quantum phase transitions and exotic 
quantum order allowed by symmetry considerations are actually energetically 
favored in realistic systems. We therefore construct explicit variational 
wavefunctions for various possible (exotic) quantum states, and carry out 
energetic calculations to find the optimal ground state. Using an in-plane 
parallel magnetic field (in addition to the perpendicular field necessary 
in the quantum Hall problem) to tune system parameters, we find a 
surprisingly rich quantum phase diagram in our systems of interest. We 
believe that the symmetry-broken states with exotic quantum order predicted 
by our theory should be experimentally observable in transport and 
inelastic light scattering experiments. We emphasize that the salient
feature of our work (making it particularly significant from the experimental
perspective) is that we not only identify spontaneous symmetry breaking
and exotic quantum order allowed by the system Hamiltonian, but also
carry out Hartree-Fock energetic calculations to obtain 
the quantum phase diagrams using realistic system 
Hamiltonians. Among the spontaneous symmetry breakings predicted in this
paper are, in additional to the usual translational and spin symmetries, 
the spontaneous breaking of discrete parity symmetry in a number of
interesting situations.
 
Very recently, magnetoresistance anisotropy 
was observed in both doped GaAs/AlGaAs \cite{pan01} and Si/SiGe 
\cite{zeitler01} based semiconductor quantum well systems at even 
integer filling factors, when a strong in-plane
magnetic field ($B_\|$), in addition to the 
perpendicular magnetic field ($B_\perp$) producing the 2D Landau levels, 
is applied. In Refs. \cite{pan01,zeitler01}, it is proposed 
that the strong in-plane 
magnetic field increases the electron Zeeman energy so much that
the highest filled Landau level (with spin down) has a level crossing with the
lowest empty Landau level (with spin up), and then a 
charge density wave (CDW) phase may be
stabilized by electron Coulomb interaction.
Our recent work \cite{wang02} on the magnetoplasmon energy dispersion
of a wide well system in the presence of a strong in-plane magnetic field
finds a near mode softening in the spin-flip channel, suggesting the
possibility of a
spin density wave (SDW) instability near the degeneracy point. As a result,
we proposed a spin-charge texture (skyrmion \cite{sondhi92}) stripe phase  
\cite{eugene01,wang02} to explain the magnetoresistance anisotropy
observed in the experiments \cite{pan01,zeitler01}. 
In an earlier theoretical work,
Brey \cite{brey91_cdw} also found a charge density wave
instability in a wide well at $\nu=1$ {\it without} any 
in-plane magnetic field.
Murthy \cite{murthy00} recently considered the the coexistence of quantum Hall
order and density wave order in a single well at $\nu=2$ with zero Zeeman
energy and strong level mixing. 
To the best of our knowledge, however, systematic theoretical analyses
of these candidate stripe phases at integer QH systems 
have not yet been carried out and need to be further developed.

A double well (bilayer) quantum Hall system is another 
system where one may see a stripe
phase order at integral filling factors. At total filling factor $\nu=4N\pm 1$,
electrons are equally distributed in the two layers
if no gate voltage or in-plane magnetic field is applied. 
We note that the bilayer systems of filling factor $\nu=4N-1$ ($N>0$)
can be understood to be equivalent to $\nu=4(N-1)+1=4N-3$ system by
electron-hole symmetry in the top filled Landau levels, if the 
Landau level mixing is negligible. Therefore
all of our results shown in this paper for bilayer $\nu=4N+1$ quantum Hall
systems can be applied to $\nu=4(N_+1)-1=4N-3$ system as well. 
We will not distinguish
these two systems and will mention $\nu=4N+1$ case only in the rest of
this paper. When the layer
separation, $d$, is small compared to the 2D magnetic length, $l_0$, 
spontaneous interlayer coherence can be generated 
by interlayer Coulomb interaction
even in the absence of any interlayer tunneling 
\cite{wen92,reviewbook}, 
and the ground state is therefore best
understood as a Halperin (1,1,1) coherent phase with finite charge gap
\cite{halperin83}.
When $d$ is comparable to or larger than $l_0$, however, 
there should be a transition to two 
decoupled compressible $\nu=2N+1/2$ systems or one of the competing many-body
phases. A detailed analysis of all
competing phases is still lacking, although it is commonly believed that 
for $N=0$ there is a direct first order transition from the
$(1,1,1)$ phase to the compressible states, although for $N>0$ there may be 
intermediate quantum Hall phases with stripe order \cite{brey00}.
Other more exotic phases, including
the ones with electron pairing \cite{kim02}, have also been proposed in the 
literature.
This interlayer coherence and the corresponding Goldstone mode
of the symmetry-broken phases have been extensively
studied theoretically \cite{yang94,yang95,stern01} as well as
experimentally \cite{murphy94,spielman01} in the literature. 
In addition to the layer separation, the in-plane magnetic field is
another important controlling parameters for the double well
QH systems. In the presence of an in-plane magnetic 
field, an Aharonov-Bohm phase factor, associated with
the gauge invariance, has to be considered for the electron tunneling
amplitude between the two layers and 
may cause a commensurate-to-incommensurate (C-I)
phase transition at a critical magnetic field strength 
\cite{yang94,hanna01,nomura02}. 
Such C-I phase transition arises as a result of competition of the 
interlayer tunneling and Coulomb interlayer exchange interaction.
The former favors a commensurate phase, in which the phase of the order
parameter winds at a rate fixed by the parallel magnetic field;
whereas the latter favors an incommensurate phase, in which 
the order parameter is
more nearly uniform in space \cite{reviewbook}.
Taking into account the stripe order associated with the layer 
separation ($d$) and
the spiral order associated with the in-plane magnetic field ($B_\|$)
the possibility exists for a very complex and rich phase diagram with
many competing phases occurring for various values of $d$ and $B_\|$
in the bilayer double well system at $\nu=4N+1$.
Following our earlier work based on the collective mode dispersion
\cite{eugene01,wang02},
in this paper we carry out the ground state energetic calculation within
the Hartree-Fock approximation to obtain and describe
these exotic phases, which
break isospin rotation, parity, and/or translational symmetries. 
We will also investigate the 
possibility of stripe formation at the even filling factor, $\nu=4N+2$,
where a nonstripe canted antiferromagnetic phase 
(CAF) has been proposed and extensively studied in the literature
\cite{canted_phase,bilayer_sym_break,bilayer_sym_break2,burkov02,caf_exp1,caf_exp2,yang99_effect_D2,anna,papa02}.
The goal of our work is to discuss possible symmetry broken exotic
quantum phases in bilayer and wide-well quantum Hall systems.

In this paper, we use the idea of isospin to label the two (nearly)
degenerate energy levels at the Fermi energy, and construct 
trial many-body wavefunctions
that have various kinds of (possibly exotic) isospin order. 
The relevant isospin quantum number
can be associated with Landau level index, layer index, spin index,
or any other good quantum numbers describing the corresponding 
noninteracting system. Within 
Hartree-Fock approximation we investigate phases with isospin spiral and/or
stripe orders and discuss their relevance for various systems.
In Table \ref{system_notation} we summarize the definition of the isospin
up(down) components for each system individually. 
For the convenience of discussion, we use following notations and 
labels to denote the systems in the rest of this paper
(see also Fig. \ref{level_fig}):
$W1(W2)$ and $W1'(W2')$ denote the intersubband and intrasubband level
crossing (or level near degeneracy in $W1'$ case) 
in a wide well system with filling factor 
$\nu=2N+1(2N+2)$ respectively;
$D1$ and $D2$ are for the level crossing (or level near degeneracy in 
$D1$ case) in a double well system at
total filling factor, $\nu=4N+1$ and $4N+2$, respectively.
(Note that the distinction between intrasubband and intersubband level 
crossing of a wide well system
in the presence of an in-plane
magnetic field is somewhat ambiguous and will be clarified
in the next section.)
By "level crossing" we mean a point in the parameter space of 
noninteracting electrons, for which the energy difference between
the highest filled Landau level and the lowest empty Landau level
vanishes, while by ``level near degeneracy'' we mean a region in the
parameter space, in which the energy difference
between the above noninteracting electron levels is not zero, 
but relatively small compared to
the interlevel interaction energy, which can strongly mix the two close 
noninteracting Landau levels in certain situations. 
In this paper,
we consider the following six quantum phases generated by our trial
wavefunction: conventional
incompressible integer QH liquid (i.e. isospin polarized), 
isospin coherent, isospin spiral,
isospin coherent stripe, isospin spiral stripe and isospin 
skyrmion stripe phases, classified by the behaviors of the expectation
values of their isospin
components, $\langle {\cal I}_\pm\rangle$ and $\langle {\cal I}_z\rangle$.
More precise definitions and the related physical properties of these phases
will be discussed in Section \ref{trial_wavefunction}.
In the last two columns of Table \ref{system_notation} we list 
the exotic many-body phases that become the ground states near the
appropriate degeneracy points and the associated broken symmetries.
For a wide well system at $\nu=2N+1$ near the intersubband level crossing
(i.e. $W1$ in small in-plane field region), we find that interactions
stabilize the isospin coherent phase, which breaks the parity (space 
inversion) symmetry and can be understood as a quantum Hall
ferroelectric state \cite{ferro}.
For the same system near the intrasubband level 
near degeneracy region (i.e. $W1'$ in large in-plane field region),
we obtain the isospin skyrmion stripe phase, which breaks both parity 
and translational symmetries of the system.
For wide well systems at even filling factors we find that many-body 
states with isospin coherence and/or stripe order are not favored
at either the intersubband ($W2$) or the intrasubband
($W2'$) degenerate points within the HF approximation, exhibiting instead
a trivial first order phase transition at the
level crossing points. However, we find the calculated HF 
energy difference between the uniform ground state and
the phases of spiral and stripe orders to be rather small,
indicating the possibility of a spontaneous symmetry breaking in
more refined approximations. Therefore,
based on our HF calculation results,
we suggest the existence of skyrmion stripe phase, 
breaking both spin rotational and translational symmetries near the
degenerate point of $W2'$ systems, may be responsible for
the resistance anisotropy observed recently in Ref. \cite{pan01}.
(As for the magnetoresistance anisotropy observed in Si/SiGe 
semiconductor 
\cite{zeitler01}, the complication of valley degeneracy
in Si makes a straightforward comparison between our theory and
the experiments difficult.) 
For a double well system at $\nu=4N+1$ ($D1$), 
several interesting phases can be stabilized in the parameter 
range of interest (see Table \ref{system_notation}), while
only a commensurate CAF phase is stabilized for even filling factor,
$\nu=4N+2$ ($D2$), being consistent with the earlier results
\cite{burkov02}.
All of our calculation are carried out at zero temperature without
any disorder or impurity scattering effects.
We suggest that these symmetry broken phases discussed in this paper should be 
observable in transport and inelastic light scattering experiments.

We note that the presence of an in-plane magnetic field $B_\|$
(in the 2D plane) automatically introduced a tilted magnetic field
of amplitude $B_{tot}=\sqrt{B_\|^2+B_\perp^2}$ at an
angle of $\tan^{-1}(B_\|/B_\perp$) tilted with respect to the 
normal vector of the 2D plane since the field $B_\perp$, perpendicular
to the 2D plane, is always present in order to produce the 2D Landau level
system. The presence of $B_\|$ (or more generally, the tilted field)
introduces qualitatively new physics to 2D quantum Hall systems by 
allowing a tuning parameter (i.e. $B_\|$) without affecting the basic 
Landau level structure. Changing $B_\|$ in a continuous manner may 
enable the 2D system to undergo various quantum phase transition 
which would not otherwise be possible. Thus the physics of quantum 
Hall systems becomes considerably richer in the presence of a 
continuously tunable parallel (or tilted) magnetic field. The 
general (and somewhat ambitious) goal of our theoretical work 
present in this paper is to study this rich quantum phase diagram 
of 2D quantum Hall systems as a function of the in-plane field 
(used as a tunning parameter). As such we concentrate in the two 
most promising candidate systems, namely the (single) wide quantum 
well system and the (bilayer) double quantum well system, where 
the tilted field configuration is expected to give rise to exotic 
symmetry-broken quantum orders in the quantum Hall ground states. 
The underlying idea is to use the in-plane magnetic field as a 
tuning parameter to cause level crossing (or almost level degeneracy) 
in the noninteracting system (around the Fermi level) to see if
the interaction effects then drive the system to non-trivial 
symmetry-broken many-body ground states (particularly with exotic 
quantum order) which can be experimentally studied. Earlier 
examples of such exotic symmetry-broken ground states (arising 
from level crossing phenomena) include, in addition to our 
prediction \cite{wang02} of a skyrmion stripe phase, the bilayer 
canted antiferromagnetic phase predicted in Ref. \cite{canted_phase} 
where a tilted field is not necessary. 

This paper is organized as follows: in Section
\ref{hamiltonian} we introduce the Hamiltonians of interest, for both  
wide well and double well systems in the presence of a finite in-plane
magnetic field.
In Section \ref{trial_wavefunction} we propose various trial many-body
wavefunctions incorporating both isospin stripe and
isospin spiral orders. 
Physical features of the six typical phases generated by the
wavefunctions will also be discussed in detail. For the systems of 
even filling factors, we propose trial wavefunctions that involve
the four Landau levels closest to the Fermi energy simultaneously. 
(Including four rather than two levels in many-body wavefunctions was shown
to be crucial for establishing the many-body canted antiferromagnetic state
in bilayer systems at $\nu=2$ \cite{canted_phase}.)
We will then divide the Hartree-Fock variational energies and 
the related numerical results into the following
four sections: wide well systems at $\nu=2N+1$ (Section \ref{sec_W1}) and
at $\nu=2N+2$ (Section \ref{sec_W2}); double well systems at total
filling factor $\nu=4N+1$ (Section \ref{sec_D1}) and at 
$\nu=4N+2$ (Section \ref{sec_D2}). For the sake of convenience 
(and relative independence of different sections), results of each
section are discussed independently and then compared with 
each other in Section
\ref{discussion}, where we also make connections 
to the earlier works in the literature.
Finally we summarize our paper in Section \ref{summary}, and discuss 
some open questions.

%
\section{Hamiltonians}
\label{hamiltonian}

In this section, we present the Hamiltonians of
the systems we will study in this paper, including both the single 
wide (parabolic) well systems and the double (thin) well (or bilayer) 
systems. Most of the formulae given in this section exist in the literature
(perhaps scattered over many publications) 
and hence we will not derive them in details unless absolutely necessary.
To make the notations consistent throughout this paper,
we use the superscript $W$ to denote physical variables or quantities 
for the wide well system and the superscript $D$ for
the double well systems. 
We will also describe the symmetry properties of these systems, which 
are crucial in formulating the many-body trial wavefunctions in the
next section and in discussing the nature of the spontaneous
symmetry breaking and the associated exotic quantum order in our
various proposed phases.
 
\subsection{Wide Well System}
\label{hamiltonian_W}

For a wide well system 
we consider a parabolic confinement potential \cite{wang02} in the growth 
direction ($z$),
$U_c(z)=\frac{1}{2}m^\ast\omega_{0}^2z^2$,
where $m^\ast$ is electron effective mass and $\omega_{0}$ is the confinement
energy. The advantage of using a parabolic well model is that we can
easily diagonalize the noninteracting Hamiltonian even
in the presence of in-plane magnetic field. 
We will use eigenstates of the non-interacting Hamiltonian as
the basis functions for writing variational Slater determinant states for
the many-body wavefunctions that may be stablized by Coulomb interactions.
We believe that our wide well results obtained in this paper 
for parabolic confinement should be
qualitatively valid even for nonparabolic quantum wells.
To incorporate both the in-plane magnetic 
field ($B_\|$), which we take to be along the $x$ axis in this section, 
and the perpendicular (along the $z$ axis) magnetic field
($B_\perp$) in the Hamiltonian, we choose two kinds of Landau gauges
for the vector potential: 
in one of them, $\vec{A}_{[y]}(\vec{r}\,)=(0,B_\bot x-B_\| z,0)$, 
particle momentum is conserved along $y$ direction, 
and in the other one, $\vec{A}_{[x]}(\vec{r}\,)=(-B_\perp y,-B_\| z,0)$,
particle momentum is conserved along $x$ direction.
The final physical results are of course 
independent of the choice of the gauge. 
However, as will be clear from the discussion below, various phases
may be more conveniently discussed in different gauges, since the
physical or the mathematical description of particular many-body 
states may be more natural in specific gquges.
For the consistency of notations,
all the explicit calculations presented in this paper will be done in the 
gauge $\vec{A}_{[y]}(\vec{r}\,)$. Since the generalization of these
equations to the 
other case is straightforward (see Appendix \ref{diff_direction}), 
we will only present the final results for the $\vec{A}_{[x]}(\vec{r}\,)$
gauge. The noninteracting (single electron)
Hamiltonian in the parabolic potential with both
perpendicular and in-plane magnetic fields in gauge 
$\vec{A}_{[y]}(\vec{r}\,)$ is (we choose $\hbar=1$ throughout this paper)
\begin{eqnarray}
{H}^W_0
&=&\frac{p_x^2}{2m^\ast}+\frac{1}{2m^\ast}\left(p_y+\frac{eB_\perp x}{c}
-\frac{eB_\| z}{c}\right)^2+\frac{p_z^2}{2m^*}
+\frac{1}{2}m^* \omega_{0}^2 z^2 -\omega_z S_z,
\label{H0_W_1}
\end{eqnarray}
where $S_z$ is the $z$ component of spin operator and $\omega_z$ 
is Zeeman energy, proportional to the total
magnetic field, $B_{tot}=\sqrt{B_\bot^2+B_\|^2}$.
It has been shown \cite{wang02} that the one electron
energies and wavefunctions of Eq. (\ref{H0_W_1})
can be obtained analytically even in the presence of an in-plane magnetic 
field by rotating to a proper coordinate.
The resulting noninteracting Hamiltonian is 
a sum of two decoupled one-dimensional 
simple harmonic oscillators with energies \cite{wang02},
\begin{eqnarray}
\omega^2_{1,2}&=&\frac{1}{2}\left[\left(\omega_b^2+\omega_\bot^2\right)\pm
\sqrt{\left(\omega_b^2-\omega_\bot^2\right)^2+4\omega_\bot^2\omega_\|^2}
\,\right],
\label{omega12}
\end{eqnarray}
where $\omega_{\bot,\|}=eB_{\bot,\|}/m^\ast c$ (i.e. $\omega_\bot$ is the
conventional cyclotron frequency), and $\omega_b\equiv\sqrt{\omega_{0}^2+
\omega_\|^2}$ is the effective confinement energy.
Using $(\vec{n},s,k)$ as the noninteracting eigenstate quantum numbers, where
$\vec{n}=(n,n')$ is the orbital Landau level index
for the two decoupled 1D simple harmonic oscillators,
$s=\pm 1/2$ the electron spin eigenvalues along the total magnetic
field, $\vec{B}_{tot}\equiv (B_\|,0,B_\perp)$ and $k$ 
the guiding center coordinate,
one can obtain the following noninteracting eigenenergies, 
$E^{0,W}_{\vec{n},s}$, for $H_0^W$ of Eq. (\ref{H0_W_1}):
\begin{equation}
E^{0,W}_{\vec{n},s}=\omega_1\left(n+\frac{1}{2}\right)+
\omega_2\left(n'+\frac{1}{2}\right)-\omega_z s,
\label{energy_levels_eqn}
\end{equation}
and the noninteracting orbital wavefunction:
\begin{eqnarray}
\phi^{W}_{\vec{n},s,k}(\vec{r})&=&
\frac{e^{iky}}{\sqrt{L_y}}
\underbrace{{\psi}^{(1)}_{n}(\cos\theta(x+l_0^2k)-\sin\theta z)
\cdot{\psi}^{(2)}_{n'}(\sin\theta(x+l_0^2k)+\cos\theta z)}_
{\displaystyle \equiv\Phi^W_{\vec{n}}(x+l_0^2k,z)},
\label{wavefunctions_W_y}
\end{eqnarray}
where $L_y$ is the system length in $y$ direction and
the function $\Phi^0_{\vec{n}}(x+l_0^2k,z)$ has $x$ and $z$ components only;
$\tan(2\theta)\equiv-2\omega_\perp\omega_\|/(\omega_b^2-\omega_\perp^2)$.
In Eq. (\ref{wavefunctions_W_y}), the
function ${\psi}^{(i)}_{n}(x)$ is
\begin{eqnarray}
{\psi}^{(i)}_{n}(x)&=&\frac{1}{\sqrt{\pi^{1/2}2^nn!l_i}}
\exp\left[-\frac{x^2}{2l_i^2}\right]
H_n\left(\frac{x}{l_i}\right),
\label{wf_i}
\end{eqnarray}
with $ l_i \equiv \sqrt{1/m^*\omega_i}$ for $ i=1,2$,
and $H_n(x)$ is Hermite polynomial.

\begin{figure}
 \vbox to 8cm {\vss\hbox to 5.5cm
  {\hss\
      {\includegraphics{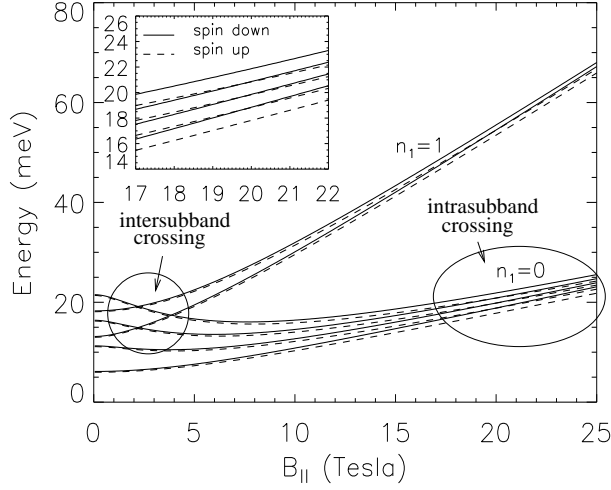}
  }
  \hss}
  }
					      
\caption{
Noninteracting Landau level energy spectra 
of a parabolic quantum well
as a function of the parallel (in-plane) magnetic field, $B_\|$. 
We choose following system parameters: $B_\perp=3$ T ($\omega_\perp=5.2$ meV), 
$\omega_0=7$ meV, and $|g|=0.44$ for GaAs material.
$n_1$ is the first Landau level index of $\vec{n}$. 
Regions of intersubband and intrasubband level crossings are indicated
by circles (see text).
}
\label{energy_levels_figure}
\end{figure}

A typical noninteracting
energy spectrum as a function of in-plane magnetic field, $B_\|$, 
is shown in Fig. \ref{energy_levels_figure} (with system parameters
similar to the experimental values of Ref. \cite{pan01}).
In Fig, \ref{energy_levels_figure}, two kinds of level crossing
can be observed, one
is in the weak $B_\|$ ($<5$ Tesla) region and the other in the 
strong $B_\|$ ($>19$ Tesla) region. In this paper we will denote the former
to be an "intersubband level crossing"
and the latter to be
an "intrasubband level crossing" (see Table \ref{system_notation})
\cite{subband_note}. 
In this paper we adopt a convention that all single particle states in 
Eqs. (\ref{energy_levels_eqn}) and (\ref{wavefunctions_W_y}) that
have the same quantum number $n$ (the first index of $\vec{n}$) correspond
to the same subband.
With this definition of ``subbands'', the characteristic energy separation
of intersubband levels, $\omega_1$, is always larger than the energy separation
of intrasubband levels, $\omega_2$.
It is useful to emphasize that the states of Eqs. 
(\ref{energy_levels_eqn})-(\ref{wf_i})
are the exact noninteracting eigenenergies for {\it arbitrary}
values of in-plane field, perpendicular field, and confinement energy.
In the second quantization notation, the noninteracting
Hamiltonian can be written to be
\begin{eqnarray}
H^{W}_0&=&\sum_{\vec{n},s,k}E^{0,W}_{\vec{n},s}
c^{W,\dagger}_{\vec{n},s,k}c^{W}_{\vec{n},s,k},
\label{H0_W}
\end{eqnarray}
where $c^{W,\dagger}_{\vec{n},s,k}$($c^{W}_{\vec{n},s,k}$) 
creates(annihilates) an electron in the state $(\vec{n},s,k)$.

Before showing the full many-body interaction Hamiltonian, it is 
convenient to define the form functions \cite{wang02},
\begin{eqnarray}
A^{W}
_{\vec{n}_i\vec{n}_j}(\vec{q}\,)
&\equiv&\int d\vec{r}\,e^{-i\vec{q}\cdot\vec{r}}
{\phi^{W}_{\vec{n}_i,s,-q_y/2}}^\dagger(\vec{r}\,)
{\phi^{W}_{\vec{n}_j,s,q_y/2}}(\vec{r}\,)
\nonumber\\
&=&\int dx\int dz\,e^{-iq_xx}e^{-iq_zz}
\tilde{\Phi}^W_{\vec{n}_1}(x-l_0^2q_y/2,z)
\tilde{\Phi}^W_{\vec{n}_2}(x+l_0^2q_y/2,z),
\label{A_def_W}
\end{eqnarray}
which are constructed from the noninteracting electron wavefunctions 
(Appendix \ref{A_function}).
In principle one could use the electron wavefunctions
obtained either from a self-consistent
Hartree-Fock approximation \cite{wang02,brey91_cdw} or
from a self-consistent local
density approximation \cite{brey89_sdw,dempsey93} in Eq. (\ref{A_def_W})
to calculate the form function. However,
in Ref. \cite{wang02} we have shown that the difference in the
form function between using
the noninteracting single electron wavefunctions and the self-consistent
Hartree-Fock wavefunctions is very small. 
Therefore we define the form
function by using the noninteracting wavefunctions 
and consider only the diagonal
part of the Hartree-Fock potential (i.e. the first order Hartree-Fock
approximation) in calculating the ground state energy.
In addition, we include the screening effect of positively charged donors
by using the statically screened Coulomb interaction, $V(\vec{q})=
e^2/\epsilon_0 (|\vec{q}|^2+(2\pi/{l}_s)^2)$ for convenience, where 
${l}_s$ is the
effective screening length by donors outside the well and $\epsilon_0=12.7$
is the static lattice dielectric constant of GaAs.
We believe this simple approximation should give reasonable
results compared with the more complicated calculation by including
the donor density in a self-consistent approximation 
\cite{brey89_sdw,dempsey93}.
Using the notations defined above, the interaction Hamiltonian can 
be expressed as follows \cite{wang02}: 
\begin{eqnarray}
H^W_1&=&\frac{1}{2\Omega}\sum_{\vec{n}_1,\cdots,\vec{n}_4}
\sum_{\sigma_1,\sigma_2}\sum_{k_1,k_2,\vec{q}}
V^W_{\vec{n}_1\vec{n}_2,\vec{n}_3\vec{n}_4}(\vec{q})\,
e^{-iq_x(k_1-k_2)l_0^2}
c^{W,\dagger}_{\vec{n}_1,\sigma_1,k_1+q_y/2}
c^W_{\vec{n}_2,\sigma_1,k_1-q_y/2}
c^{W,\dagger}_{\vec{n}_3,\sigma_2,k_2-q_y/2}
c^W_{\vec{n}_4,\sigma_2,k_2+q_y/2} \nonumber\\
&=&\frac{1}{2\Omega_\perp}\sum_{\vec{n}_1,\cdots,\vec{n}_4}
\sum_{\sigma_1,\sigma_2}\sum_{k_1,k_2,\vec{q}_\perp}
\tilde{V}^W_{\vec{n}_1\vec{n}_2,\vec{n}_3\vec{n}_4}(\vec{q}_\perp)\,
e^{-iq_x(k_1-k_2)l_0^2}
c^{W,\dagger}_{\vec{n}_1,\sigma_1,k_1+q_y/2}
c^W_{\vec{n}_2,\sigma_1,k_1-q_y/2}
c^{W,\dagger}_{\vec{n}_3,\sigma_2,k_2-q_y/2}
c^W_{\vec{n}_4,\sigma_2,k_2+q_y/2},
\label{H1_W}
\end{eqnarray}
where $V^W_{\vec{n}_1\vec{n}_2,\vec{n}_3\vec{n}_4}(\vec{q}\,)
\equiv V(\vec{q}\,)A^{W}_{\vec{n}_1\vec{n}_2}(-\vec{q}\,)
A^{W}_{\vec{n}_3\vec{n}_4}(\vec{q}\,)$ and
$\tilde{V}^W_{\vec{n}_1\vec{n}_2,\vec{n}_3\vec{n}_4}(\vec{q}_\perp)\equiv
L_z^{-1}\sum_{q_z}V^W_{\vec{n}_1\vec{n}_2,\vec{n}_3\vec{n}_4}(\vec{q})$.
$\Omega=L_xL_yL_z$ is the usual normalization volume and 
$\Omega_\perp=L_xL_y$ is the 2D normalization area.
If we choose the alternate Landau gauge, 
$\vec{A}_{[x]}(\vec{r}\,)$, the expression of $H^W_1$ will have the
phase factor, $e^{-iq_y(k_1-k_2)l_0^2}$, rather than
$e^{iq_x(k_1-k_2)l_0^2}$ above.

For the convenience of later discussion, we can express the Hamiltonians
shown in Eq. (\ref{H0_W}) and Eq. (\ref{H1_W}) as follows:
\begin{eqnarray}
H^{W}_{tot}&=&\sum_{\vec{n},\sigma}E^{0,W}_{\vec{n},\sigma}
\rho^W_{\vec{n}\sigma,\vec{n}\sigma}(0)
+\frac{1}{2\Omega_\perp}\sum_{\vec{q}_\perp}\sum_{\vec{n}_1,\cdots,\vec{n}_4}
\sum_{\sigma_1,\sigma_2}
\tilde{V}^W_{\vec{n}_1\vec{n}_2,\vec{n}_3\vec{n}_4}(\vec{q}_\perp)\,
\rho^W_{\vec{n}_1\sigma_1,\vec{n}_2\sigma_1}(-\vec{q}_\perp)
\rho^W_{\vec{n}_3\sigma_2,\vec{n}_4\sigma_2}(\vec{q}_\perp),
\label{H_W_rho}
\end{eqnarray}
where 
\begin{eqnarray}
\rho^W_{\vec{n}_i\sigma_1,\vec{n}_j\sigma_2}(\vec{q}_\perp)
\equiv \sum_k e^{iq_xkl_0^2}c^{W,\dagger}_{\vec{n}_i,\sigma_1,k-q_y/2}
c^{W}_{\vec{n}_j,\sigma_2,k+q_y/2}
\label{widewell_rho}
\end{eqnarray}
is the density operator for the wide well system.

\subsection{Double Well System}
\label{Hamiltonian_D}

For a double well (bilayer) system, we assume both wells are of zero width 
in their growth direction so that the in-plane magnetic field 
does not change the electron orbital
wavefunctions as in wide well systems. 
For most bilayer problems of physical interest, neglecting the
indivitual layer width is an extremely reasonable approximation.
In the presence of an in-plane (parallel) magnetic field, 
the tunneling amplitude in the Hamiltonian
acquires an Aharonov-Bohm phase factor to satisfy the gauge invariance
of the whole system \cite{reviewbook,burkov02,yang95}.
This, in fact, is the main effect of the in-plane field in the double well
system for our purpose.
To describe this important effect, we first express the Hamiltonians
of a double well system in a special gauge \cite{yang94}:
$\vec{A}_{[y]}'(\vec{r}\,)=(0,B_\perp x, B_x y- B_y x)$, where 
$B_x$ and $B_y$ are the $x$ and $y$ components of the in-plane field
(i.e. $\vec{B}_\|=(B_x,B_y,0)$), by using a conventional 
basis of electron states, $(n,l,s,k)$, where $n$ 
is Landau level index in a single layer, 
$l=\pm$ is the layer index for right(left) layer
(sometimes also called up/down layers in analogy with electron spin), and 
$s$ and $k$ are the same as before.
The Aharonov-Bohm phase factor, 
$\exp[-i\Phi_0^{-1}\int_{\pm d/2}^{\mp d/2} 
{A}_{[y],z}'(\vec{r}_\perp,z) dz]$
(where $d$ is layer separation and $\Phi_0=hc/e$ is fundamental 
flux quantum), 
is then introduced for electron tunneling
from one layer to the other.
In the bilayer Hamiltonian 
we need to keep the tunneling term only for the
electrons in the highest filled Landau level, and therefore
the noninteracting Hamiltonian in the second quantization 
representation becomes:
\begin{eqnarray}
{H}^D_0&=&\sum_{n,l,k,s}\left[\left(n+\frac{1}{2}\right)\omega_\perp
-\omega_z s\right] c^{D,\dagger}_{n,l,s,k}c^{D}_{n,l,s,k}
\nonumber\\
&&-t_{N,P}\sum_{s,k}
\left[e^{-ikP_yl_0^2}c^{D,\dagger}_{N,+1,s,k-P_x/2}c^{D}_{N,-1,s,k+P_x/2}
+e^{ikP_yl_0^2}c^{D,\dagger}_{N,-1,s,k+P_x/2}
c^{D}_{N,+1,s,k-P_x/2}\right],
\label{H0_D}
\end{eqnarray}
where $c^{D,\dagger}_{n,l,s,k}(c^{D}_{n,l,s,k})$ is the electron
creation(annihilation) operator of state $(n,l,s,k)$, described by
$L_y^{-1/2}e^{iky}\psi^{(0)}_{n}(x+kl_0^2)\delta(z\mp d/2)$, where
$\psi^{(0)}_{n}(x)$ is the same as Eq. (\ref{wf_i}) with $l_i$ replaced
by magnetic length $l_0$.
$\vec{P}_\perp=(P_x,P_y)\equiv 2\pi d \vec{B}_\|/\Phi_0$ is 
a characteristic wavevector introduced by the in-plane field.
The effective tunneling amplitude $t_{N,P}$ is
$t\,e^{-P^2l_0^2/4}L_N^0(P^2l_0^2/2)$ with $t$ being a 
parameter for interlayer tunneling, where
$P=|\vec{P}_\perp|$ and $L_n^0(x)$ is the generalized Laguerre polynomial.
Following the convention in most of the existing literature,
we will treat $t_{N,P}$ as a whole to be an independent variable in this paper
and hence neglect all its level and magnetic field dependence for 
simplicity \cite{cote02,footnote_t}. 

Similar to Eq. (\ref{A_def_W}) we define the following form function
for the double well system:
\begin{eqnarray}
A^{D}
_{{n}_i{n}_j}(\vec{q}_\perp)&=&\int d\vec{r}_\perp
e^{-i\vec{q}_\perp\cdot\vec{r}_\perp}
{\phi^{D,\dagger}_{{n}_i,-q_y/2}}(\vec{r}_\perp)
{\phi^{D}_{{n}_j,q_y/2}}(\vec{r}_\perp),
\label{A_def_D}
\end{eqnarray}
where ${\phi^{D}_{{n},k}}(\vec{r}_\perp)=
L_y^{-1/2}e^{iky}\psi^{(0)}_{n}(x+kl_0^2)$ is the one electron wavefunction
of Landau level $n$ in a single well. 
The $z$ component of the wavefunction can be integrated first and
the resulting explicit formula of
$A^{D}_{{n}_i{n}_j}(\vec{q}_\perp)$ is 
shown in Appendix \ref{A_function}.
The interaction Hamiltonian in this basis is then
\begin{eqnarray}
H^D_1&=&\frac{1}{2\Omega_\perp}
\sum_{n_1\cdots n_4}\sum_{l_1,l_2}\sum_{s_1,s_2}\sum_{k_1,k_2,\vec{q}_\perp}
V^{D,l_1,l_2}_{n_1n_2,n_3n_4}(\vec{q}_\perp)\,e^{-iq_x(k_1-k_2)l_0^2}
c^{D,\dagger}_{n_1,l_1,s_1,k_1+q_y/2}c^D_{n_2,l_1,s_1,k_1-q_y/2}
c^{D,\dagger}_{n_3,l_2,s_2,k_2-q_y/2}c^D_{n_4,l_2,s_2,k_2+q_y/2}
\nonumber\\
&=&\frac{1}{2\Omega_\perp}\sum_{\vec{q}_\perp}\sum_{l_1,l_2}
V_{n_1n_2,n_3n_4}^{D,l_1l_2}
(\vec{q}_\perp)\rho^{D}_{n_1 l_1, n_2 l_1}(-\vec{q}_\perp)
{\rho}^{D,}_{n_3 l_2,n_4 l_2}(\vec{q}_\perp)
\label{H1_D}
\end{eqnarray}
where $V^{D,l_1,l_2}_{n_1n_2,n_3n_4}(\vec{q}_\perp)=V(\vec{q}_\perp)A^{D}
_{n_1n_2}(-\vec{q}_\perp)A^{D}_{n_3n_4}(\vec{q}_\perp)
\exp(-\frac{1}{2}|l_1-l_2|d\,|\vec{q}_\perp|)$, 
including the contribution of the $z$ 
component of the electron wavefunction; ${\rho}^{D}_{n_1l, n_2l}
(\vec{q}_\perp)=\sum_{s,k}e^{iq_xkl_0^2}
c^{D,\dagger}_{n_1,l,s,k-q_y/2}c^{D}_{n_2,l,s,k+q_y/2}$ is the electron 
density operator of layer $l$ and spin $s$.

Sometimes it is more convenient to rewrite the Hamiltonian in terms
of the eigenstates of the noninteracting Hamiltonian, which
are defined as:
\begin{eqnarray}
{c}_{n,+1,s,k}^\dagger&=&\frac{e^{i(k+P_x/2)P_yl_0^2/2}}{\sqrt{2}}
\left(a_{n,+1,s,k+P_x/2}^\dagger+a_{n,-1,s,k+P_x/2}^\dagger\right)
\nonumber\\
{c}_{n,-1,s,k}^\dagger&=&\frac{e^{-i(k-P_x/2)P_yl_0^2/2}}{\sqrt{2}}
\left(a_{n,+1,s,k-P_x/2}^\dagger-a_{n,-1,s,k-P_x/2}^\dagger\right),
\label{change_basis}
\end{eqnarray}
where $a^\dagger_{n,\alpha,s,k}$ is the new electrons creation operator
of a symmetric ($\alpha=+1$) or an antisymmetric ($\alpha=-1$) 
noninteracting state,
because they are also the eigenstates of parity (space inversion) tranformation
($c^\dagger_{n,l,s,k}\to c^\dagger_{n,-l,s,-k}$).
Note that the description in terms of the left/right layer
index $l$ (in Eq. (\ref{H0_D})) or the symmetric/antisymmetric state
$\alpha$ are completely equivalent since they are simple linear combinations
of each other (cf. Eq. (\ref{change_basis})).

In the $\alpha$ basis, the noninteracting 
Hamiltonian of Eq. (\ref{H0_D}) is
\begin{eqnarray}
{H}^D_0&=&
\sum_{n,\alpha,k,s}E^{0,D}_{n,\alpha,s}a^\dagger_{n,\alpha,s,k}
a^{}_{n,\alpha,s,k},
\label{H0_Da}
\end{eqnarray}
where the noninteracting electron eigenenergy, $E^{0,D}_{n,\alpha,s}$, is
\begin{eqnarray}
E^{0,D}_{n,\alpha,s}=(n+1/2)\omega_\perp
-\alpha\Delta_{SAS}/2-s\omega_z,
\label{E0_Da}
\end{eqnarray}
and $\Delta_{SAS}=2t$ is the tunneling energy.
The interacting Hamiltonian in Eq. (\ref{H1_D}) then becomes
\begin{eqnarray}
{H}^D_1&=&\frac{1}{2\Omega_\perp}\sum_{\alpha1,\alpha_2,\vec{q}_\perp}
\sum_{n_1\cdots n_4}\left[\tilde{V}_{n_1n_2,n_3n_4}^I(\vec{q}_\perp)
{\rho}^{\alpha_1\alpha_1}_{n_1n_2}(-\vec{q}_\perp)
{\rho}^{\alpha_2\alpha_2}_{n_3n_4}(\vec{q}_\perp)
+\tilde{V}_{n_1n_2,n_3n_4}^o(\vec{q}_\perp)
{\rho}^{-\alpha_1\alpha_1}_{n_1n_2}(-\vec{q}_\perp)
{\rho}^{-\alpha_2\alpha_2}_{n_3n_4}(\vec{q}_\perp)
\right.\nonumber\\
&&\hspace{1.5cm}\left.
+\tilde{V}_{n_1n_2,n_3n_4}^s(\vec{q}_\perp)\sum_{\lambda=\pm 1}\lambda
\rho^{-\lambda\alpha_1\alpha_1}_{n_1n_2}(-\vec{q}_\perp)
\rho^{\lambda\alpha_2\alpha_2}_{n_3n_4}(\vec{q}_\perp)\right],
\label{H1_Da}
\end{eqnarray}
where the new density operator, $\rho^{\alpha_1\alpha_2}_{n_1n_2}
(\vec{q}_\perp)=
\sum_{s,k}e^{iq_xkl_0^2}a^\dagger_{n_1,\alpha_1,s,k-q_y/2}
a^{}_{n_2,\alpha_2,s,k+q_y/2}$ and 
the three interaction matrix elements in Eq. (\ref{H1_Da}) 
are as follows: 
\begin{eqnarray}
\tilde{V}_{n_1n_2,n_3n_4}^{I,o}(\vec{q}_\perp)&\equiv&\frac{1}{2}
\left(\tilde{V}_{n_1n_2,n_3n_4}^{D,++}(\vec{q}_\perp)\pm
\cos((\vec{q}_\perp\cdot\vec{P}_\perp)l_0^2)\tilde{V}_{n_1n_2,n_3n_4}^{D,+-}
(\vec{q}_\perp)\right)
\\
\tilde{V}_{n_1n_2,n_3n_4}^s(\vec{q}_\perp)&\equiv&\frac{i}{2}
\sin((\vec{q}_\perp\cdot\vec{P}_\perp)l_0^2)\tilde{V}_{n_1n_2,n_3n_4}
^{D,+-}(\vec{q}_\perp).
\label{V_I_o_s}
\end{eqnarray}
As will become clear from the later discussion in this paper 
the basis described by
Eqs. (\ref{H0_Da})-(\ref{V_I_o_s}))
is often more convenient for
constructing trial wavefunctions, especially for an even filling system
in the presence of in-plane magnetic field.

\subsection{Symmetry properties of the systems}
\label{symmetry}

We now discuss the symmetry properties of the wide well systems
and the double well systems in the presence of in-plane field.
We first consider parity (full space inversion) symmetry, translational 
symmetry and spin rotational symmetry individually, which 
exist in both the systems of interest (i.e. wide well and double well).
Then we discuss the $z$-parity (reflection about $x-y$ plane) 
and the isospin rotation symmetries, which exist only 
in the double well system in certain situations. 

In this paper,
we define the parity operator, $\hat{\cal P}$, so that it 
reverses the {\it full} space 
coordinates from $\vec{r}=(x,y,z)$ to $-\vec{r}=(-x,-y,-z)$.
Therefore, in the second quantization notation, we have
\begin{eqnarray}
\hat{\cal P}c^W_{\vec{n},s,k}\hat{\cal P}^{-1}=(-1)^{n+n'}c^W_{\vec{n},s,-k}
\label{parity_W}
\end{eqnarray}
for the wide well system (see Eq. (\ref{wavefunctions_W_y})), 
and 
\begin{eqnarray}
\hat{\cal P}c^D_{n,l,s,k}\hat{\cal P}^{-1}
=(-1)^n c^D_{n,-l,s,-k}
\mbox{    or    }
\hat{\cal P}a_{n,\alpha,s,k}\hat{\cal P}^{-1}
=(-1)^n \alpha a_{n,\alpha,s,-k}
\label{parity_D}
\end{eqnarray}
for the double well system in the two different basis.
It is easy to see that the Hamiltonians of these two 
systems shown above 
are not changed under the parity transformation by using the identities
(see Eq. (\ref{A_explicit})):
$A^W_{\vec{n}_1\vec{n}_2}(-\vec{q}\,)=(-1)^{n_1+n_1'}(-1)^{n_2+n_2'}
A^W_{\vec{n}_1\vec{n}_2}(\vec{q}\,)$ and $A^D_{n_1n_2}(-\vec{q}_\perp)
=(-1)^{n_1+n_2}A^D_{n_1n_2}(\vec{q}_\perp)$.
This simple result is valid even in the presence 
of the tilted magnetic field, if only we consider
a symmetrically confined (but not necessary parabolic) potential 
well in the wide well system and no external bias voltage in
the double well systems.
Thus, parity is rather general symmetry property of the physical systems 
under consideration.

To consider the 2D translational symmetry in the $x-y$ plane,
it is more convenient to consider the many-body system in 
the first quantization representation, and introduce
the total ``momentum'' operator, 
$\hat{\cal M}$ \cite{kallin,wang02}:
\begin{eqnarray}
\hat{\cal M} \equiv \sum_i\left[\vec{p}_{i}+
\frac{e}{c}\vec{A}(\vec{r}_i)-\frac{e}{c}\vec{B}_{tot}\times\vec{r}_i\right],
\label{momentum_m}
\end{eqnarray}
where $\vec{p}_{i}$ and 
$\vec{r}_i$ are the momentum and position operators of the $i$th electron.
$\vec{A}$ is the vector potential and 
$B_{tot}=(B_\|,0,B_\perp)$ is the total magnetic field.
Then it is straightforward to see that the $x$ and $y$ components of 
$\hat{\vec{\cal M}}$ commute with the Hamiltonian,
$H=\frac{1}{2m^\ast}\sum_i(\vec{p}_i+e\vec{A}(\vec{r}_i)/c)^2
+\sum_i U_c(z_i)+\frac{1}{2}\sum_{i\neq j}V(\vec{r}_i-\vec{r}_j)$.
This result is true even
in the presence of an in-plane magnetic field and for arbitrary shape of
the electron confinement potential in $z$ direction. Therefore it applies
equally well to wide well systems and double well systems.
We can then define 
the 2D translation operators in the $x$ and $y$ directions as
\begin{eqnarray}
\hat{\cal T}_x(R_x)&\equiv&
\exp\left(-i{R}_x\hat{\cal M}_x\right),
\label{translation_x}
\\
\hat{\cal T}_y({R}_y)&\equiv&
\exp\left(-iR_y\hat{\cal M}_y\right),
\label{translation_y}
\end{eqnarray}
which form closed translational symmetry group individually
\cite{footnote_group} (they do not commute with each other
due to the presence of the magnetic field). 
We first study how $\hat{\cal T}_{x,y}(R_{x,y})$
can transform a noninteracting electron eigenstate
of a wide well system, 
$\phi^W_{\vec{m},k}(\vec{r}\,)=L_y^{-1/2}e^{iky}
\Phi^W_{\vec{m}}(x+l_0^2k,z)$,
with a shift, $\vec{R}_\perp=(R_x,R_y)$, in the $x-y$ plane.
In the Landau gauge, $\vec{A}_{[y]}(\vec{r}\,)$), we obtain
\begin{eqnarray}
\hat{\cal T}_x(R_x)\phi^W_{\vec{m},k}(\vec{r}\,)
&=& e^{-iR_xy/l_0^2}\phi^W_{\vec{m},k}(x-R_x,y,z)
=\phi^W_{\vec{m},k-R_x/l_0^2}(\vec{r}\,),
\label{T_x}
\\
\hat{\cal T}_y(R_y)\phi^W_{\vec{m},k}(\vec{r}\,)
&=&\phi^W_{\vec{m},k}(x,y-R_y,z)
=e^{-ikR_y}\phi^W_{\vec{m},k}(\vec{r}\,).
\label{T_y}
\end{eqnarray}
In the second quantization representation, it is equivalent to
\begin{eqnarray}
\hat{\cal T}_x(R_x)c^{W}_{\vec{m},s,k}\hat{\cal T}_x(R_x)^{-1}
=c^{W}_{\vec{m},s,k+R_x/l_0^2},
\label{translation_cx}
\\
\hat{\cal T}_y(R_y)c^{W}_{\vec{m},s,k}\hat{\cal T}_y(R_y)^{-1}
=e^{-ikR_y}c^{W}_{\vec{m},s,k}.
\label{translation_cy}
\end{eqnarray}
In other words, $\hat{\cal T}_x(R_x)$ shifts the guiding center coordinate
and $\hat{\cal T}_y(R_y)$ adds an additional phase factor of the
electron operators respectively.
As discussed in Section \ref{Hamiltonian_D}, in the case of 
double well systems we use the gauge $\vec{A}_{[y]}'(\vec{r}\,)
=(0,B_\perp x, B_x y- B_y x)$ (see the paragraph above Eq. (\ref{H0_D})).
Therefore the ``momentum'' operator in this gauge becomes
$\hat{\cal M}=(p_x+y/l_0^2-zP_y/d,p_y+zP_x/d,p_z)$ and the translational
operators are
\begin{eqnarray}
\hat{\cal T}_x(R_x)\phi^D_{n,k}(\vec{r}_\perp)\delta(z-ld/2)
&=&e^{ilR_xP_y/2}\phi^D_{n,k-R_x/l_0^2}(\vec{r}_\perp)\delta(z-ld/2),
\label{T_x2}
\\
\hat{\cal T}_y(R_y)\phi^D_{n,k}(\vec{r}_\perp)\delta(z-ld/2)
&=&e^{-ilR_yP_x/2}e^{-ikR_y}\phi^D_{n,k}(\vec{r}_\perp)\delta(z-ld/2).
\label{T_y2}
\end{eqnarray}
In the second quantization representation, they are  equivalent to
\begin{eqnarray}
\hat{\cal T}_x(R_x)c^{D}_{n,l,s,k}\hat{\cal T}_x(R_x)^{-1}
=e^{ilR_xP_y/2}c^{D}_{n,l,s,k+R_x/l_0^2},
\label{translation_cx2}
\\
\hat{\cal T}_y(R_y)c^{D}_{n,l,s,k}\hat{\cal T}_y(R_y)^{-1}
=e^{-ilR_yP_x/2}e^{-ikR_y}c^{D}_{n,l,s,k}.
\label{translation_cy2}
\end{eqnarray}
Comparing Eqs. (\ref{translation_cx2})-(\ref{translation_cy2}) to
Eqs. (\ref{translation_cx})-(\ref{translation_cy}), we find 
that an additional
phase factor appears in the layer index basis, which is related to
the Aharonov-Bohm phase factor and hence confirms that
the tunneling Hamiltnonain in Eq. (\ref{H0_D}) commutes 
with the translational
operator, ${\cal T}_{x,y}$. If we use 
Eq. (\ref{change_basis}) to change the
above results (Eqs. (\ref{translation_cx2})-(\ref{translation_cy2})) to the
symmetric-antisymmetric basis, we obtain 
\begin{eqnarray}
\hat{\cal T}_x(R_x)a_{n,\alpha,s,k}\hat{\cal T}_x(R_x)^{-1}
=a_{n,\alpha,s,k+R_x/l_0^2},
\label{translation_ax}
\\
\hat{\cal T}_y(R_y)a_{n,\alpha,s,k}\hat{\cal T}_y(R_y)^{-1}
=e^{-ikR_y} a_{n,\alpha,s,k},
\label{translation_ay}
\end{eqnarray}
which are the same
as Eqs. (\ref{translation_cx})-(\ref{translation_cy}) obtained by replacing 
$c^{W}_{\vec{m},s,k}$ by $a^{}_{n,\alpha,s,k}$.

As for the spin rotational symmetry about the total magetic field
direction with an angle $\chi$, we can simply apply the following operator
on the spin wavefunction,
\begin{eqnarray}
\hat{\cal U}(\chi)=e^{-i\chi S_z},
\label{spin_rotation}
\end{eqnarray}
where $S_z$ is the $z$ component of spin operator with the spin
$z$ axis being along the direction of total magnetic field.
Since our noninteracting eigenstate is always the eigenstate of $S_z$,
the spin rotational operator, $\hat{\cal U}(\chi)$, just gives an additional
prefactor, $e^{\mp i\chi/2}$ for spin up(down) electron annihilation
operators. The spin rotational symmetry is  
conserved by the Hamiltonians that we consider.

Now we discuss the symmetry properties that exist only in the double
well system but {\it not} in the wide well system. First,
if there is no in-plane magnetic field ($B_\|=0$), the double 
well system has
two separate parity symmetries: one is the $z$-parity symmetry,
$\hat{\cal P}_z$ (changing
$z$ to $-z$) and the other one is the in-plane parity symmetry,
$\hat{\cal P}_{xy}$ (changing $(x,y)$ to $(-x,-y)$).
In the second quantization representation, we have
$\hat{\cal P}_z c^D_{n,l,s,k}\hat{\cal P}_z^{-1}=c^D_{n,-l,s,k}$
and $\hat{\cal P}_{xy} c^D_{n,l,s,k}\hat{\cal P}_{xy}^{-1}=
(-1)^nc^D_{n,l,s,-k}$. It is easy to show that
both $\hat{\cal P}_z$ and $\hat{\cal P}_{xy}$ commute with
the total Hamiltonian of the double well system in the absence of
in-plane magnetic field. When an in-plane magnetic field is applied,
however, none of them is conserved due to the Aharonov-Bohm   
phase shift in the tunneling amplitude (see Eq. (\ref{H0_D})), while
their product, the full space inversion $\hat{\cal P}=\hat{\cal P}_z
\hat{\cal P}_{xy}$, is still conserved as discussed above.
Second, if there is no electron interlayer tunneling
($\Delta_{SAS}=0$) in the double well system, 
the electron number in each layer is also a 
constant of motion. In the isospin language of the layer index basis,
such conservation is described by an isospin rotational
symmetry about the isospin $z$ axis, $\hat{\cal U}_{iso}(\xi)
\equiv \exp[-i\xi(c^{D,\dagger}_{n,+,s,k}c^{D}_{n,+,s,k}
-c^{D,\dagger}_{n,-,s,k}c^{D}_{n,-,s,k})/2]$.
The spontaneous interlayer coherence 
will break such continuous symmetry and give rise to a
Goldstone mode. Such $U(1)$ symmetry breaking
has been extensively studied in the literature (see Ref. 
\cite{reviewbook} and references therein), and hence
we will not seriously address this issue in this paper.
However, since our interest in this paper is to study the charge
and spin density wave phases induced by the in-plane magnetic field,
we will always consider systems with finite tunneling and 
finite in-plane magnetic field throughout, so that 
the above unique symmetries ($\hat{\cal P}_z$, $\hat{\cal P}_{xy}$,
and $\hat{\cal U}_{iso}$) actually do not exist 
in the double well systems we consider in this paper. As a result, both
the wide well systems and the double well systems we study will
have full space parity symmetry ($\hat{\cal P}$), two-dimensional 
translational symmetry ($\hat{\cal T}_{x,y}$), and 
spin rotational symmetry about the total magnetic field ($\hat{\cal U}$). 
We will then discuss how a stablized many-body coherent wavefunction can break
these symmetries near the level crossing (or level near degeneracy) regions.

\section{Variational Wave Functions}
\label{trial_wavefunction}

In this section, we propose trial many-body wavefunctions to variationally
minimize the Hartree-Fock energy of the integer quantum Hall systems
in the level degeneracy region: a wide parabolic well and a double 
well system at both even and odd filling factors. 
Stripe phases, that we are primarily interested in, have translational
symmetry along the longitudinal direction of the stripes. They are
conveniently discussed in the Landau gauge, for which particle momentum
is a good quantum number in the direction of the stripe. 
In this paper we only consider stripes that are parallel or 
perpendicular to the direction of the in-plane magnetic field. Therefore
for the in-plane field in the $x$ direction we choose gauges
$\vec{A}_{[y]}(\vec{r}\,)=(0,B_\perp x-B_\| z,0)$ and 
$\vec{A}_{[x]}(\vec{r}\,)=(-B_\perp y, -B_\|z, 0)$ to describe
phases with stripes along $y$ and $x$ axes respectively. By comparing
the HF energies of these two stripes phases
(one is along $y$ direction obtained by using gauge 
$\vec{A}_{[y]}(\vec{r}\,)$, and the other one is along $x$ direction obtained
by using gauge $\vec{A}_{[x]}(\vec{r}\,)$) and 
the HF energies of other nonstripe
phases (discussed later), we can determine if a stripe phase can 
be stabilized and which direction is energetically more favorable
for stripe formation.
If no stripe phase is energetically favorable, the nonstripe phases 
obtained in these two gauges are identical (as they should be due to 
symmetry). In the rest of this paper,
for the sake of brevity, we will show equations
and formulae for the trial wavefunction and the related HF energy of a wide
well system only for stripes along $y$ axis (perpendicular to
the in-plane field) in the gauge $\vec{A}_{[y]}(\vec{r}\,)$, 
although (we emphasize that) we always consider both stripe directions
in our variational calculations.

For a zero width double well (bilayer) system, 
the trial wavefunction for a stripe phase
along the in-plane field direction ($x$) can be studied in a much easier way.
Using the fact that the in-plane magnetic field does not
change the electron orbital wavefunctions in the zero width wells,
we can simply rotate the in-plane magnetic field, 
$\vec{B}_\|$, from the direction along $x$ axis to the direction
along $y$, keeping all the trial wavefunction obtained by the conventional
gauge, $\vec{A}_{[y]}(\vec{r}\,)$, the same. We then equivalently obtain
the HF energies of a stripe phase with a conserved momentum either 
perpendicular or parallel to the in-plane magnetic field.
Their energy difference results from the Aharonov-Bohm phase factor 
in the tunneling amplitude, and that is the reason we keep both $P_x$ and 
$P_y$ in the bilayer Hamiltonian in Eq. (\ref{H0_D}).

In the rest of this section, we first show the trial 
wavefunction for the two level crossing (or level near degeneracy) situations 
in Section \ref{trial_wf_2l}.
We also consider several special many-body phases
it generates and study their physical properties.
In Section \ref{trial_wf_4l} we propose wavefunctions 
for the four level degeneracy, involving the second highest
filled level and the second lowest empty level about the Fermi energy
for the double well system at $\nu=4N+2$
and for the wide well system at $\nu=2N+2$, leaving out all
other lower filled levels and higher empty levels as irrelevant core states. 
In Section
\ref{pert_stripe} we develop a perturbation technique to investigate the 
possible instability toward the stripe formation from
a uniform coherent phase.
We can use this method to study the existence of a stripe phase
and its basic properties without
optimizing the whole HF energy if
the stripe formation is a second order phase transition.

\subsection{Two level degeneracy: variational wavefunctions
for odd filling factors}
\label{trial_wf_2l}

\subsubsection{Wavefunction}

We propose the following trial many-body
wavefunction to incorporate both isospin
stripe and isospin spiral orders simultaneously:
\begin{eqnarray}
|\Psi_G(\psi_k,\vec{Q}_\perp,\gamma)\rangle 
&=& \prod_k\tilde{c}^\dagger_{1,k}|LL\rangle
\nonumber\\
\left[\begin{array}{c}
       \tilde{c}_{1,k}^\dagger \\ \tilde{c}_{2,k}^\dagger
      \end{array}\right]
&=&\left[\begin{array}{cc}
       e^{ikQ_xl_0^2/2+i\gamma/2}\cos(\psi_k/2)& 
            e^{-ikQ_xl_0^2/2-i\gamma/2}\sin(\psi_k/2) \\
      -e^{ikQ_xl_0^2/2+i\gamma/2}\sin(\psi_k/2)& 
            e^{-ikQ_xl_0^2/2-i\gamma/2}\cos(\psi_k/2)
       \end{array}\right]
\cdot
\left[\begin{array}{c}
       c_{\Uparrow,k-Q_y/2}^\dagger \\ c_{\Downarrow,k+Q_y/2}^\dagger
      \end{array}\right],
\label{wavefunction_general}
\end{eqnarray}
where we use $\Uparrow (\Downarrow)$ to denote the isospin up(down) state,
which can represent subband level, spin and/or layer indices depending
on the systems we are considering (see Table \ref{system_notation}).
The spiral winding wavevector, $\vec{Q}_\perp$, 
stripe phase function, $\psi_k$, and the additional phase, $\gamma$,
are the variational parameters to minimize the total Hartree-Fock energy.
If the system has isospin rotation symmetry around the 
isospin $z$ axis (e.g. double well systems at $\nu=4N+1$ in the absence
of tunneling), 
the new ground state energy obtained by the
trial wavefunction above will be independent of $\gamma$.
If the system has no such isospin rotation symmetry
(e.g. double well systems at $\nu=4N+1$ in the presence of tunneling or
the wide well systems at $\nu=2N+1$), the new ground state
energy will then depend on $\gamma$, selecting some specific values
of $\gamma$ to minimize the total variational energy. We will discuss both
of there cases in details in our HF analysis in Sections 
\ref{sec_W1}-\ref{sec_D2}.
In general this two level coherence approximation should be  
good for a system near the degeneracy point, especially for an odd filling
factor. For systems with even filling factor, the inclusion of the
next nearest two levels about Fermi energy becomes crucial as will be
discussed later.

For the convenience of later discussion, 
we define following functions:
\begin{eqnarray}
\Theta_1(q_x)&\equiv&\frac{1}{N_\phi}
\sum_k e^{ikq_xl_0^2}\cos^2(\psi_{k}/2) \nonumber\\
\Theta_2(q_x)&\equiv&\frac{1}{N_\phi}
\sum_k e^{ikq_xl_0^2}\sin^2(\psi_{k}/2)
=\delta_{q_x,0}-\Theta_1(q_x) \nonumber\\
\Theta_{3}(q_x)&\equiv&\frac{1}{N_\phi}
\sum_k e^{ikq_xl_0^2}\sin(\psi_{k}/2)\cos(\psi_{k}/2),
\label{def_theta}
\end{eqnarray}
Note that when $\psi_k=\psi_0$ as a constant,
$\Theta_i(q_x)=\Theta_i(0)\delta_{q_x,0}$ (for $i=1,2,3$) 
and $\Theta_{3}(0)^2=\Theta_1(0)
\Theta_2(0)$. The physical meaning of these functions are the following: 
$\Theta_1(0)$ describes the density of isospin-up electrons, 
$\Theta_2(0)$ describes the density of isospin-down electrons, and
$\Theta_{3}(0)$ 
measures the coherence between isospin up and isospin down electrons.
For the periodic function, $\psi_k$, without any loss of generality,
we can assume that it is a real and even function 
of the guiding center coordinate $k$ so that
$\Theta_i(q_x)$ are all real quantities. 

\subsubsection{Isospin phases and their physical properties}
\label{diff_phases}

To understand the physical properties of the trial wavefunction proposed in
Eq. (\ref{wavefunction_general}), we first define the
following generalized density operators
(isospin index, $I=\pm 1=\Uparrow(\Downarrow)$):
\begin{eqnarray}
\rho^{}_{I_1I_2}(\vec{q}_\perp)&\equiv&\sum_k e^{ikq_xl_0^2}
c_{I_1,k-q_y/2}^\dagger c^{}_{I_2,k+q_y/2},
\label{rho_I}
\end{eqnarray}
which generalizes 
Eqs. (\ref{widewell_rho}), (\ref{H1_D}) and (\ref{H1_Da})).
The isospin operator at the in-plane
momentum, $\vec{q}_\perp$, can be defined as
\begin{eqnarray}
\vec{\cal I}(\vec{q}_\perp)&\equiv&\frac{1}{2}\int dz
\int d\vec{r}_\perp\,e^{-i\vec{q}_\perp\cdot\vec{r}_\perp}
\sum_{I_1,I_2}\sum_{k_1,k_2}{c}_{I_1,k_1}^{\dagger}
\vec{\sigma}^{}_{I_1,I_2} c^{}_{I_2,k_2}
{\phi}^{}_{I_1,k_1}
(\vec{r}\,)^\ast {\phi}^{}_{I_2,k_2}(\vec{r}\,)
\nonumber\\
&=&\sum_{I_1,I_2}\vec{\sigma}^{}_{I_1,I_2}\rho^{}_{I_1I_2}(\vec{q}_\perp)
{A}_{I_1 I_2}(\vec{q}_\perp,0),
\label{S_def}
\end{eqnarray}
where $\vec{\sigma}^{}_{I_1,I_2}$ are the Pauli matrix elements, and
$A^{}_{I_1 I_2}(\vec{q}\,)$ is the generalized form function as 
has been specifically defined in Eqs. (\ref{A_def_W}) and (\ref{A_def_D})
for wide well systems and double well systems respectively.

Using the trial wavefunction, $|\Psi_G\rangle$, in 
Eq. (\ref{wavefunction_general}),
we obtain the following expectation value of isospin components:
\begin{eqnarray}
\langle {\vec I}_z(\vec{q}\,)\rangle &=& \frac{1}{2}\left[
A_{\Uparrow\Uparrow}(\vec{q}_\perp,0)\langle\Psi_G| 
\rho_{\Uparrow\Uparrow}(\vec{q}_\perp) |\Psi_G\rangle
- A_{\Downarrow\Downarrow}(\vec{q}_\perp,0)\langle\Psi_G|
\rho_{\Downarrow\Downarrow}(\vec{q}_\perp)|\Psi_G\rangle
\right]
\nonumber\\
&=&\frac{1}{2}N_\phi\delta_{q_y,0}\left[
A_{\Uparrow\Uparrow}(\vec{q}_\perp,0)e^{-iq_xQ_yl_0^2/2}\Theta_1(q_x)
- A_{\Downarrow\Downarrow}(\vec{q}_\perp,0)e^{iq_xQ_yl_0^2/2}\Theta_2(q_x)
\right]
\label{I_z}
\end{eqnarray}
and
\begin{eqnarray}
\langle {\cal I}_+(\vec{q}_\perp)\rangle&=&
A_{\Uparrow\Downarrow}(\vec{q}_\perp,0)
\langle\Psi_G|\rho_{\Uparrow\Downarrow}(\vec{q}_\perp)|\Psi_G\rangle
=N_\phi \delta_{q_y,Q_y} A_{\Uparrow\Downarrow}(\vec{q}_\perp,0)
\Theta_{3}(q_x-Q_x)\, e^{-i\gamma}.
\label{I+}
\end{eqnarray}
There are two classes of interesting phases one can find from
Eqs. (\ref{I_z}) and (\ref{I+}). First, if
$\psi_k=\psi_0$ is a constant, the $z$ component of
the mean value of isospin, $\langle {\cal I}_z(\vec{q}_\perp)\rangle
\propto\delta_{q_x,0}
\delta_{q_y,0}$, is uniform over the 2D well plane,
while its transverse components, 
$\langle {\cal I}_\pm(\vec{q}_\perp)\rangle\propto\delta_{q_x,Q_x}
\delta_{q_y,Q_y}$, select a certain wavevector, $\vec{Q}_\perp=(Q_x,Q_y)$, 
in the guiding center coordinate for winding, i.e. the isospin $x$ and 
$y$ components oscillate
in the real space at wavevector, $\vec{Q}_\perp$ (see Fig. 
\ref{isospin_angle}(a)). 
Therefore we define 
$\vec{Q}_\perp$ as the wavevector of isospin 
spiral order. Secondly 
if $\psi_k$ is a periodic function of $k$,
the isospin polarization, $\langle {\cal I}_z(\vec{r}\perp)\rangle$, 
will oscillate 
along $x$ direction (but uniform in $y$ direction)
according to Eq. (\ref{I_z}), characterizing a stripe phase with normal 
vector $\hat{n}\|\hat{x}$.
Choosing the other gauge, $\vec{A}_{[x]}(\vec{r}\,)$, in which the particle
momentum is conserved along $x$ direction,
we can construct a stripe phase along $x$ direction and
isospin $\langle {\cal I}_z(\vec{r}\perp)\rangle$ then 
modulates in $y$ direction
as shown in Appendix \ref{diff_direction}.

It is also interesting to investigate the
local charge density distribution in the top Landau level, 
$\rho^{}_{local}(\vec{r}_\perp)$, by using 
the trial wavefunction, $|\Psi_G\rangle$, in Eq. (\ref{wavefunction_general}).
It has contributions from both the isospin up and isospin down 
electrons:
\begin{eqnarray}
\rho^{}_{local}(\vec{r}_\perp)&=&\frac{1}{\Omega_\perp}\sum_{\vec{q}_\perp}
e^{i\vec{q}_\perp\cdot\vec{r}_\perp}
\langle\Psi_G| A^{}_{\Uparrow\Uparrow}(\vec{q}_\perp,0)
\rho^{}_{\Uparrow\Uparrow}(\vec{q}_\perp)+
A^{}_{\Downarrow\Downarrow}(\vec{q}_\perp,0)
\rho^{}_{\Downarrow\Downarrow}(\vec{q}_\perp)|\Psi_G\rangle
\nonumber\\
&=&\frac{1}{2\pi l_0^2}-\frac{1}{2\pi l_0^2}\sum_{q_x}\Theta_2(q_x)
\left[A^{}_{\Uparrow\Uparrow}(q_x,0,0)\cos(q_x(x-Q_yl_0^2/2))
-A^{}_{\Downarrow\Downarrow}(q_x,0,0)\cos(q_x(x+Q_yl_0^2/2))\right],
\nonumber\\
&\equiv&\rho_0+\rho_{ex}(\vec{r}_\perp)
\label{rho_local}
\end{eqnarray}
where 
$N_\phi/\Omega_\perp=(2\pi l_0^2)^{-1}$ is the average electron density 
in each Landau level.
Eq. (\ref{rho_local}) shows that if $\Theta_2(q_x)$ selects a specific
wavevector (i.e. $\psi_k$ oscillates periodically with a characteristic
wavevector) at $q_x=q_n=2\pi n/a$, where $a$ is the period of stripe, 
the extra charge density, $\rho_{ex}(\vec{r}_\perp)$, will be nonzero and
a periodic function in the real space.
In Appendix \ref{topo_charge} we will show that 
this extra change density is related to the
charge density induced by the topological isospin density
by generalizing the theory of skyrmion excitations 
developed in Ref. \cite{sondhi92}
for a double well system.
Therefore in the rest of this paper we will
denote such a CDW state as a "skyrmion stripe" phase.
In general we should have skyrmion stripe phases whenever the stripe normal 
vector $\hat{n}$ (i.e. the modulation direction of $\langle {\cal I}_z\rangle$)
is perpendicular to the spiral wavevector $\vec{Q}_\perp$ \cite{eugene01}.

\begin{figure}

 \vbox to 6cm {\vss\hbox to 5.cm
 {\hss\
   {\includegraphics{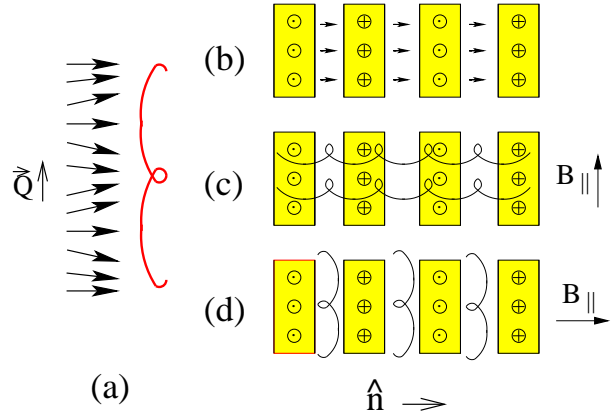}
   }
  \hss}
 }
\caption{
(a) A cartoon for the isospin spiral structure. 
Horizontal arrows denote the isospin
direction and the spiral curve indicate the transverse isospin 
$({\cal I}_x,{\cal I}_y)$ order parameter. 
(b), (c) and (d) are the isospin coherent stripe phase,
isospin spiral stripe phase, and isospin skyrmion stripe phase respectively.
Shaded areas show isospin up ($\odot$)
and down ($\oplus$) domains, and arrows in the right hand side
show the directions of $B_\parallel$. Note that the isospin spiral
direction is always perpendicular to $B_\parallel$.
$\hat{n}$ is the normal vector of the stripes, which denotes the direction
of isospin $\langle {\cal I}_z \rangle$ modulation.
For isospin coherent stripe in (b), there is no spiral order and therefore,
in general, the stripe can have arbitrary direction with respect 
to the in-plane field. 
} 
\label{stripe_fig}
\end{figure}

According to the above analysis, we can 
consider the following six different phases
obtained from Eq. (\ref{wavefunction_general}):
when $\psi_k$ is a constant, the wavefunction of Eq.
(\ref{wavefunction_general}) can describe three nonstripe
phases: (i) fully (un)polarized uniform quantum Hall
phase for $\psi_k=(0)\pi$ (with $\vec{Q}_\perp$ being arbitrary), (ii)
coherent phase for $\vec{Q}_\perp=0$ and $\psi_k\neq 0,\pi$, and
(iii) spiral phase for finite $\vec{Q}_\perp$ and $\psi_k\neq 0,\pi$.
When $\psi_k$ changes periodically with $k$, three kinds of stripe
phases arises: (i) when $\vec{Q}_\perp$ is perpendicular to the stripe
normal direction, we have a skyrmion stripe described above;
(ii) when $\vec{Q}_\perp$ is parallel to the stripe normal direction, 
we have a spiral stripe phase, and (iii)
when $\vec{Q}_\perp=0$ we have a coherent stripe phase. 
For simplicity, in this paper we will only consider stripe and spiral orders
(if they exist) with characteristic 
wavevector $\hat{n}$ and $\vec{Q}_\perp$ being 
either perpendicular or parallel to the 
in-plane magnetic field, which is fixed to be in the positive ${x}$ axis.
A simple visualized cartoon of the spiral winding and the three
stripe phases is shown in Fig. \ref{stripe_fig}.
We calculate the HF energy given by the six different phases generated from
the trial wavefunction, Eq. (\ref{wavefunction_general}), and compare them
to get the minimum energy for the true ground state. 

\subsubsection{Symmetry properties of the trial wavefunction}
\label{symmetry_wavefunction}

In Section \ref{symmetry} we have shown that the systems we are considering
have three kinds of symmetries: parity (full space inversion) symmetry, 
two-dimensional translational symmetry, and spin rotational symmetry. Here 
we show how a general isospin trial wavefunction in 
Eq. (\ref{wavefunction_general}) is transformed under these symmetry 
operators defined in Section \ref{symmetry}. This will enable us
to identify the broken symmetries in each of the isospin phases discussed
above.

First it is easy to see that the conventional integer QH state (i.e. 
isospin polarized with $\psi_k=0$ or $\pi$), 
$|\Psi_0\rangle=\prod_k c^\dagger_{\Uparrow(\Downarrow),k}|LL\rangle$,
is an eigenstate of all the three symmetry operators ($\hat{\cal P}$, 
$\hat{\cal T}_{x(y)}$, and $\hat{\cal U}$), by applying the 
equations in Section \ref{symmetry} directly. (For simplicity, in our
discussion below, the isospin basis of double well system at $\nu=4N+1$
is chosen to be the symmetric-antisymmetric basis, i.e. the noninteracting
eigenstate basis. The symmetry properties of the isospin coherent phases
constructed in the layer index basis will be discussed later
in Section \ref{symmetry_CI}). Therefore, the isospin polarized state,
$|\Psi_0\rangle$, does not break any symmetry properties as expected.

Now we consider a general many-body state described by the wavefunction
Eq. (\ref{wavefunction_general}) with variational parameters 
obtained from minimizing the HF energy: $\psi_k=\psi_k^\ast\neq 0, \pi$;
$\vec{Q}_\perp=\vec{Q}_\perp^\ast$, and $\gamma=\gamma^\ast$.
Applying the parity operator, $\hat{\cal P}$, translation
operator, $\hat{\cal T}_{x(y)}$, and spin rotation operator, $\hat{\cal U}$,
on $|\Psi_G(\psi_0^\ast,\vec{Q}_\perp^\ast,
\gamma^\ast)\rangle$, we can obtain respectively
\begin{eqnarray}
\hat{\cal P}|\Psi_G(\psi_k^\ast,\vec{Q}_\perp^\ast,
\gamma^\ast)\rangle &=& 
(-i)^{N_\phi}|\Psi_G(\psi_{-k}^\ast,-\vec{Q}_\perp^\ast,
\gamma^\ast+\pi)\rangle,
\label{uniform_parity}
\\
\hat{\cal T}_\alpha (R_\alpha)|\Psi_G(\psi_k^\ast,
\vec{Q}_\perp^\ast,\gamma^\ast)\rangle &=& 
|\Psi_G(\psi_{k-\hat{n}\cdot\vec{R}_\perp/l_0^2}^\ast,
\vec{Q}_\perp^\ast,
\gamma^\ast-Q_\alpha R_\alpha)\rangle,
\label{uniform_translation}
\\
\hat{\cal U}(\chi)|\Psi_G(\psi_k^\ast,\vec{Q}_\perp^\ast,
\gamma^\ast)\rangle &=& 
|\Psi_G(\psi_k^\ast,\vec{Q}_\perp^\ast,
\gamma^\ast-\chi(s_1-s_2))\rangle,
\label{uniform_spin}
\end{eqnarray} 
where $\alpha=x$ or $y$, $\vec{R}_\perp=(R_x,R_y)$, 
$s_{1(2)}=\pm 1/2$ is the spin quantum number of 
isospin up(down) state,
and $\hat{n}$ is the stripe oscillation direction (it is perpendicular
to the direction of stripes, e.g. $\hat{n}=\hat{x}$ 
for stripes along $y$ direction described by 
the Landau gauge, $\vec{A}_{[y]}(\vec{r}\,)$).
In Eq. (\ref{uniform_parity}) we 
have assumed that the signs for the isospin up state
and isospin down state are opposite after parity operation
(we will show that this is always true in the level 
crossings considered in this paper). According to Eqs. 
(\ref{uniform_parity})-(\ref{uniform_spin}), we find that 
only parity symmetry is broken
in a coherent phase ($\psi_k^\ast=\psi_0^\ast\neq 0,2\pi$ and 
$\vec{Q}_\perp^\ast=0$), if $s_1=s_2$. Spin rotation symmetry is also 
broken if the spin quantum number of the two crossing
levels (isospin up and down) are different ($s_1\neq s_2$), which is true
only for the level crossings in the even filling systems.
When we consider the spiral phase with $\psi^\ast_k=\psi_0^\ast\neq 0,\pi$
and $\vec{Q}_\perp^\ast\neq 0$, we find that in addition to the broken 
symmetries discussed above (parity symmetry and spin rotational symmetry
if $s_1\neq s_2$), it breaks translational symmetry in the direction
of isospin winding (i.e. along $\vec{Q}_\perp^\ast$). 
From Eq. (\ref{uniform_translation}) we immediately see that the
wavefunction has a period $2\pi/|\vec{Q}_\perp^\ast|$.
For the coherent stripe phase ($\psi_k^\ast$ modulated periodically and
$\vec{Q}_\perp^\ast=0$) we have broken parity, spin rotational
symmetry if $s_1\neq s_2$, and translational symmetry in the direction
of $\hat{n}$.
For a spiral stripe, the spiral winding direction is parallel
to the stripe oscillation direction 
($\vec{Q}_\perp^\ast\|\hat{n}$), and therefore the translational
symmetry is broken only in one direction, while a skyrmion
stripe ($\vec{Q}_\perp^\ast\perp\hat{n}$) 
breaks translational symmetries in both $x$ and $y$
directions. Both the spiral stripe and the skyrmion stripe also
break the  parity symmetry and spin rotational symmetry if $s_1\neq s_2$.

A fundamental quantum mechanical principle stipulates that
when a system
undergoes a quantum phase transition to break a symmetry (i.e. the
ground state wavefunction is not an eigenstate of the symmetry
operator), the new symmetry-broken ground state will have additional
degeneracy associated with the spontaneously broken symmetry. This result
is also obtained in our HF energy calculation shown later.
In Table \ref{system_notation} we list the many-body states
obtained by our HF variational calculation and the resulting 
broken symmetries. We will discuss each of them individually
in the following sections for different systems and then
compare the results with each other in Section \ref{discussion}.

\subsection{Four levels near a degeneracy: variational wavefunctions
for even filling factors}
\label{trial_wf_4l}

\subsubsection{Wavefunction for double well systems}
\label{4l_D}

\begin{figure}

 \vbox to 4cm {\vss\hbox to 5.cm
 {\hss\
   {\includegraphics{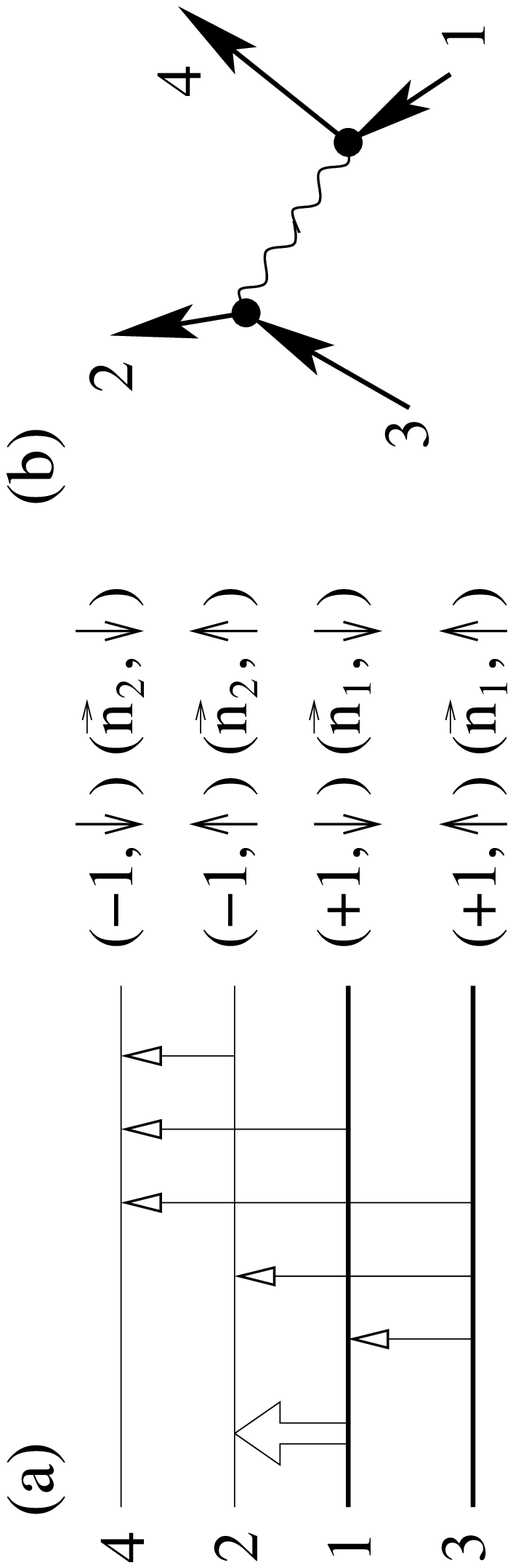}
   }
  \hss}
 }
\caption{
(a) A schematic noninteracting energy configuration
for a general four level
degeneracy of an even filling system.
Level 1 is the highest filled level,
level 2 is the lowest empty level, level 3 is the second highest level,
and level 4 is the second lowest empty level as defined in
Section \ref{trial_wf_4l}.
$(\alpha,s)$, in the right hand side
denotes the quantum numbers of each level in the double well system
($\alpha$ is the parity quantum number and $s$ is the electron spin), 
while $(\vec{n}_i,s)$ is the quantum number for levels in 
the wide well systems (see Table \ref{system_notation}).
Thick(thin) horizontal lines denote the filled(empty) levels.
The thick upward arrow represents the density operator,
$\rho_{2,1}$, which annihilates one electron in level 1 and creates another
one in level 2. This is the main mechanism and the order parameter 
for the many-body phases of the four level system 
near the level crossing region.
The other upward arrows denote the density operators for
$\rho_{1,3}$, $\rho_{2,3}$, $\rho_{4,3}$, $\rho_{4,1}$ and $\rho_{4,2}$
respectively from the left to the right.
(b) The exchange interaction diagram for the coupling between
$\rho_{2,1}$ and $\rho_{4,3}$. The parity quantum number, $\alpha$,
has sign changed at the vertex. This is the main mechanism to stablize
the canted antiferromagnetic phase in the double well system at $\nu=4N+2$.
}
\label{level_coupling}
\end{figure}

For a double well system at $\nu=4N+2$, the complete
filled core levels are the lowest $4N$ levels, while the top two
filled levels may coherently hybridize in some situations
with the empty levels above the Fermi energy (we consider 
only the two lowest empty Landau levels in the context motivated by
the scenario originally discussed in Ref. \cite{canted_phase}).
We label the states involved in forming a many-body state as follows:
the second highest filled
level is denoted to be level 3, the highest (top) filled level
is level 1, the lowest 
empty level above the Fermi energy is level 2, 
and the second lowest
empty level is level 4 (see Fig. \ref{level_coupling}(a)),
and the other higher empty levels are assumed to be irrelevant.
Naively one may think that we may construct a trial wavefunction
similar to Eq. (\ref{wavefunction_general}) to consider interlevel
coherence only between level 1 (the highest filled level) and 
level 2 (the lowest empty level), because the single 
electron energy separation between these two levels is the smallest energy
scale near level crossing. However, 
it was explicitly shown that in a double well system at
$\nu=4N+2$, the contribution of
the second highest filled level (level 3) and the second lowest empty
level (level 4) could make crucial contribution to the 
coherent hybridization between level 1 and level 2,
resulting a novel canted antiferromagnetic phase (CAF) with broken 
spin symmetry \cite{canted_phase}. 
This symmetry-broken coherent phase cannot be obtained if one considers
the coherence between the most degenerate pair only
(i.e. level 1 and level 2) via Eq. (\ref{wavefunction_general}). 
Another way to understand why it is natural to consider four rather than two
levels around the Fermi energy in creating a many-body state 
for the double well system
at $\nu=4N+2$ is to note that these four levels 
are separated from the other ones by a large cyclotron energy, whereas
they are separated from each other only by (much) 
smaller energies of Zeeman and tunneling splittings.
Therefore in this section we will consider the mixing of
the four (rather than two) levels closest to the Fermi energy to 
study the trial wavefunctions for the double well $\nu=4N+2$ system
near the level crossing region in the presence of the in-plane field. 

Let us first consider the uniform ground state of a 
double well system at $\nu=4N+2$ {\it without} any
in-plane magnetic field. We assume 
that the cyclotron resonance energy, $\omega_\perp$,
is much larger than the tunneling energy, $\Delta_{SAS}$, and the 
Zeeman (spin-splitting) energy, $\omega_z$, so that the two highest
filled levels and the two lowest empty levels have the same orbital
Landau level and all other levels can be treated as incoherent core states
not actually participating in the level-crossing hybridization process.
Single particle states for noninteracting electrons are shown in Fig.
\ref{level_coupling}(a) in the case when the 
tunneling energy, $\Delta_{SAS}$,
is larger than the Zeeman energy, $\omega_z$.
We use $(\alpha,s)$ to label
the four levels around Fermi level under 
consideration, where $\alpha=\pm 1$ is the 
quantum number of parity symmetry 
(which in this case is the reflection symmetry
about $x-y$ plane, $z\to -z$, or the interchange of layers, see 
Section \ref{symmetry}),
and $s=\uparrow(\downarrow)$ is the spin quantum number. 
For convenience of comparison and later discussion, we define level 1 
($(\alpha,s)=(+,\downarrow)$) and level 2 
($(\alpha,s)=(-,\uparrow)$) to be the isospin up and isospin down state
respectively as shown in Table \ref{system_notation}, because 
these two levels, being closest to the Fermi energy, are 
obviously the most energetically relevant ones compared
to the other two levels in this four level scenario 
(see Fig. \ref{level_fig}(f)). Many-body states that appear 
in this system correspond to mixing some of these single particle states
and describe the breaking of certain symmetries in this problem. As discussed
by Das Sarma {\it et. al.} \cite{canted_phase} in the context of the canted
antiferromagnetic state in bilayer systems, the states that are most likely
to be hybridized by Coulomb interaction are the ones that are closest to
each other for the noninteracting system: state $1=(+,\downarrow)$, and
$2=(-,\uparrow)$. When the expectation value 
$\langle c^\dagger_{1,k} c^{}_{2,k}\rangle$ is finite
and independent of $k$, 
the system has canted antiferromagnetic spin order.
In the two layers the transverse components of the
spin point in the opposite directions. Such an 
order parameter breaks the
spin rotational symmetry around $z$-axis and and the parity symmetry
$\hat{\cal P}=\hat{\cal P}_z\hat{\cal P}_{xy}$. (We use the full parity
symmetry, $\hat{\cal P}$, rather than $\hat{\cal P}_z$, since
the latter is not conserved when we include an 
in-plane magnetic field later,
see Section \ref{symmetry}.) More precisely,
under the spin rotation, $\hat{\cal U}(\chi)$ (defined in 
Eq. (\ref{spin_rotation})), we have 
$\hat{\cal U}(\chi)a^\dagger_{1,k} a^{}_{2,k}\hat{\cal U}^{-1}(\chi)
=e^{-i\chi}a^\dagger_{1,k} a^{}_{2,k}$, and under parity transformation,
we have $\hat{\cal P}a^\dagger_{1,k} a^{}_{2,k}\hat{\cal P}^{-1}
=-a^\dagger_{1,-k} a^{}_{2,-k}$. So the order parameter has spin $S_z=1$
and is odd under parity. However, 
the operator $a^\dagger_{4,k} a^{}_{3,k}$ has 
exactly the same symmetry properties as $a^\dagger_{1,k} a^{}_{2,k}$, so in a
canted antiferromagnetic phase both of them acquire finite expectation
values. Therefore it would be
insufficient to consider
mixing of the states $1$ and $2$ only and treat levels $3$ and $4$
as frozen \cite{canted_phase} in the double well system at even filling
factors. On the other hand hybridization between any other pair of levels
does not take place since the appropriate expectation values would have 
symmetry properties different from the transverse CAF Neel order.
For example $a^\dagger_1 a^{}_4$ has the right parity symmetry but
wrong spin symmetry, and $a^\dagger_1 a^{}_3$ has the correct spin
symmetry but wrong parity. As was discussed earlier 
\cite{canted_phase,bilayer_sym_break,bilayer_sym_break2} 
and as we will demonstrate below, the 
physical origin of the CAF phase is the lowering of the exchange energy
due to the additional spin correlations associated with 
$a^\dagger_{4,k} a^{}_{3,k}$, present above.

To construct an appropriate trial wavefunction {\it in the presence of} 
in-plane field, we first consider a uniform phase. If we make the assumption
that the order parameter breaks the same symmetries as in the case $B_\|=0$,
it has to have spin $S_z=1$ and be odd under parity. The operators,
$a^{}_{1\cdots 4,k}$ defined in Eq. (\ref{change_basis}), have the 
same transformation properties under $\hat{\cal U}(\chi)$ and $\hat{\cal P}$
(see Section \ref{symmetry}) with or without $B_\|$. The 
identical nature of broken symmetries in the cases 
$B_\|=0$ and $B_\|\neq 0$ implies that $a^\dagger_{1,k}a^{}_{2,k}$
and $a^\dagger_{3,k}a^{}_{4,k}$ acquire expectation values even in the
presence of a finite in-plane magnetic field.
Therefore we can still apply the ansatz of Ref. 
\cite{canted_phase} and consider the
hybridization of level 1 with level 2 and level 3
with level 4 separately to construct 
a trial wavefunction wavefunction for the {\it uniform} phase
in the presence of in-plane field.
In addition, as suggested by the wavefunction of odd filling system 
in Eq. (\ref{wavefunction_general}),
this hybridization may be further 
extended to a slightly more complicated form to
reflect the possible uniform winding of the transverse Neel
order \cite{burkov02} and/or the stripe order to break 
the translational symmetry. We then propose a trial 
wavefunction for a double well system at $\nu=4N+2$ in the presence
of in-plane magnetic field as follows:
\begin{eqnarray}
&&|\Psi^{D2}_G(\psi_k,\psi_k';\vec{Q}_\perp,\vec{Q}_\perp';
\gamma,\gamma')\rangle =
\prod_k\tilde{a}^\dagger_{1,+,k}\tilde{a}^\dagger_{2,+,k}
|LL\rangle 
\nonumber\\
&&\left[\begin{array}{c}
       \tilde{a}_{1,+,k}^\dagger \\ \tilde{a}_{1,-,k}^\dagger \\
       \tilde{a}_{2,+,k}^\dagger \\ \tilde{a}_{2,-,k}^\dagger
      \end{array}\right]
=\left[\begin{array}{cccc}
e^{ikQ_xl_0^2/2+i\gamma/2}\cos(\psi_k/2)& e^{-ikQ_xl_0^2/2-i\gamma/2}
            \sin(\psi_k/2) &0&0 \\
-e^{ikQ_xl_0^2/2+i\gamma/2}\sin(\psi_k/2)& e^{-ikQ_xl_0^2/2-i\gamma/2}
             \cos(\psi_k/2) &0&0 \\
0&0&  e^{ikQ'_xl_0^2/2+i\gamma'/2}\cos(\psi'_k/2) & 
            e^{-ikQ'_xl_0^2/2-i\gamma'/2}\sin(\psi'_k/2) \\
0&0&  -e^{ikQ'_xl_0^2/2+i\gamma'/2}\sin(\psi'_k/2) & 
            e^{-ikQ'_xl_0^2/2-i\gamma'/2}\cos(\psi'_k/2)
       \end{array}\right]
\nonumber\\
&&\cdot
\left[\begin{array}{c}
 a_{N,+1,\downarrow,k-Q_y/2}^\dagger \\ a_{N,-1,\uparrow,k+Q_y/2}^\dagger \\
a_{N,+1,\uparrow,k-Q_y'/2}^\dagger \\ a_{N,-1,\downarrow,k+Q_y'/2}^\dagger
      \end{array}\right],
\label{wavefunction_4l_D2}
\end{eqnarray}
where $\psi_k'$, $\vec{Q}_\perp'=(Q_x',Q_y')$, and $\gamma'$ 
are four additional parameters
to be determined variationally. We allow $\psi_k$ and $\psi_k'$ to be
arbitrary periodic functions of $k$, in order
to consider the possible stripe formation.
Note that the original
$4\times 4$ matrix representation of an unitary transformation 
has been reduced to an effective block-diagonalized matrix form,
and reduces to the uniform wavefunction originally
proposed in Ref. \cite{canted_phase} if we put $\vec{Q}_\perp=0$ and 
$\psi_k=\psi_0$, a constant. The fact that
the uniform wavefunction ($Q_\perp=0$, $\psi_k=\psi_0$) turns out to
be an excellent description \cite{canted_phase,bilayer_sym_break,bilayer_sym_break2}
for the corresponding $B_\|=0$ case leads us to believe that the
variational symmetry-broken wavefunction defined by
Eq. (\ref{wavefunction_4l_D2}) 
should be a reasonable generalization to study the many-body phases in 
the {\it presence} of an in-plane magnetic field, for the $\nu=4N+2$
bilayer system.

The specific symmetry-broken form
of the wavefunction in Eq. (\ref{wavefunction_4l_D2}) allows us
to introduce the concept of "double isospinors" to describe
the coherence in the four levels near degeneracy, because
each degenerate pair (i.e. levels 1 and 2, and levels 3 and 4) 
form two distinct isospinors in the state defined by 
Eq. (\ref{wavefunction_4l_D2}). 
The exchange energy between the two isospinors may stabilize a
symmetry-broken phase, and it reaches its maximum value if the
two isospinors have the same stripe period but have opposite 
spiral winding wavevectors. (This result is explicitly
obtained later in our numerical calculations of Section \ref{sec_D2}.)
The symmetry properties of this four level mixing coherent wavefunction
(Eq. (\ref{wavefunction_4l_D2})) are identical
to those discussed in Section \ref{symmetry_wavefunction}, and
we do not discuss it further here.

To develop a deeper understanding of the trial wavefunction proposed
in Eq. (\ref{wavefunction_4l_D2}), it is instructive to 
transform Eq. (\ref{wavefunction_4l_D2}) back
into the layer index basis, which, while not being a noninteracting
energy eigenbasis, is 
physically more appealing and easier to visualize conceptually.
For the convenience of comparison, we let $P_x=Q_y=Q_y'=0$ 
and $\gamma=\gamma'=0$, i.e.
the in-plane magnetic field chosen to be in the $y$ direction. 
(Here we have used a known result that the winding vector is 
always perpendicular
to the in-plane magnetic field in the double well system, which 
is justified by the numerical results 
shown later in this paper.)
Combining Eq. (\ref{change_basis}) and Eq. (\ref{wavefunction_4l_D2}) 
we obtain the ground state wavefunction as follows (the Landau level index, 
$N$ is omitted):
\begin{eqnarray}
|\Psi_G^{D2}\rangle&=&
\prod_{i=1,2}\prod_k\left(\sum_{l,s}z^{(i)}_{l,s,k}\,
e^{ikQ^{(i)}_{l,s}l_0^2}c^{D,\dagger}_{l,s,k}\right)|LL\rangle,
\label{trial_wf_l_index}
\end{eqnarray}
where $z^{(1)}_{\pm,\uparrow,k}=\cos(\psi_k/2)$, 
$z^{(1)}_{\pm,\downarrow,k}=-\sin(\psi_k/2)$,
$z^{(2)}_{\pm,\uparrow,k}=\pm\sin(\psi'_k/2)$, and
$z^{(2)}_{\pm,\downarrow,k}=\cos(\psi'_k/2)$, and their phases are
respectively
\begin{eqnarray}
Q^{(1)}_{\pm,\uparrow}&=&(\mp P_y+ Q_x)/2 \nonumber\\
Q^{(1)}_{\pm,\downarrow}&=&(\mp P_y-Q_x)/2 \nonumber\\
Q^{(2)}_{\pm,\uparrow}&=&(\mp P_y- Q'_x)/2 \nonumber\\
Q^{(2)}_{\pm,\downarrow}&=&(\mp P_y+Q'_x)/2.
\label{Q_P_relation}
\end{eqnarray}
One can see that the phase difference between 
the right ($l=+$) and the left ($l=-$) layers 
of the same spin is always $-P_y$ as in a commensurate phase,
while it is $Q_x$ or $Q_x'$ between up spin electrons and down
spin electrons within the same layer. 
In other words the wavefunction proposed in Eq. (\ref{wavefunction_4l_D2})
can only give a commensurate phase, 
because the effect of in-plane magnetic field 
has been automatically taken into account by transforming 
the layer index basis
into the noninteracting energy eigenstate basis as shown
in Section \ref{Hamiltonian_D}. 
More precisely, following Ref. \cite{burkov02}, we can define three 
different states according to the phase difference of electrons in
different layer and spin quantum states: 
(i) fully commensurate state, if $Q_x=Q_x'=0$, and hence spin up and spin 
down electrons in the same layer
have the same winding phase determined by the in-plane field;
(ii) partially commensurate/incommensurate state, if $Q_x\neq 0$ or
$Q_x'\neq 0$,
and hence spin up and spin down electrons in the same layer
have different winding frequency, although the phase difference
between electrons
of the same spin but in different layers still oscillates with
a wavevector determined by the
in-plane field $P_y$; (iii) fully incommensurate state, if
$Q^{(1,2)}_{+,\uparrow(\downarrow)}
-Q^{(1,2)}_{-,\uparrow(\downarrow)}\neq -P_y$, 
and therefore the tunneling energy becomes ineffective.  
According to Eq. (\ref{Q_P_relation}), the fully incommensurate
cannot be obtained from the trial wavefunction of
Eq. (\ref{wavefunction_4l_D2}), and
only the fully commensurate and partially commensurate/incommensurate
phases are the possible solutions.
We note, however, that the fully incommensurate phase can be formally
described by Eq. (\ref{wavefunction_4l_D2}) if we use operators defined
in Eq. (\ref{change_basis}) with $P_x=P_y=0$ and set $\Delta_{SAS}=0$
in the HF energy.
Symmetry properties of the wavefunction, Eq. (\ref{wavefunction_4l_D2}),
may be easily described in analogy with Section \ref{symmetry_wavefunction}
(see Eqs. (\ref{uniform_parity})-(\ref{uniform_spin})).

\subsubsection{Wavefunction for wide well systems}
\label{4l_W}

\begin{figure}

 \vbox to 8cm {\vss\hbox to 5.cm
 {\hss\
   {\includegraphics{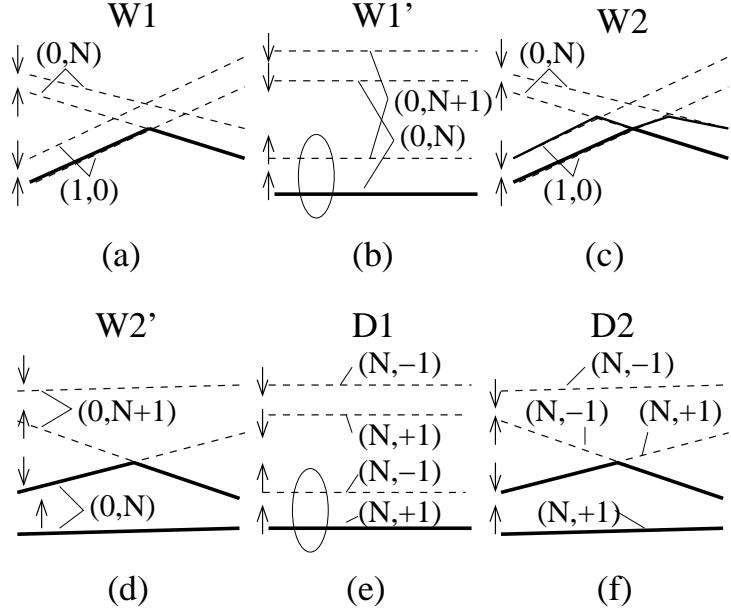}
   }
  \hss}
 }
\caption{Schematic pictures of the noninteracting energy 
configuration in the regions of level crossing (or level near degeneracy)
for the six systems
discussed in this paper: (a) intersubband level crossing of a wide
well at $\nu=2N+1$, 
(b) intrasubband level near degeneracy of a wide well at $\nu=2N+1$,
(c) intersubband level crossing of a wide well at $\nu=2N+2$,
(d) intrasubband level crossing of a wide well at $\nu=2N+2$,
(e) interlayer level near degeneracy of a double well at $\nu=4N+1$,
and (f) level crossing of a double well at $\nu=4N+2$.
Solid(dashed) lines are for filled(empty) levels. The horizontal axis is 
the strength of in-plane magnetic field. Up(down) arrows 
denote the electron spin states.
Note that the level indices for (a)-(d) are 
$\vec{n}=(n,n')$ for a parabolic wide well system,
while they are $(n,l)$ for (e) and $(n,\alpha)$ for (f) (see Table 
\ref{system_notation}) for different definitions of isospin indices. 
The circles in (b) and (e) denote 
the two near degenerate levels we consider. 
}
\label{level_fig}
\end{figure}

For a wide well system at $\nu=2N+2$, we have to separate 
the two kinds of level 
crossing possibilities in different regimes of the in-plane magnetic field:
(i) the intersubband level crossing ($W2$) at small $B_\|$
(see Figs. \ref{energy_levels_figure} and \ref{level_fig}(c)),
and (ii) the intrasubband level crossing ($W2'$) at larger $B_\|$
(see Figs. \ref{energy_levels_figure} and \ref{level_fig}(d)).
In the first case, the intersubband level crossing for noninteracting
electrons at even filling
factors, there are two separate possible
level crossings close to each other:
$((1,0),\downarrow)$ with $((0,N),\uparrow)$ and 
$((1,0),\uparrow)$ with $((0,N),\downarrow)$.
The in-plane magnetic field (and hence Zeeman energy) is so small 
(for realistic situations in GaAs-based 2D systems)
in this case that the two level crossings
are actually very close, and therefore it is
better to consider both of them simultaneously 
in a single trial wavefunction (constructed by the four degenerate levels)
rather than consider them separately as two independent crossings. 
The simplest wavefunction in this case should be similar to
Eq. (\ref{wavefunction_4l_D2}),
where the $4\times 4$ matrix is block-diagonalized as two 
separate isospinors for each pair of the crossing levels. 
According to the symmetry arguments of the last section and the
parity symmetry properties shown in Eq. (\ref{parity_W}),
such simple block-diagonalized $4\times 4$ matrix representation
of the trial wavefunction is further justified when considering
a uniform phase only (i.e. $\vec{Q}_\perp=\vec{Q}_\perp'=0$, and $\psi_k$ 
and $\psi_k'$ are constants),
if the two crossing levels have different parity symmetries.
As a result, we just consider the even-$N$ case in 
this paper (so that level (1,0) and (0,N) have different parity and
spin rotational symmetries), 
and speculate that the results for the odd-$N$ case should be similar.
Therefore, for the 
intersubband level crossing of a wide well system at even filling factors,
we propose the following trial wavefunction, similar to
Eq. (\ref{wavefunction_4l_D2}):
\begin{eqnarray}
&&|\Psi^{W2}_G(\psi_k,\psi_k';\vec{Q}_\perp,\vec{Q}_\perp';
\gamma,\gamma')\rangle
=\prod_k\tilde{c}^{W,\dagger}_{1,+,k}
\tilde{c}^{W,\dagger}_{2,+,k}|LL\rangle
\nonumber\\
&&\left[\begin{array}{c}
       \tilde{c}_{1,+,k}^\dagger \\ \tilde{c}_{1,-,k}^\dagger \\
       \tilde{c}_{2,+,k}^\dagger \\ \tilde{c}_{2,-,k}^\dagger
      \end{array}\right]
=\left[\begin{array}{cccc}
e^{ikQ_xl_0^2/2+i\gamma/2}\cos(\psi_k/2)& e^{-ikQ_xl_0^2/2-i\gamma/2}
         \sin(\psi_k/2) &0&0 \\
-e^{ikQ_xl_0^2/2+i\gamma/2}\sin(\psi_k/2)& e^{-ikQ_xl_0^2/2-i\gamma/2}
         \cos(\psi_k/2) &0&0 \\
0&0&  e^{ikQ'_xl_0^2/2+i\gamma'/2}\cos(\psi'_k/2) & 
         e^{-ikQ'_xl_0^2/2-i\gamma'/2}\sin(\psi'_k/2) \\
0&0&  -e^{ikQ'_xl_0^2/2+i\gamma'/2}\sin(\psi'_k/2) & 
         e^{-ikQ'_xl_0^2/2-ii\gamma'/2}\cos(\psi'_k/2)
       \end{array}\right]
\nonumber\\
&&\cdot
\left[\begin{array}{c}
   c_{\vec{n}_1,\downarrow,k-Q_y/2}^{W,\dagger} \\
   c_{\vec{n}_2,\uparrow,k+Q_y/2}^{W,\dagger} \\
   c_{\vec{n}_1,\uparrow,k-Q_y'/2}^{W,\dagger} \\
   c_{\vec{n}_2,\downarrow,k+Q_y'/2}^{W,\dagger}
      \end{array}\right],
\label{wavefunction_4l_W2}
\end{eqnarray}
where $\vec{n}_1=(1,0)$ and $\vec{n}_2=(0,N)=(0,2)$ are the Landau level
indices of the crossing levels (see Fig. \ref{level_fig}(c)). 

For the intrasubband level crossing at large $B_\|$ region ($W2'$), only
two noninteracting levels participate in level crossing (see Figs.
\ref{level_fig}(d) and \ref{level_coupling}(a)): level 
$1=((0,N),\downarrow)$ and level $2=((0,N+1),\uparrow)$, and all the other
levels remain separated by a finite gap. In this situation one is allowed
to consider only two degenerate levels when discussing the formation of a 
many-body state created by hybridization of levels 1 and 2.
We will, however, still include mixing the next nearest levels, 
levels 3 $=((0,N),\uparrow)$ and 4 $=((0,N+1),\downarrow)$, 
in the theory, and consider 
the block-diagonal wavefunctions of the type given in Eq. 
(\ref{wavefunction_4l_W2}) (but with $\vec{n}_1=(0,N)$ and
$\vec{n}_2=(0,N+1)$). In the lowest order in Coulomb interaction
they agree with the two-level coupling wavefunction 
(c.f. Eq. (\ref{wavefunction_general})), but have an advantage that they
allow us to discuss $W2'$ level crossing point at the same footing as $W2$.
The parity symmetry argument for a {\it uniform} phase
can also be applied in this case, since
levels $(0,N)$ and $(0,N+1)$ are always of different parity symmetry
and the corresponding Zeeman-split levels obviously have 
different spin polarizations. 
Therefore we believe Eq. (\ref{wavefunction_4l_W2}) to be 
a reasonable trial wavefunction to study possible isospin winding and/or
stripe order, although we cannot exclude the possibility that the ground 
state maybe stabilized by other more general wavefunctions.
We do point out, however, that if we include the mixing of 
$((0,N),\uparrow)$ with 
$((0,N+1),\downarrow)$, then $((0,N-1),\uparrow)$ mixing with 
$((0,N+2),\downarrow)$ arises exactly in the 
same order in Coulomb interaction, but with
a numerically larger energy gap for the unperturbed levels. Therefore, 
unlike the double well system at $\nu=4N+2$ (where the 
cyclotron resonance energy,
$\omega_\perp$, can be assumed to be much larger than the tunneling energy 
and the Zeeman splitting), considering only 
states 3 and 4 and neglecting all the 
other levels in the wide well $W2'$ case can not be energetically 
justified.
In any case, our four-level coupling wavefunction agrees with the simple
two-level coupling in the lowest order in Coulomb interaction, and
we will show (in Section \ref{sec_W2}) that its 
HF energy also gives the correct magnetoplasmon dispersion 
enabling us to study 
the possibility of a second order quantum phase transition to a state of 
broken spin symmetry \cite{wang02}. 
Therefore we believe that Eq. (\ref{wavefunction_4l_W2}) 
provides a reasonable
trial wavefunction to describe the single wide well system for
even filling factors ---- the fact that the numerical implementation
of the Hartree-Fock calculation using Eq. (\ref{wavefunction_4l_W2})
is relative easy is an additional motivation to study it in details.
 
\subsection{Stripe formation in the isospin coherent phase}
\label{pert_stripe}

\begin{figure}

 \vbox to 5cm {\vss\hbox to 5.cm
 {\hss\
   {\includegraphics{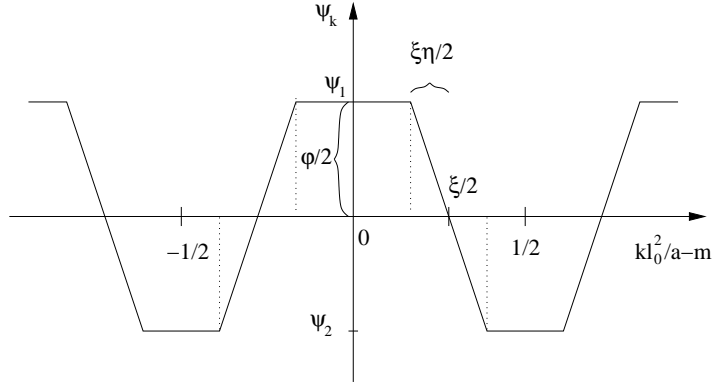}
   }
  \hss}
 }
\caption{A trial periodic function of the stripe phase function, $\psi_k$,
in Eq. (\ref{wavefunction_general}).
$\psi_1$, $\psi_2$, $\varphi$, $\xi$, $\eta$, and stripe period, $a$, are
variational parameters (not all of them are independent). 
Its mathematical expression is shown in Appendix \ref{psi_k}.}
\label{fig_psi}
\end{figure}

In Section \ref{trial_wf_2l} we have discussed some physical properties
of a stripe phase, where $\psi_k$ can be
a periodic function of the guiding center
coordinate, hence providing an oscillatory isospin polarization, 
$\langle {\cal I}_z\rangle$. 
A generic stripe phase discussed in this paper 
is provided by the hybridization of two crossing levels of different
parity symmetries near the level crossing (or near degenerate) region,
and therefore (at least) both parity (full space inversion) symmetry
and translational symmetry are broken when a stripe phase is stabilized
by Coulomb interaction. (Note that the spiral phase may also break
translational symmetry without any stripe order;
spin rotational symmetry around the total magnetic field direction
may also be broken if the two coherent levels are of different spin
directions in an even filling factor system.)
Parity is broken in the
non-fully polarized regions between
the stripes that choose a spatial direction, and the translational
symmetry is broken when the stripes 
choose their positions. Transitions
between states of no-broken symmetries (i.e. fully isospin polarized 
states in our case) and states that break parity and translational 
symmetries simultaneously may happen in two ways. The first
possibility is a direct first order transition when the in-plane
magnetic field exceeds some critical value.
To calculate the ground state wavefunction for this first order
phase transition, we 
need to include many variational parameters in the theory (as shown in 
Fig. \ref{fig_psi} and Appendix \ref{psi_k}) for the stripe phase function
in addition to the spiral wavevector, $\vec{Q}_\perp$.
The numerical calculation for this first order transition 
is very time-consuming in general. The second possibility
is two consecutive transitions that break symmetries one by one: 
at the first transition (which could be either first order or second order) 
parity symmetry is broken through a
{\it uniform} superposition of the two crossing Landau levels, 
which have different parity symmetries, and no stripe order is present; 
at the second transition, the stripe order appears
spontaneously with a concommitent breaking of
the transitional symmetry via a second order phase transition. 
In this section we concentrate on the second scenario (two consecutive
transitions) and develop a formalism for 
studying instabilities of an
interlevel coherent phase toward the formation of stripe order.
In some situations a spiral order may be 
stabilized in the first step, which, strictly speaking, 
breaks the translational symmetry even
without a stripe order, see Section \ref{symmetry_wavefunction}. 
Appearance of the stripe order from such s spiral phase can also be
described by the perturbation theory that we develop below
as long as the stripe
order appears via a second order phase transition.
In Appendix \ref{goldstone_mode} we will show that our perturbation method
for probing the existence of a stripe phase is actually equivalent to studying
the finite wavevector mode softening of a 
collective mode inside the uniform isospin coherent phase.
We mention, however, that in general the perturbation calculation is
easier to carry out than the mode softening calculation.

The perturbation method consists of the following steps.
First, we use the trial wavefunction in Eq. (\ref{wavefunction_general})
to search for non-stripe phases (i.e. use $\vec{Q}^\ast_\perp$, $\gamma$,
and $\psi_k=\psi_0$ as the variational parameters) that minimize
the energy of the system. If the optimal configuration has $\psi_0=0$
or $\pi$, then no uniform many-body phases are stabilized near the level 
crossing (or near degeneracy) region. Formation of the stripe phase
is still possible, but if it does occur, it happens via the first order
transition, and we need to consider the explicit variational forms for
the stripe phases, and compare their energies to the energies of the
uniform isospin polarized phases (see Appendix \ref{psi_k}). Alternatively,
if the optimal nonstripe configuration has $\psi_0\neq 0,\pi$
(and possibly finite $\vec{Q}_\perp^\ast$), the stripe phase may appear
via a second order phase transition, which may be understood as the 
appearance of small oscillations in $\psi_k$.
Therefore, we can choose the oscillation
amplitude of $\psi_k$ to be an order parameter, and approximate
$\psi_k$ by the following formula:
\begin{eqnarray}
\psi_k&=&\psi^\ast_0+4\Delta\cos(k\tilde{q}l_0^2),
\label{perturb_psi}
\end{eqnarray}
where $|\Delta|\ll\psi^\ast_0\neq 0$; $\tilde{q}$, the
characteristic wavevector of the stripe (the stripe period 
$a=2\pi/\tilde{q}$), is the only additional parameter
we need in the perturbation theory. We then obtain the following
expansion of $\Theta_i(q_n)$ to the second order in $\Delta$
using the definition given in Eq. (\ref{def_theta}):
\begin{eqnarray}
\Theta_1(q_n)&=&
\delta_{q_n,0}\left[\Theta^\ast_1(0)-2\Delta^2\cos(\psi^\ast_0)\right]
-\Delta\sin(\psi^\ast_0)[\delta_{q_n,-\tilde{q}}+
\delta_{q_n,\tilde{q}}]-\Delta^2\cos(\psi^\ast_0)[\delta_{q_n,-2\tilde{q}}
+\delta_{q_n,2\tilde{q}}]+{\cal O}(\Delta^3)
\label{Theta1_pert}
\\
\Theta_2(q_n)&=&
\delta_{q_n,0}\left[\Theta^\ast_2(0)+2\Delta^2\cos(\psi^\ast_0)\right]
+\Delta\sin(\psi^\ast_0)[\delta_{q_n,-\tilde{q}}+
\delta_{q_n,\tilde{q}}]+\Delta^2\cos(\psi^\ast_0)[\delta_{q_n,-2\tilde{q}}
+\delta_{q_n,2\tilde{q}}]+{\cal O}(\Delta^3)
\label{Theta2_pert}
\\
\Theta_{3}(q_n)&=&
\delta_{q_n,0}\left[\Theta^\ast_{3}(0)-2\Delta^2\sin(\psi^\ast_0)\right]+
\Delta\cos(\psi^\ast_0)[\delta_{q_n,-\tilde{q}}+\delta_{q_n,\tilde{q}}]
-\Delta^2\sin(\psi^\ast_0)[\delta_{q_n,-2\tilde{q}}+\delta_{q_n,2\tilde{q}}]
+{\cal O}(\Delta^3),
\label{Theta3_pert}
\end{eqnarray}
where $\Theta^\ast_i(0)$ ($i=1,2,3$) are their extreme values. 
Putting Eqs. (\ref{Theta1_pert})-(\ref{Theta3_pert})
in the expression of our HF energy as shown in the latter sections, 
we obtain the leading order (quadratic terms of $\Delta$ only)
energy perturbation of a stripe phase from the HF energy of the
uniform phase, $E^{HF}_{nonstripe}(\psi^\ast_0,\vec{Q}_\perp^\ast)$.
This result can be expressed as follows:
\begin{eqnarray}
E^{HF}_{stripe}(\tilde{q})=E^{HF}_{nonstripe}(\psi^\ast_0,\vec{Q}_\perp^\ast)
+E_{pert}^{HF}(\tilde{q};\psi^\ast_0,\vec{Q}_\perp^\ast)\Delta^2
+{\cal O}(\Delta^4),
\label{E_pert}
\end{eqnarray}
where the sign of $E_{pert}^{HF}(\tilde{q};\psi^\ast_0,\vec{Q}_\perp^\ast)$
determines the existence of a stripe phase:
if the minimum value of $E_{pert}^{HF}(\tilde{q};
\psi^\ast_0,\vec{Q}_\perp^\ast)$ 
is negative {\it and} at a finite value of $\tilde{q}=\tilde{q}^\ast$,
we can claim that the original uniform phase  
is not energetically favorable compared to the stripe phase,
and the ground state can then be a stripe phase with iso(spin) winding vector
$\vec{Q}^\ast_\perp$ and stripe oscillation wavevector $\tilde{q}^\ast$ 
along $x$ axis (we are using 
Landau gauge, $\vec{A}_{[y]}(\vec{r}\,)$).
On the other hand, if
$E_{pert}^{HF}(\tilde{q};\psi^\ast_0,\vec{Q}_\perp^\ast)$ 
is positive for all $\tilde{q}$, then a stripe phase along $y$ 
cannot be
formed through a second order phase transition. 
If we want to study the possibility of stripe formation along $x$ direction
(i.e. stripe modulation is along $y$ axis), we can do the same analysis
as above, but using the Landau gauge, $\vec{A}_{[x]}(\vec{r}\,)$. If both
stripe phases are possible, we need to compare them and find the one
of the lowest energy.

Finally we note that the same approach can be also applied to
study possible stripe formation via a continuous transition in the double
well systems at $\nu=4N+2$ and wide well system at $\nu=2N+2$. In
these cases we start with Eqs. (\ref{wavefunction_4l_D2})
and (\ref{wavefunction_4l_W2}) respectively, and use Eq. (\ref{perturb_psi})
and 
\begin{eqnarray}
\psi_k'&=&\psi_0'{}^\ast+4\Delta'\cos(k\tilde{q}'l_0^2),
\label{perturb_psi'}
\end{eqnarray}
and expand the HF energy for small $\Delta$ and $\Delta'$:
\begin{eqnarray}
E^{HF}_{stripe}(\tilde{q},\tilde{q}')=
E^{HF}_{nonstripe}(\psi_0^\ast,\psi_0'{}^\ast,
\vec{Q}_\perp^\ast,\vec{Q}_\perp'{}^\ast)
+[\Delta,\Delta']\cdot {\bf E}_{pert}^{HF}
(\tilde{q},\tilde{q}')\cdot
\left[\begin{array}{c}
      \Delta \\ \Delta'
      \end{array}\right]
+{\cal O}(\Delta^4),
\label{E_pert_4l}
\end{eqnarray}
where ${\bf E}^{HF}_{pert}(\tilde{q},\tilde{q}')$ is a $2\times 2$ matrix
(we have suppressed all other fixed 
parameters, $\psi_0^\ast$, etc. for notational simplicity).
Therefore if the lowest eigenvalue of 
${\bf E}^{HF}_{pert}(\tilde{q},\tilde{q}')$
is negative and located at finite $(\tilde{q}^\ast$,$\tilde{q}'{}^\ast)$, 
we can obtain a stripe phase with total HF energy lower than
the uniform coherent phases. On the other hand, if both eigenvalues of
${\bf E}^{HF}_{pert}(\tilde{q},\tilde{q}')$ are positive for the whole
range of $(\tilde{q},\tilde{q}')$ or its minimum value is at 
$(\tilde{q},\tilde{q}')=(0,0)$, then we conclude that no stripe phase
should arise via a second order phase transition.
Once again we emphasize that, in general, it is possible 
that stripe phases at large value of $\Delta$
are more favorable, for which the lowest order expansion in 
Eqs. (\ref{Theta1_pert})-(\ref{Theta3_pert}) is not sufficient.
But this would correspond to the first order transition to the stripe phase,
and we do not have a better method to study its existence except for a direct
numerical variational calculation (Appendix \ref{psi_k}).

\section{Wide well systems at $\nu=2N+1$}
\label{sec_W1}

In this section we calculate the Hartree-Fock variational energy obtained
by the trial wavefunction of Eq. (\ref{wavefunction_general}),
and show the numerical results for a wide well system 
at odd filling factors, $\nu=2N+1$. 
As mentioned earlier, there are two classes of level coherence in a wide 
well system (see Fig. \ref{energy_levels_figure}): 
one is for the intersubband level crossing at smaller $B_\|$ 
(denoted by $W1$), and
the other one is the intrasubband ``level near degeneracy''
at larger $B_\|$ region (denoted by $W1'$). 
By ``level near degeneracy'', we mean a small (but non-zero) gap
between energy levels of noninteracting electrons, which is much smaller
than the Coulomb interaction energy. This is not a true level crossing 
(which would imply a zero gap rather than a ``small'' gap),
but as we will show below, it is sufficient for interlevel hybridization 
leading to non-trivial many-body ground states. For simplicity, 
we assume that the lowest $2N$ levels do not have any
interlayer coherence while
the spin polarized top level is allowed to have 
interlevel coherence with the lowest empty 
level of the same spin polarization.
In Figs. \ref{level_fig}(a) and (b) we show the corresponding quantum 
numbers for the relevant Landau levels of these two kinds of level coherence
we consider in this section:
in the small $B_\|$ region, we consider 
the intersubband level crossing between $\vec{n}_1=(1,0)$ and 
$\vec{n}_2=(0,N)$, and in the large $B_\|$ region, 
we consider the level near degeneracy between 
$\vec{n}_1=(0,N)$ and $\vec{n}_2=(0,N+1)$.
We note that this two level approximation is easily
justified energetically in the $W1$ case,
where a level crossing always ensures the noninteracting energy gap
between the two crossing levels is smaller than than their energy
separation with other levels. It is, however, less justifiable for the
$W1'$ case, 
where the finite gap between the
two coherent levels is just numerically smaller 
than their energy separation from other lower filled or higher empty 
levels.
For simplicity, in this paper we will restrict our analysis 
in the $W1'$ case (intrasubband level near degeneracy) to $N=0$ ($\nu=1$) only,
and speculate that the results for other odd filling factors ($N\ge 1$)
should be qualitatively similar.
Note that in Fig. \ref{energy_levels_figure} there are 
more level crossings in the small $B_\|$ region, e.g. crossing between
$(1,1)$ and $(0,3)$, and also more crossing in the
large $B_\|$ region (not shown in the figure), e.g. the 
between $(0,N-1)$ (spin down) 
and $(0,N+2)$ (spin up). For the sale of brevity, 
we will not discuss these additional level crossings in this paper. 
We believe the level crossings or level near degeneracy situations 
we consider in this paper are the most
typical realistic ones for a wide well system, and the 
results for other level crossing situations should not be qualitatively
different from the ones we discuss in this paper.

We use the trial state proposed in Eq. (\ref{wavefunction_general}), 
and obtain the following expectation values:
\begin{eqnarray}
&&\langle\Psi_G^{W1}|c^{W,\dagger}_{\vec{m}_1,s_1,k_1}
c^{W}_{\vec{m}_2,s_2,k_2}|\Psi_G^{W1}\rangle
\nonumber\\
&=&\delta_{s_1,s_2}\delta_{k_1,k_2}\delta_{\vec{m}_1,\vec{m}_2}\left[
\cos^2(\psi_{k_1+Q_y/2}/2)\delta_{\vec{m}_1,\vec{n}_1}\delta_{s_1,1/2}+
\sin^2(\psi_{k_1-Q_y/2}/2)\delta_{\vec{m}_1,\vec{n}_2}\delta_{s_1,-1/2}+
\delta_{m_1,0}\theta(N-m'_1)\right] \nonumber\\
&&+\delta_{s_1,s_2}\delta_{s_1,1/2}\delta_{k_1,k_2-Q_y}
\delta_{\vec{m}_1,\vec{n}_1}\delta_{\vec{m}_2,\vec{n}_2}
e^{-iQ_x(k_2-Q_y/2)l_0^2}\sin(\psi_{k_2-Q_y/2}/2)\cos(\psi_{k_2-Q_y/2}/2)
\,e^{-i\gamma}
\nonumber\\
&&+\delta_{s_1,s_2}\delta_{s_1,1/2}\delta_{k_2,k_1-Q_y}
\delta_{\vec{m}_1,\vec{n}_2}\delta_{\vec{m}_2,\vec{n}_1}
e^{iQ_x(k_1-Q_y/2)l_0^2}\sin(\psi_{k_1-Q_y/2}/2)\cos(\psi_{k_1-Q_y/2}/2)
\,e^{i\gamma},
\label{exp_value_W1}
\end{eqnarray}
where $\vec{m}_i=(m_i,m_i')$ $(i=1,2$), and 
$\theta(x)$ is the Heaviside step function ($=1$ if $x>0$ and $=0$ 
otherwise).

\subsection{Hartree-Fock variational energy}
\label{HF_W1}

Using Eq. (\ref{exp_value_W1}) we can calculate
the single electron
noninteracting energy from the noninteracting Hamiltonian 
of Eq. (\ref{H0_W}):
\begin{eqnarray}
E^{W1}_0&=&E^{0,W}_{\vec{n}_1,\uparrow}\Theta_1(0)+
E^{0,W}_{\vec{n}_2,\uparrow}\Theta_2(0)+\sum_{\vec{m},s}{}'E^{0,W}_{\vec{m},s},
\label{E_0^W1}
\end{eqnarray}
where $\sum{}'$ means a summation over the core state.
The Hartree (direct) energy per electron can also be obtained from the direct
term of Eq. (\ref{H1_W}):
\begin{eqnarray}
E_H^{W1}
&=&\frac{N_\phi}{2\Omega_\perp}\sum_{q_n}\left\{
\tilde{V}^W_{\vec{n}_1\vec{n}_1,\vec{n}_1\vec{n}_1}(q_n,0)
\Theta_1(q_n)^2+
\tilde{V}^W_{\vec{n}_2\vec{n}_2,\vec{n}_2\vec{n}_2}(q_n,0)
\Theta_2(q_n)^2
+2\tilde{V}^W_{\vec{n}_1\vec{n}_1,\vec{n}_2\vec{n}_2}(q_n,0)
\cos(q_nQ_yl_0^2)\Theta_1(q_n)\Theta_2(q_n)\right.
\nonumber\\
&&+\left.4\sum_{\vec{m}}{}'
\left[\tilde{V}^W_{\vec{n}_1\vec{n}_1,\vec{m}\vec{m}}
(0,0)\Theta_1(0)
+\tilde{V}^W_{\vec{n}_2\vec{n}_2,\vec{m}\vec{m}}(0,0)\Theta_2(0)\right]
\right\}
+\frac{2N_\phi}{2\Omega_\perp}\sum_{q_n}
\tilde{V}^W_{\vec{n}_2\vec{n}_1,\vec{n}_1\vec{n}_2}(q_n,Q_y)
\Theta_{3}(q_n-Q_x)^2
\nonumber\\
&&+\delta_{Q_y,0}\frac{2N_\phi}{2\Omega_\perp}\sum_{q_n}
Re\left[\tilde{V}^W_{\vec{n}_1\vec{n}_2,\vec{n}_1\vec{n}_2}(q_n,0)
\,e^{i2\gamma}\right]\Theta_{3}(q_n-Q_x)\Theta_{3}(q_n+Q_x)
\nonumber\\
&&+\frac{4N_\phi\delta_{Q_y,0}}{2\Omega_\perp}\sum_{q_n}\left\{Re\left[
\tilde{V}^W_{\vec{n}_1\vec{n}_2,\vec{n}_1\vec{n}_1}(q_n,0)\,e^{i\gamma}\right]
\Theta_{3}(q_n+Q_x)\Theta_1(q_n)
+Re\left[\tilde{V}^W_{\vec{n}_1\vec{n}_2,\vec{n}_2\vec{n}_2}(q_n,0)
\,e^{i\gamma}\right]\Theta_{3}(q_n+Q_x)\Theta_2(q_n)\right\}
\nonumber\\
&=&\frac{1}{2}\sum_{q_n}\left\{E^{W1}_{H1}(q_n,0)\Theta_1(q_n)^2+
E^{W1}_{H2}(q_n,0)\Theta_2(q_n)^2
+2E^{W1}_{H3}(q_n,0\,;Q_y)\Theta_1(q_n)\Theta_2(q_n)\right\}
\nonumber\\
&&+2\left[E^{W1}_{H4}(0,0)\Theta_1(0)
+E^{W1}_{H5}(0,0)\Theta_2(0)\right]
+\sum_{q_n}E^{W1}_{H6}(q_n,Q_y)\Theta_{3}(q_n-Q_x)^2
\nonumber\\
&&+\delta_{Q_y,0}\sum_{q_n}{\cal E}^{W1}_H(q_n,0\,;2\gamma)
\Theta_3(q_n-Q_x)\Theta_3(q_n+Q_x)
\nonumber\\
&&+2\delta_{Q_y,0}\sum_{q_n}\left\{ 
\tilde{E}^{W1}_{H1}(q_n,0\,;\gamma)\Theta_{3}(q_n+Q_x)\Theta_1(q_n)
+\tilde{E}^{W1}_{H2}(q_n,0\,;\gamma)\Theta_{3}(q_n+Q_x)\Theta_2(q_n)\right\},
\label{E_H^W1}
\end{eqnarray}
where in the last equation we have introduced $E^{W1}_{Hi}$ ($i=1\cdots 6$),
${\cal E}^{W1}_{H}$ and $\tilde{E}^{W1}_{Hj}$ ($j=1,2$)
to label the Hartree energies contributed by each 
corresponding term for the convenience 
of later discussion. Their definition is obvious from
Eq. (\ref{E_H^W1}). Here $q_n=2\pi n/a$
is the stripe wavevector with $a$ being the period of the stripe. 
Note that there are three kinds of Hartree energies shown in 
Eq. (\ref{E_H^W1}): (i) the term $E^{W1}_{Hi}$ ($i=1\cdots 6$), which
have no explicit phase ($\gamma$) dependence, are finite for
all value of $\vec{Q}_\perp$ in general; (ii) the term 
${\cal E}^{W1}_{H}$, which has explicit $e^{i2\gamma}$ dependence,
is nonzero only when $Q_y=0$; (iii) the terms 
$\tilde{E}^{W1}_{Hj}$ ($j=1,2$), which have 
explicit $e^{i\gamma}$ dependence, are nonzero only when $Q_y=0$
{\it and} for finite stripe order (when considering nonstripe phase,
$q_n=0$, $\tilde{V}^W_{\vec{n}_1\vec{n}_2,\vec{m}\vec{m}}(0,0)=0$).
Since our explicitly numerical results show that the third kind of
contribution ($\tilde{E}^{W1}_{Hj}$) is always zero in the ground states
we obtain near the level degeneracy region, we will neglect them 
throughout in our discussion. 
Comparing the $E^{W1}_{Hi}$ ($i=1\cdots 6$) terms with the 
${\cal E}^{W1}_{H}$ term, we find that
their distinction arises from the fundamental difference of the ordered phases
at finite $\vec{Q}_\perp$ and at $\vec{Q}_\perp=0$ when no stripe
order is present (note that when $\psi_k$ is a constant, 
$\Theta_i(q_n)=\delta_{q_n,0}\Theta_i(0)$ and
$\Theta_i(q_n\pm Q_x)=\delta_{q_n,\mp Q_x}\Theta_i(0)$ so that
${\cal E}^{W1}_H$ term is proportional to $\delta_{Q_y,0}\delta_{Q_x,0}$).
For $\vec{Q}_\perp$ finite, the state breaks translational symmetry and 
the invariance of energy with respect to $\gamma$ reflects a freedom
of choice of the origin (see Eq. (\ref{uniform_translation}) and
the discussion in Section \ref{symmetry_wavefunction}).
It also signals the presence of a gapless Goldstone
mode coming from the spontaneously broken continuous (translational)
symmetry. On the other
hand, for $\vec{Q}_\perp=0$, the many-body state breaks only the discrete
parity symmetry (see Eq. (\ref{uniform_parity})). 
As a result there is an explicit dependence of energy on
$e^{i2\gamma}$ with $\gamma=0,\pi$ being the two degenerate minima (see also
the discussion below and in Section \ref{effective_theory}). 
The ground state selects either $\gamma=0$ or $\pi$ via the 
Ising type transition, which describes the breaking of parity symmetry.
No Goldstone mode exists in this case since the broken symmetry is
discrete (i.e. Ising type).

As for the Fock (exchange) energy per electron, we can have
\begin{eqnarray}
E_F^{W1}
&=&\frac{-1}{2\Omega_\perp}\sum_{\vec{q}_\perp}\left\{
\tilde{V}^W_{\vec{n}_1\vec{n}_1,\vec{n}_1\vec{n}_1}(\vec{q}\,)
\sum_{q_n}\cos(q_nq_yl_0^2)\Theta_1(q_n)^2
+\tilde{V}^W_{\vec{n}_2\vec{n}_2,\vec{n}_2\vec{n}_2}(\vec{q}\,)
\sum_{q_n}\cos(q_nq_yl_0^2)\Theta_2(q_n)^2\right.
\nonumber\\
&&+2\tilde{V}^W_{\vec{n}_1\vec{n}_2,\vec{n}_2\vec{n}_1}(\vec{q}\,)
\sum_{q_n}\cos(q_n(q_y+Q_y)l_0^2)
\Theta_1(q_n)\Theta_2(q_n)
\nonumber\\
&&+2\sum_{\vec{m}}{}'\left[
\tilde{V}^W_{\vec{n}_1,\vec{m},\vec{m}\vec{n}_1}(\vec{q}\,)\Theta_1(0)
+\tilde{V}^W_{\vec{n}_2,\vec{m},\vec{m}\vec{n}_2}(\vec{q}\,)\Theta_2(0)\right]
\nonumber\\
&&+2\tilde{V}^W_{\vec{n}_1\vec{n}_1,\vec{n}_2\vec{n}_2}(\vec{q}\,)
\cos((q_xQ_y-q_yQ_x)l_0^2)\sum_{q_n}\cos(q_nq_yl_0^2)
\Theta_{3}(q_n)^2
\nonumber\\
&&+2\delta_{Q_y,0}
Re\left[\tilde{V}^W_{\vec{n}_1\vec{n}_2,\vec{n}_1\vec{n}_2}(\vec{q}\,)
\,e^{i2\gamma}\right]
\sum_{q_n}\cos(q_nq_yl_0^2)\Theta_{3}(q_n-Q_x)\Theta_{3}(q_n+Q_x)
\nonumber\\
&&+4\delta_{Q_y,0}\left[\sum_{q_n}
Re[e^{-ip_nq_yl_0^2}\tilde{V}^W_{\vec{n}_1\vec{n}_2,\vec{n}_1\vec{n}_1}
(\vec{q}_\perp)\,e^{i\gamma}]\Theta_1(q_n)\Theta_{3}(q_n+Q_x)
\right.\nonumber\\
&&+\left.\sum_{p_n}Re[e^{-ip_nq_yl_0^2}
\tilde{V}^W_{\vec{n}_2\vec{n}_2,\vec{n}_1\vec{n}_2}(\vec{q}_\perp)
\,e^{i\gamma}]\Theta_2(q_n)\Theta_{3}(q_n+Q_x)\right\}
\nonumber\\
&=&\frac{1}{2}\sum_{q_n}
\left\{E^{W1}_{F1}(q_n,0)\Theta_1(q_n)^2+
E^{W1}_{F2}(q_n,0)\Theta_2(q_n)^2
+2E^{W1}_{F3}(q_n,0;Q_y)\Theta_1(q_n)\Theta_2(q_n)\right\}
\nonumber\\
&&+\left[E^{W1}_{F4}(0,0)\Theta_1(0)+E^{W1}_{F5}(0,0)\Theta_2(0)\right]
+\sum_{q_n}E^{W1}_{F6}(q_n,0;Q_x,Q_y)\Theta_{3}(q_n)^2
\nonumber\\
&&+\delta_{Q_y,0}\sum_{q_n}{\cal E}^{W1}_F(q_n,0\,;2\gamma)
\Theta_3(q_n-Q_x)\Theta_3(q_n+Q_x)
\nonumber\\
&&+2\delta_{Q_y,0}\sum_{q_n}\left\{ 
\tilde{E}^{W1}_{F1}(q_n,0\,;\gamma)\Theta_{3}(q_n+Q_x)\Theta_1(q_n)
+\tilde{E}^{W1}_{F2}(q_n,0\,;\gamma)\Theta_{3}(q_n+Q_x)\Theta_2(q_n)\right\}
\label{E_F^W1}
\end{eqnarray}
where $E^{W1}_{Fi}$ $(i=1\cdots 6)$, ${\cal E}^{W1}_{F}$, and
$\tilde{E}^{W1}_{Fj}$ $(j=1,2)$ are introduced to label each 
contribution of the exchange energy.
The dependence of these exchange terms on the phase $\gamma$ associated
with possible broken symmetry behavior is the
same as discussed earlier for the corresponding Hartree energy terms.

We think that it is worthwhile to emphasize again that 
Eqs. (\ref{E_H^W1}) and (\ref{E_F^W1})
are based on a specific choice of the Landau gauge for the vector potential,
$\vec{A}_{[y]}(\vec{r}\,)$, in which the particle momentum is conserved 
along $y$ axis (perpendicular to the in-plane magnetic field). 
To obtain the HF variational energy for a stripe phase
along $x$ direction, we can choose the alternate 
gauge, $\vec{A}_{[x]}(\vec{r}\,)$,
in which particle momentum is conserved along $x$ direction, to construct
a many-body wavefunction similar to 
Eq. (\ref{wavefunction_general}) (see details in Appendix 
\ref{diff_direction}).
If no stripe phase is stabilized, the results obtained in these two
gauges are identical. To save space, we will not 
show the HF variational energy obtained in the second gauge
throughout this paper, 
although we take it into consideration in our numerical calculations.

\subsection{Magnetoplasmon excitations}
\label{plasmon_W1}

Before showing the results of minimizing the HF energy, 
it is instructive to address the close relationship between the HF 
variational energy shown in
Eqs. (\ref{E_0^W1})-(\ref{E_F^W1}) and the collective 
magnetoplasmon excitations
of the conventional incompressible quantum Hall states
(i.e. the isospin polarized states).
In an integer quantum Hall system, magnetoplasmons are collective modes
associated with magneto-exciton excitations above the Fermi energy that can be 
theoretically studied by using the generalized Hartree-Fock (or 
time-dependent Hartree-Fock) approximation \cite{kallin,wang02,canted_phase},
which is correct to the leading order of the ratio of the electron interaction
energy to the noninteracting Landau energy separation. 
The softening of the magnetoplasmon mode indicates that the
system may undergo
a second order phase transition from a usual isospin polarized 
state (i.e. the uniform quantum Hall state) to a new 
symmetry-broken ground state,
which is precisely the same as that 
obtained by minimizing the variational HF energy shown
in Eqs. (\ref{E_0^W1})-(\ref{E_F^W1}).
Moreover, the full analytical expression of the magnetoplasmon 
dispersion can be obtained
from the uniform variational HF energy (i.e. same as Eqs. (\ref{E_0^W1})-
(\ref{E_F^W1}) but considering $\psi_k=\psi_0$ or equivalently $\Theta_i(q_n)=
\Theta_i(0)\delta_{q_n,0}$) by taking small $|\psi_0|$ expansion
from the isospin up ground state or by taking small $|\pi-\psi_0|$
expansion from the isospin down ground state.
For example, if we consider the $W1$ case with 
isospin up state ($\vec{n}_1=(1,0)$) being the highest filled level,
the HF energy of Eqs. (\ref{E_0^W1})-(\ref{E_F^W1}) can be expanded
to the leading order of $\Theta_2(0)$ (i.e. small $\psi_0$) to
obtain (using $\Theta_{3}(0)^2=\Theta_1(0)\Theta_2(0)$)
the following total HF energy:
\begin{eqnarray}
E^{W1}_{HF}(\psi_0)=E^{W1}_{\vec{n}_1,\uparrow}+
E^{pl,W1}_{\vec{n}_1\vec{n}_2}(\vec{Q}_\perp)
\Theta_2(0)+\delta_{Q_x,0}\delta_{Q_y,0}{\cal E}^{a,W1}_{\vec{n}_1\vec{n}_2}
(\gamma)\Theta_2(0)+{\cal O}(\Theta_2(0)^2),
\label{E_pl^W1_0}
\end{eqnarray}
where the first term is the total electron energy
of the isospin up state (the ground state),
including the HF self-energy correction, the second term is the plasmon
dispersion shown below, and the third one is the additional point energy
shift associated with the broken parity symmetry. 
More explicitly we have
\begin{eqnarray}
E^{W1}_{\vec{n}_1,s}&=&E^{0,W}_{\vec{n}_1,s}+\frac{1}{2}
\left[E^{W1}_{H1}(0,0)+E^{W1}_{F1}(0,0)\right]+E^{W1}_{H4}(0,0)
+E^{W1}_{F4}(0,0),
\label{E_sp_W1}
\end{eqnarray}
where the first term is the 
noninteracting energy, the second term is the
self-energy produced by electrons within the top level ($\frac{1}{2}$ is
for double counting), and the third term is the self-energy produced
by electrons in the core state (see the definition of $E^{W1}_{H(F)i}(0,0)$
in Eqs. (\ref{E_H^W1}) and (\ref{E_F^W1})). The
magnetoplasmon excitation energy in the right
hand side of Eq. (\ref{E_pl^W1_0}) gives 
\begin{eqnarray}
E^{pl,W1}_{\vec{n}_1\vec{n}_2}(\vec{Q}_\perp)
&=&E^{0,W}_{\vec{n}_2,\uparrow}-E^{0,W}_{\vec{n}_1,\uparrow}
+\Sigma^{W1}_{\vec{n}_2,\uparrow}-\Sigma^{W1}_{\vec{n}_1,\uparrow}
+E^{W1}_{H6}(Q_x,Q_y)+E^{W1}_{F6}(0,0;Q_x,Q_y),
\label{E_pl^W1}
\end{eqnarray}
where $\Sigma^{W1}_{\vec{n},s}$ is the Hartree-Fock
self-energy of level $\vec{n}$ and spin $s$. Eq. (\ref{E_pl^W1}) is
exactly the same as the magnetoplasmon excitation energy of 
the incompressible (isospin up) quantum Hall state obtained directly
from the time-dependent Hartree-Fock approximation (TDHFA) \cite{wang02}.
The contributions from the bubble diagrams (the direct term) and
from the ladder diagrams (the exchange term) 
correspond to the last two terms of Eq. (\ref{E_pl^W1}) respectively.
Note that the energy of the $\vec{Q}_\perp=0$ point is disconnected 
from the rest of the spectrum due to the last term in Eq. (\ref{E_pl^W1_0}),
${\cal E}^{a,W1}_{\vec{n}_1\vec{n}_2}(\gamma)={\cal E}^{W1}_H(0,0\,;\gamma)
+{\cal E}^{W1}_F(0,0\,;\gamma)<0$.
This reflects the fact that a many-body state at $\vec{Q}_\perp=0$
breaks only a discrete symmetry and should not have Goldstone modes.
Alternatively continuous dispersion for $\vec{Q}_\perp\neq 0$ would break
the continuous translational symmetry and lead to Goldstone modes.
The plasmon dispersion obtained are based on the uniform integer quantum Hall
state, or equivalently, the
isospin polarized ground state,
therefore Eq. (\ref{E_pl^W1}) has to be changed if we want to study 
the dispersion of the collective modes 
{\it inside} the symmetry-broken ground state.
We mention that such a complete equivalence between the Hartree-Fock 
ground state energetic calculation and the corresponding collective mode
dispersion follows from the Ward identities, and has also been used extensively
in Ref. \cite{canted_phase} in discussing the canted anti-ferromagnetic 
state in bilayer systems.

\subsection{Results I: intersubband level crossing ($W1$ case)}
\label{result_W1}

\begin{figure}

 \vbox to 5.8cm {\vss\hbox to 5.cm
 {\hss\
   {\includegraphics{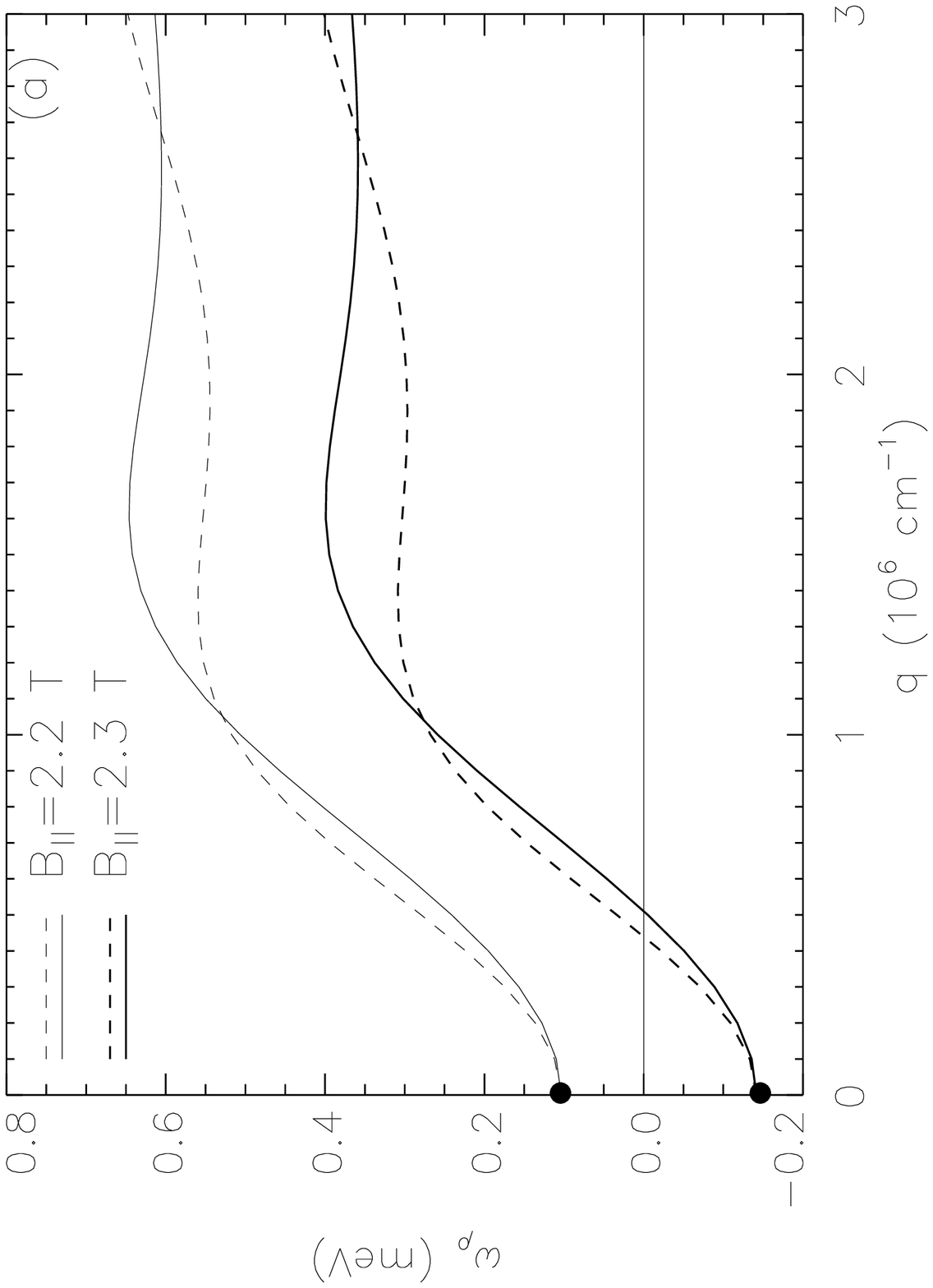}
   }
  \hss}
 }
 \vbox to 5.8cm {\vss\hbox to 5.cm
 {\hss\
   {\includegraphics{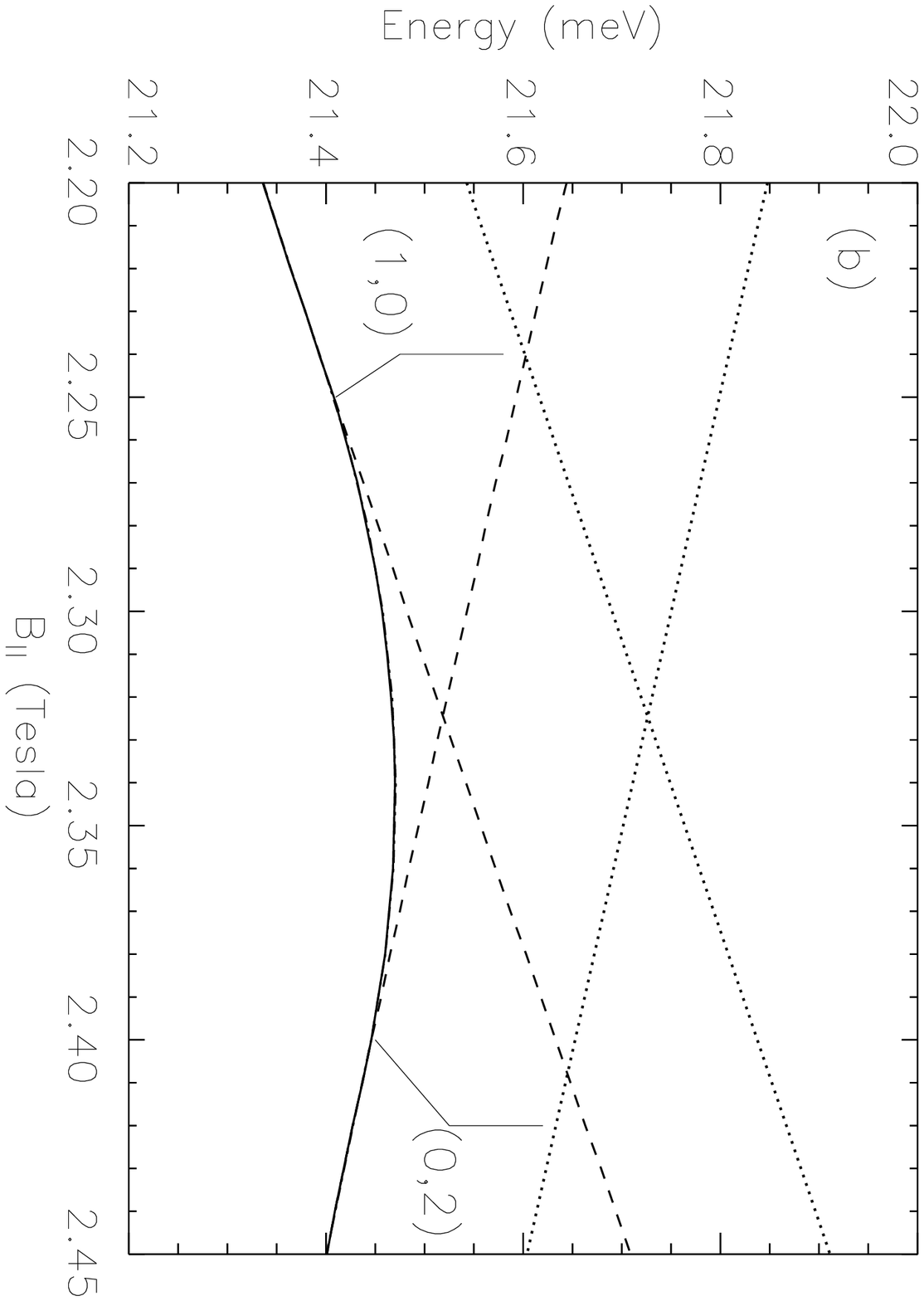}
   }
  \hss}
 }

\caption{(a) Magnetoplasmon dispersion near intersubband 
level crossing region of
a wide well system at $\nu=5$ (system parameters are the same
as in Fig. \ref{energy_levels_figure}). The upper(lower)
curves are for $B_\|=2.2$ and 2.3 Tesla respectively.
(The latter is calculated based on the isospin polarized basis 
and its negative energy at zero wavevector indicates a mode softening
in a symmetry breaking phase.
The correct curve for such plasmon mode should be modified based on 
the new coherent state as mentioned in the text.)
Solid(dashed) lines
are for dispersion along $y$ and $x$ axes respectively.
The filled circle at $q=0$ denotes the energy of disconnected excitations,
softening of which is a signature of parity symmetry breaking.
Note that the energy of the circle 
is just slightly lower ($<0.01$ meV) than the asymptotic 
magnetoplasmon energy in the long wavelength ($q\to 0^+$).
(b) Single particle energy as a function of in-plane magnetic field, $B_\|$,
of the same system.
The dashed(dotted) lines are the spin up(down) levels of the
two crossing isospin polarized states (including HF self-energy correction), 
while the solid lines are the energies of the many-body 
isospin coherent state, which breaks the parity symmetry of the system
for $2.2\mbox{ Tesla}<B_\|<2.40\mbox{ Tesla}$.
}
\label{W1_cross}
\end{figure}

In Fig. \ref{W1_cross}(a) we
first show the energy dispersion of the magnetoplasmon mode (in charge
channel only) obtained from Eq. (\ref{E_pl^W1}) for $\nu=5$, near
the intersubband level crossing
point at $B^\ast_\|\sim 2.33$ Tesla (after including the self-energy 
correction) for realistic GaAs 2D systems.
We also use a filled circle to denote the excitation energy
at $\vec{q}_\perp=0$, that are disconnected from the rest of the spectrum
by a negative energy shift, ${\cal E}^{a,W1}_{\vec{n}_1\vec{n}_2}$, 
according to Eq. (\ref{E_pl^W1}).
In the system parameter range we consider here ($B_\perp=3$ T and 
the bare confinement energy, $\omega_0$, is 7 meV), this disconnected
energy shift is very small ($<0.01$ meV).
When the in-plane magnetic field is above $B_\|^\ast=2.25$ Tesla,
the magnetoplasmon mode is softened at $\vec{q}_\perp=0$, indicating 
a second order phase transition toward a many-body coherent state
breaking the parity symmetry.
Strictly speaking, only the disconnected point at $\vec{q}_\perp=0$
is softened at $B_\|^\ast$, and the whole collective mode dispersion
will be modified for $B_\|>B_\|^\ast$ inside the new symmetry-broken phase.

In Fig. \ref{W1_cross}(b), we show the HF energy calculated
from Eqs. (\ref{E_0^W1})-(\ref{E_F^W1}) around the intersubband level
crossing point. We find an isospin coherent phase ($\psi_k=\psi_0^\ast
\neq 0,\pi$ with no spiral
order ($\vec{Q}^\ast_\perp=0$) and no stripe order) in addition to
the isospin polarized quantum Hall states within a small
range of the in-plane magnetic field ($2.25 < B_\| < 2.40$ T) for the
chosen system parameters. According to the symmetry analysis discussed
in Section \ref{symmetry_wavefunction}, the 
coherent phase only breaks the parity symmetry
of the system and therefore has no Goldstone mode.
This is consistent with the result studied by the mode softening of 
the collective excitations shown in Fig. \ref{W1_cross}(a).

We note that the new coherent phase, breaking the discrete parity
symmetry, is similar to the ferroelectric state observed in
ferroelectric crystals \cite{ferro}. More precisely, the electric 
dipole moment, $\langle \vec{r}\,\rangle$, is obviously zero
if the ground state has a definite parity, while it can be nonzero
if the ground state mixes two states of different parities.
Therefore we think the simple coherent state we find above in the intersubband
level crossing region of odd filling systems is
a ``ferroelectric'' quantum Hall state with finite electric
dipole moment. 
The recent experiments observing anomalies in 
the Shubnikov-de Haas oscillations of 
a wide parabolic well in the presence of a tilted magnetic field 
may be due to
the existence of such coherent states \cite{sergio_expW1}, but more definite
experimental work would be needed to settle this point.

\subsection{Results II: intrasubband level near degeneracy ($W1'$ case)}
\label{result_W1'}

\begin{figure}
 \vbox to 10cm {\vss\hbox to 5.5cm
 {\hss\
    {\includegraphics{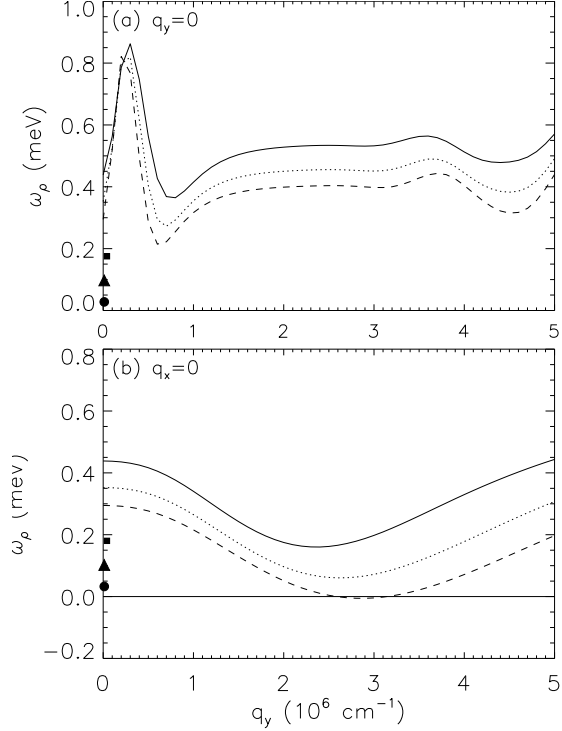}
  }
  \hss}
  }
\caption{
Magnetoplasmon dispersion of a wide well system at $\nu=1$ 
with large in-plane magnetic field. The perpendicular magnetic
field ($B_\perp$) is 3 Tesla and the bare parabolic confinement potential
($\omega_0$) is 3 meV.
Solid, dotted, and dashed lines are for $B_\|=20$, 25, and 30 Tesla
respectively. When $B_\|>30$ Tesla, the plasmon mode is softened at a 
finite wavevector in $y$ direction (perpendicular to the in-plane field).
The filled squares, triangles, and circles denote the energies of
the disconnected excitation energy at $\vec{q}_\perp=0$ for $B_\|=20$,
25 and 30 Tesla respectively.
}
\label{pl_nu1}
\end{figure}

In Fig. \ref{pl_nu1} we show a typical magnetoplasmon
mode dispersion in the charge channel of a wide well 
at $\nu=1$ in the large $B_\|$ region near the intrasubband 
level near degeneracy point, where the interaction energy is of the
same order as the noninteracting energy separation.
(Note that the bare confinement energy $\omega_0$ is 3 meV here.) 
When $B_\|$ is larger
than 30 Tesla, we find a mode softening 
at a finite wavevector perpendicular to the
in-plane magnetic field (i.e. along $y$ axis).
In the same figure, we use filled squares, triangles, and circles to
denote the energies of the zero momentum excitation, which is different
from the long wavelength limit of the plasmon curve by an
energy $|{\cal E}^{a,W1}_{(0,0),(0,1)}(0)|\sim 0.25$ meV
(see Eq. (\ref{E_pl^W1_0})).
In sharp contrast to the $W1$ case shown in Fig. \ref{W1_cross}(a),
the collective mode softening occurs here at a finite wavevector, 
$\vec{Q}_\perp=(0,\pm Q_y^\ast)$ rather than at $\vec{Q}_\perp=0$,
showing an isospin spiral order in this system. Therefore
the ground state can be a spiral phase if only one of the ordering
wavevectors,  $(0,\pm Q_y^\ast)$, is present, or it can be a collinear
spin density wave if there is an ordering at {\it both} wavevectors with
equal amplitude. We have not been able to 
write a wavefunction for such a collinear
phase to compare its energy with the spiral phase, and therefore
we cannot rule out the possibility that a collinear phase can also 
be a ground state in the $W1'$ case.

\begin{figure}

 \vbox to 5.8cm {\vss\hbox to 5.cm
 {\hss\
   {\includegraphics{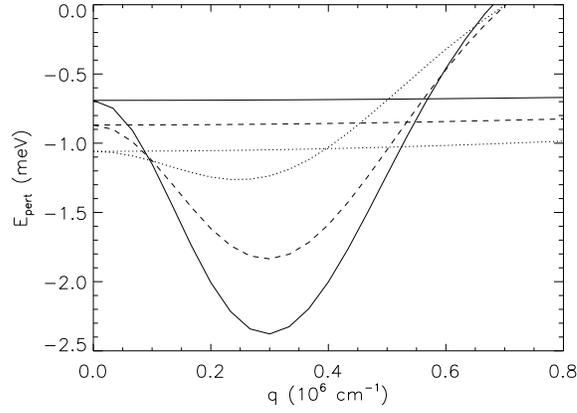}
   }
  \hss}
 }
\caption{Perturbative energy, $E^{HF}_{pert}(q)$, for the stripe formation
in a wide well system at $\nu=1$ in large magnetic field region.
Dotted, dashed, and solid lines are for 
in-plane magnetic field, $B_\|=31$, 35, and 40 Tesla respectively. 
Thick and thin lines are obtained in Landau gauges, $\vec{A}_{[y]}(\vec{r}\,)$
and $\vec{A}_{[x]}(\vec{r}\,)$, respectively. This result clearly shows that
a stripe phase can always be stablized to be along $y$ direction
(stripe modulation is in $x$ direction) for $B_\| > 30$ Tesla.
System parameters are the same as used in Fig. \ref{pl_nu1}.
}
\label{W1_E_pert}
\end{figure}

Now we have to investigate if such uniform coherent isospin 
spiral phase is stable against the formation of a stripe phase.
We use the perturbation method developed in Section \ref{pert_stripe}
and calculate the perturbative energy $E_{pert}^{HF}(q)$.
In Fig. \ref{W1_E_pert} we show our numerical 
results for $E_{pert}^{HF}(q)$
as a function of $q$ for several different values of $B_\|$. 
Both gauges of the vector potential, $\vec{A}_{[x]}(\vec{r}\,)$ and
$\vec{A}_{[y]}(\vec{r}\,)$, are considered in calculating $E_{pert}^{HF}(q)$ 
as indicated in the figure caption.
When $B_\|$ is larger than a critical value 
(it is also about 30 Tesla in this 
situation), the minimum of $E_{pert}^{HF}(q)$ is located at a 
finite wavevector ($q^\ast\sim 0.3 \times 10^6$ cm$^{-1}$) along $x$ axis,
showing a stripe order with isospin
$I_z$ modulating in the $x$ direction with a period 
$2\pi/q^\ast\sim 2000$ \AA.
Therefore, combining the two results above,
we conclude that an isospin skyrmion stripe
phase (see Section \ref{trial_wf_2l}) can be stabilized, with the stripe
normal vector along $x$ direction and the spiral winding vector along 
$y$ direction. In our numerical calculation, we do not
see signature for any intermediate phase (e.g. 
isospin spiral phase without stripe order)
between the isospin polarized (incompressible) quantum Hall state
and the isospin skyrmion stripe phase --- the local
minimum of $E_{pert}$ occurs at finite wavevector  
simultaneously with the plasmon mode softening. 
Therefore, following the results of Section \ref{symmetry_wavefunction},
the spiral order breaks the translational
symmetry along $y$ direction, while the stripe order breaks the translational
symmetry in $x$ direction (parallel to the in-plane field).
As discussed in our earlier paper \cite{eugene01}, 
such skyrmion stripe has finite topological isospin density
that leads to charge stripe order with stripes perpendicular
to the in-plane field. This should lead to anisotropy in charge transport
with larger conductivity along the stripes, i.e. perpendicular to
$B_\|$.

\begin{figure}

 \vbox to 5.8cm {\vss\hbox to 5.cm
 {\hss\
   {\includegraphics{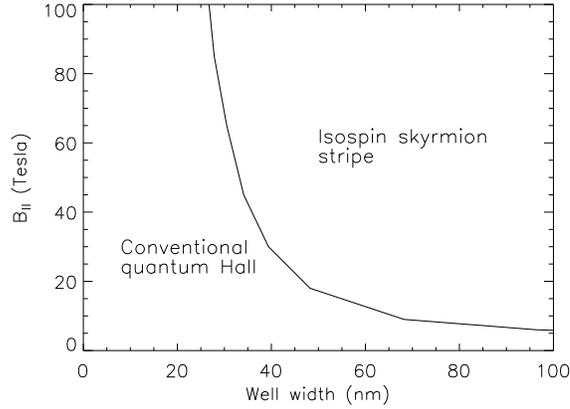}
   }
  \hss}
 }
\caption{Phase diagram of a wide well system at $\nu=1$. System parameters
are the same as used in Figs. \ref{pl_nu1} and \ref{W1_E_pert} but with 
different confinement potential, $\omega_0$, leading to different well widths.
The well width is estimated from the size of the single electron wavefunction
in the lowest subband of the parabolic confinement potential in the absence
of in-plane magnetic field, i.e. $2(m^\ast\omega_0)^{-1/2}$.
}
\label{W1_phase_diag}
\end{figure}

In Fig. \ref{W1_phase_diag} we show the phase diagram of the wide well
system at $\nu=1$ in a strong in-plane field.
The usual incompressible integer quantum Hall state is favored at small
well width (large bare confinement energy) and/or small $B_\|$ 
values. At larger
well width and/or stronger $B_\|$ field, the system undergoes a second order
phase transition toward an isospin skyrmion stripe phase 
with translational symmetries broken in both $x$ and $y$ directions 
(parity symmetry is of course also broken).
In extremely large $B_\|$ and large well width, we expect the 
isospin skyrmion stripe phase to evolve toward the Wigner
crystal phase, which, however, is not included in our present theory.

\section{Single wide well system at $\nu=2N+2$}
\label{sec_W2}

For a wide well system at even filling factor, $\nu=2N+2$, we 
also consider two kinds of level crossings (see Figs. 
\ref{energy_levels_figure}, \ref{level_fig}(c) and (d)) :
one is the intersubband level crossing between
levels $((1,0),\downarrow(\uparrow))$ and 
$((0,N),\uparrow(\downarrow))$ in the small 
$B_\|$ region ($W2$), and the other is the 
intrasubband level crossing between levels $((0,N),\downarrow)$ 
and $((0,N+1),\uparrow)$) in the large $B_\|$ region ($W2'$).
For simplicity, we do not discuss the system
behavior at yet higher fields, for example, when there is a crossing 
between levels $((0,N-1),\downarrow)$ and $((0,N+2),\uparrow)$.
The main difference between a level crossing in an odd filing system
and the one in an even filling system is the spin degree of freedom.
In the odd filling situation, the two crossing levels are of the same spin 
so that only isospin degree of freedom affects the existence of a
novel coherent phase via their different orbital wavefunctions
(and the spin degree of freedom is essentially frozen). 
In even filling situation, however, the two crossing levels are 
of opposite spin polarization, so that a stabilized many-body 
coherent phase must also break the spin rotational symmetry.
This fact leads important consequences, since spin 
and isospin are not equivalent in their roles: the Coulomb 
interaction does not flip spin polarization but may flip the isospin
polarization of each scattered electron (see Fig. 
\ref{level_coupling}(b)).
We will discuss this subject in more details in Section \ref{comparison}.

As shown in Section \ref{4l_W}, 
when discussing possible many-body states around level crossings
at $\nu=2N+2$, we will consider trial wavefunctions that mix
the four closest levels around the Fermi energy, the two highest 
filled Landau levels and two lowest empty
Landau levels, and will assume that the lower $2N$ (core) levels are 
completely
filled (frozen) and do not participate in 
the coherent hybridization process.
Using Eq. (\ref{wavefunction_4l_W2}), we obtain
the following expectation value similar to Eq. (\ref{exp_value_W1}):
\begin{eqnarray}
&&\langle{\Psi}_G^{W2}|c^{W,\dagger}_{\vec{m}_1,\sigma_1,k_1}
c^{W}_{\vec{m}_2,\sigma_2,k_2}|{\Psi}_G^{W2}\rangle
\nonumber\\
&=&\delta_{\vec{m}_1,\vec{m}_2}\delta_{\sigma_1,\sigma_2}\delta_{k_1,k_2}\left[
\cos^2(\psi_{k+Q_y/2}/2)\delta_{\vec{m}_1,\vec{n}_1}\delta_{\sigma_1,-1/2}+
\sin^2(\psi_{k-Q_y/2}/2)/2)\delta_{\vec{m}_1,\vec{n}_2}\delta_{\sigma_1,1/2}
\right.
\nonumber\\
&&\hspace{2cm}+\left.\cos^2(\psi'_{k+Q'_y/2}/2)\delta_{\vec{m}_1,\vec{n}_1}
\delta_{\sigma_1,1/2}+
\sin^2(\psi'_{k-Q'_y/2}/2)\delta_{\vec{m}_1,\vec{n}_2}\delta_{\sigma_1,-1/2}+
\delta_{m_1,0}\theta(N-m_1')\right]
\nonumber\\
&&+\delta_{\vec{m}_1,\vec{n}_1}\delta_{\vec{m}_2,\vec{n}_2}
\delta_{\sigma_1,-1/2}
\delta_{\sigma_2,1/2}\delta_{k_1,k_2-Q_y}e^{-iQ_x(k_2-Q_y/2)l_0^2}
\sin(\psi_{k_2-Q_y/2}/2)\cos(\psi_{k_2-Q_y/2}/2)\,e^{-i\gamma/2}
\nonumber \\
&&+\delta_{\vec{m}_1,\vec{n}_2}\delta_{\vec{m}_2,\vec{n}_1}
\delta_{\sigma_1,1/2}
\delta_{\sigma_2,-1/2}\delta_{k_2,k_1-Q_y}e^{iQ_x(k_1-Q_y/2)l_0^2}
\sin(\psi_{k_1-Q_y/2}/2)\cos(\psi_{k_1-Q_y/2}/2)\,e^{i\gamma/2}
\nonumber\\
&&+\delta_{\vec{m}_1,\vec{n}_1}\delta_{\vec{m}_2,\vec{n}_2}
\delta_{\sigma_1,1/2}
\delta_{\sigma_2,-1/2}\delta_{k_1,k_2-Q'_y}e^{-iQ'_x(k_2-Q'_y/2)l_0^2}
\sin(\psi'_{k_2-Q'_y/2}/2)\cos(\psi'_{k_2-Q'_y/2}/2)\,e^{-i\gamma'/2}
\nonumber \\
&&+\delta_{\vec{m}_1,\vec{n}_2}\delta_{\vec{m}_2,\vec{n}_1}
\delta_{\sigma_1,-1/2}
\delta_{\sigma_2,1/2}\delta_{k_2,k_1-Q'_y}e^{iQ'_x(k_1-Q'_y/2)l_0^2}
\sin(\psi'_{k_1-Q'_y/2}/2)\cos(\psi'_{k_1-Q'_y/2}/2)\,e^{i\gamma'/2},
\label{exp_value_W2}
\end{eqnarray}
where $\vec{n}_1=(1,0)$ and $\vec{n}_2=(0,N)$ for $W2$ case, while
$\vec{n}_1=(0,N)$ and $\vec{n}_2=(0,N+1)$ for $W2'$ case.

\subsection{Hartree-Fock variational energy}

The noninteracting single electron energy can be obtained from
Eq. (\ref{H0_D}):
\begin{eqnarray}
E^{W2}_0 &=&
E^{0,W}_{\vec{n}_1,\downarrow}\Theta_1(0)+
E^{0,W}_{\vec{n}_2,\uparrow}\Theta_2(0)
+E^{0,W}_{\vec{n}_1,\uparrow}{\Theta}'_1(0)
+E^{0,W}_{\vec{n}_2,\downarrow}{\Theta}'_2(0)
+\sum_{l,\sigma}{}' E^{0,W}_{l,\sigma},
\label{E_0_W2}
\end{eqnarray}
where the last term is a constant energy shift from the frozen core states.
The Hartree (direct) and the Fock 
(exchange) energies per electron are respectively
\begin{eqnarray}
E^{W2}_H&=&\frac{N_\phi}{2\Omega_\perp}\sum_{q_n}\left\{
\tilde{V}^W_{\vec{n}_1\vec{n}_1,\vec{n}_1\vec{n}_1}(q_n)
\left[\Theta_1(q_n)^2+\Theta_1'(q_n)^2\right]+
\tilde{V}^W_{\vec{n}_2\vec{n}_2,\vec{n}_2\vec{n}_2}(q_n)
\left[\Theta_2(q_n)^2+\Theta'_2(q_n)^2\right]
\right.\nonumber\\
&&+2\tilde{V}^W_{\vec{n}_1\vec{n}_1,\vec{n}_2\vec{n}_2}(q_n)
\left[\Theta_1(q_n)\Theta_2(q_n)
\cos(q_nQ_yl_0^2)+\Theta'_1(q_n)\Theta'_2(q_n)\cos(q_nQ'_yl_0^2)\right]
\nonumber\\
&&+2\tilde{V}^W_{\vec{n}_1\vec{n}_1,\vec{n}_1\vec{n}_1}(q_n)\Theta_1(q_n)
\Theta'_1(q_n)\cos(q_n(Q_y-Q_y')l_0^2/2)
\nonumber\\
&&+2\tilde{V}^W_{\vec{n}_2\vec{n}_2,\vec{n}_2\vec{n}_2}(q_n)
\Theta_2(q_n)\Theta'_2(q_n)
\cos(q_n(Q_y-Q_y')l_0^2/2)
\nonumber\\
&&+2\tilde{V}^W_{\vec{n}_1\vec{n}_1,\vec{n}_2\vec{n}_2}(q_n)
\left[\Theta_1(q_n)\Theta'_2(q_n)
+\Theta_2(q_n)\Theta'_1(q_n)\right]\cos(q_n(Q_y+Q_y')l_0^2/2)
\nonumber\\
&&+\left.4\delta_{q_n,0}\sum_{\vec{m}}{}'
\left[\tilde{V}^W_{\vec{n}_1\vec{n}_1,\vec{m}\vec{m}}(0)\left(\Theta_1(0)+
\Theta'_1(0)\right)+\tilde{V}^W_{\vec{n}_2\vec{n}_2,\vec{m}\vec{m}}(0)
\left(\Theta_2(0)+\Theta'_2(0)\right)
+\sum_{\vec{l}'}{}'\tilde{V}^W_{\vec{m}\vec{m},\vec{m}'\vec{m}'}(0)
\right]\right\},
\label{E_H_W2}
\end{eqnarray}
and
\begin{eqnarray}
E^{W2}_{F}\nonumber
&=&\frac{-1}{2\Omega}\sum_{\vec{q}_\perp}
\left\{\tilde{V}^W_{\vec{n}_1\vec{n}_1,\vec{n}_1\vec{n}_1}(\vec{q}_\perp)
\sum_{q_n}\cos(q_nq_yl_0^2)
\left[\Theta_1(q_n)^2+\Theta'_1(q_n)^2\right]
+\tilde{V}^W_{\vec{n}_2\vec{n}_2,\vec{n}_2\vec{n}_2}(\vec{q})
\sum_{q_n}\cos(q_nq_yl_0^2)
\left[{\Theta_2(q_n)}^2+{\Theta'_2(q_n)}^2\right]\right.
\nonumber\\
&&+2\tilde{V}^W_{\vec{n}_1\vec{n}_2,\vec{n}_2\vec{n}_1}(\vec{q}_\perp)\sum_{q_n}
\cos(q_nq_yl_0^2)\cos(q_n(Q_y+Q_y')l_0^2/2)
\left[\Theta_1(q_n)\Theta'_2(q_n)+\Theta_2(q_n)\Theta'_1(q_n)\right]
\nonumber\\
&&+\left.2\sum_{\vec{m}}{}'
\tilde{V}^W_{\vec{n}_1\vec{m},\vec{m}\vec{n}_1}(\vec{q}_\perp)
\left[\Theta_1(0)+\Theta_1'(0)\right]
+2\sum_{\vec{m}}{}'
\tilde{V}^W_{\vec{n}_2\vec{m},\vec{m}\vec{n}_2}(\vec{q}_\perp)\left[
\Theta_2(0)+\Theta'_2(0)\right]
+2\sum_{\vec{m}_1,\vec{m}_2}
\tilde{V}^W_{\vec{m}_1\vec{m}_2,\vec{m}_2\vec{m}_1}(\vec{q})\right\}
\nonumber\\
&&+\frac{-2}{2\Omega_\perp}\sum_{\vec{q}_\perp}
\tilde{V}^W_{\vec{n}_1\vec{n}_1,\vec{n}_2\vec{n}_2}(\vec{q}_\perp)
\cos((Q_xq_y-Q_yq_x)l_0^2)
\sum_{q_n}\cos(q_nq_yl_0^2)\Theta_{3}(q_n)^2
\nonumber\\
&&+\frac{-2}{2\Omega_\perp}\sum_{\vec{q}_\perp}
\tilde{V}^W_{\vec{n}_1\vec{n}_1,\vec{n}_2\vec{n}_2}(\vec{q}_\perp)
\cos((Q'_xq_y-Q'_yq_x)l_0^2)
\sum_{q_n}\cos(q_nq_yl_0^2)\Theta'_{3}(q_n)^2
\nonumber\\
&&+\frac{-4\delta_{Q_y,-Q_y'}}{2\Omega_\perp}\sum_{\vec{q}_\perp}
Re[\tilde{V}^W_{\vec{n}_1\vec{n}_2,\vec{n}_1\vec{n}_2}(\vec{q}_\perp)
\,e^{i(\gamma+\gamma')}]\sum_{q_n}
\Theta_{3}(q_n-Q_x)\Theta'_{3}(q_n+Q'_x)\cos((q_xQ_y-q_nq_y)l_0^2),
\label{E_F_W2}
\end{eqnarray}
where the last term in the exchange energy shows the interplay between
the two isospinors as mentioned in Section \ref{4l_D}, 
and this term is nonzero only when
$\vec{Q}_\perp'=-\vec{Q}_\perp$ (for simplicity, we have chosen their
stripe wavevectors, $\tilde{q}_n$ and $\tilde{q}_n'$, to be the same).
Note that this is also the only term which 
depends on the phase, $\gamma+\gamma'$, in the
Hartree-Fock energy. This is because, in the even filling systems,
the two crossing levels are of different spin polarizations. Therefore,
spin symmetry breaking of the coherent state gives a 
continuous energy degeneracy (i.e. $\gamma-\gamma'$ is arbitrary),
while the breaking of parity symmetry selects
$\gamma+\gamma'=2m\pi$, where $m$ is an integer.

\subsection{Magnetoplasmon excitations}
\label{plasmon_W2}

Using the arguments similar to those in
Section \ref{plasmon_W1}, we can also obtain
the magnetoplasmon excitation energy of 
the even filling system in the usual
incompressible quantum Hall ground state by taking small
$\Theta_2(0)$ and $\Theta_2'(0)$ limits in the above Hartree-Fock variational
energy, Eqs. (\ref{E_0_W2})-(\ref{E_F_W2}). The result is equivalent to
solving the eigenvalue problem of the following $2\times 2$ matrix: 
\begin{eqnarray}
{\bf E}^{pl,W2}_{\vec{n}_1\vec{n}_2}(\vec{Q}_\perp)=
\left[\begin{array}{cc}
\Delta E^{0,W}_{\vec{n}_2\uparrow,\vec{n}_1\downarrow}
+\Delta\Sigma^{HF,W2}_{\vec{n}_2\uparrow,\vec{n}_1\downarrow}
+E^{W2}_{X}(\vec{Q}_\perp) & E^{W2}_{X'}(\vec{Q}_\perp) \\
\left[E^{W2}_{X'}(\vec{Q}_\perp)\right]^\ast &
\Delta E^{0,W}_{\vec{n}_2\downarrow,\vec{n}_1\uparrow}
+\Delta\Sigma^{HF,W2}_{\vec{n}_2\downarrow,\vec{n}_1\uparrow}
+E^{W2}_{X}(\vec{Q}_\perp)
\end{array}
\right],
\label{E_pl_W2}
\end{eqnarray}
where we have used $\vec{Q}_\perp'=-\vec{Q}_\perp$;
$\Delta E^{0,W}_{\vec{n}_2 \uparrow(\downarrow),\vec{n}_1 
\downarrow(\uparrow)}\equiv
E^{0,W}_{\vec{n}_2,\uparrow(\downarrow)}-
E^{0,W}_{\vec{n}_1,\downarrow(\uparrow)}$ and
$\Delta\Sigma^{HF,W}_{\vec{n}_2,\uparrow(\downarrow),
\vec{n}_1\downarrow(\uparrow)}\equiv
\Sigma^{H,W}_{\vec{n}_2,\uparrow(\downarrow)}
-\Sigma^{H,W}_{\vec{n}_1,\downarrow(\uparrow)}
+\Sigma^{F,W}_{\vec{n}_2,\uparrow(\downarrow)}
-\Sigma^{F,W}_{\vec{n}_1,,\downarrow(\uparrow)}$
are respectively the noninteracting energy and the HF self-energy 
difference between the two relevant levels.
The definition and the explicit expression of the HF self-energies,
$\Sigma^{H/F,W2}_{\vec{m},s}$, are similar to those 
in the odd filling systems. The two electron-hole 
(exciton) binding energies are respectively
\begin{eqnarray}
E^{W2}_{X}(\vec{Q}_\perp)=\frac{-1}{\Omega_\perp}\sum_{\vec{q}_\perp}
\tilde{V}^W_{\vec{n}_1\vec{n}_1,\vec{n}_2\vec{n}_2}(\vec{q}_\perp)
\cos((q_xQ_y-q_yQ_x)l_0^2)
\label{E_W2_X}
\end{eqnarray}
and 
\begin{eqnarray}
E^{W2}_{X'}(\vec{Q}_\perp)=\frac{-1}{\Omega_\perp}\sum_{\vec{q}_\perp}
\tilde{V}^W_{\vec{n}_1\vec{n}_2,\vec{n}_1\vec{n}_2}(\vec{q}_\perp)
\cos((q_xQ_y-q_yQ_x)l_0^2).
\label{E_W2_bi_X}
\end{eqnarray}
We note that Eq. (\ref{E_pl_W2}) is exactly the same as the 
magnetoplasmon dispersion
matrix derived for the triplet spin channel in the 
time-dependent-Hartree-Fock
approximation \cite{wang02}, demonstrating that our proposed four level
trial wavefunction, Eq. (\ref{wavefunction_4l_W2}), is adequate in 
investigating the existence of new broken 
symmetry phases. On the other hand, we note that
the other two spin singlet magnetoplasmon modes \cite{wang02} 
cannot be obtained in our theory, since 
the trial wavefunction of 
Eq. (\ref{wavefunction_4l_W2}) is still not of the most general 
form for the four level degeneracy. 
As mentioned earlier, this fact will not affect any of our
results or conclusions shown in this section, because these two
singlet excitations are relatively higher energy excitations 
and are 
of different symmetries from the lowest one we consider here.
For the purpose of understanding quantum phase transitions in the
system, it is crucial to have the correct description for the low
energy sector of the relevant Hilbert space, and clearly our
four level trial wavefunction of Eq. (\ref{wavefunction_4l_W2})
accomplished that very well.
Since the 2D magnetoplasmon dispersion calculation of the 
integer quantum Hall system has been
reported before \cite{wang02}, we will not further discuss the magnetoplasmon
dispersion and just focus on the HF variational energy calculation.
\subsection{Results I: intersubband level crossing ($W2$ case)}
\label{result_W2_1}

For level crossing in the small $B_\|$ region, our numerical calculation
shows that there is {\it no} many-body coherent phase
with total energy lower
than the (uniform) isospin polarized states within our Hartree-Fock  
approximation. In other words, 
we find that such level crossing always introduces a (trivial) first order 
phase transition with a sharp polarization change
in the narrow region of level crossing tuned by $B_\|$ (see Fig. 
\ref{energy_levels_figure}). This
is related to the resistance
hysteresis recently observed and discussed in Refs. 
\cite{hysteretic,jungwirth01} --- small domain walls may occur during
the first order phase transition separating the two polarizations
so that the resistance shows a hysteretic
behavior when the external electric field is swapt.
Although such domain wall physics associated with intersubband level crossing
induced first order transition is of intrinsic interest, we do not include this possibility in our theory since our interest here is to classify the (second order) quantum phase transitions between nontrivial quantum Hall phases (see a brief discussion in Section \ref{bubble}.

\subsection{Results II: intrasubband level crossing ($W2'$ case)}

\begin{figure}

 \vbox to 5.8cm {\vss\hbox to 5.cm
 {\hss\
   {\includegraphics{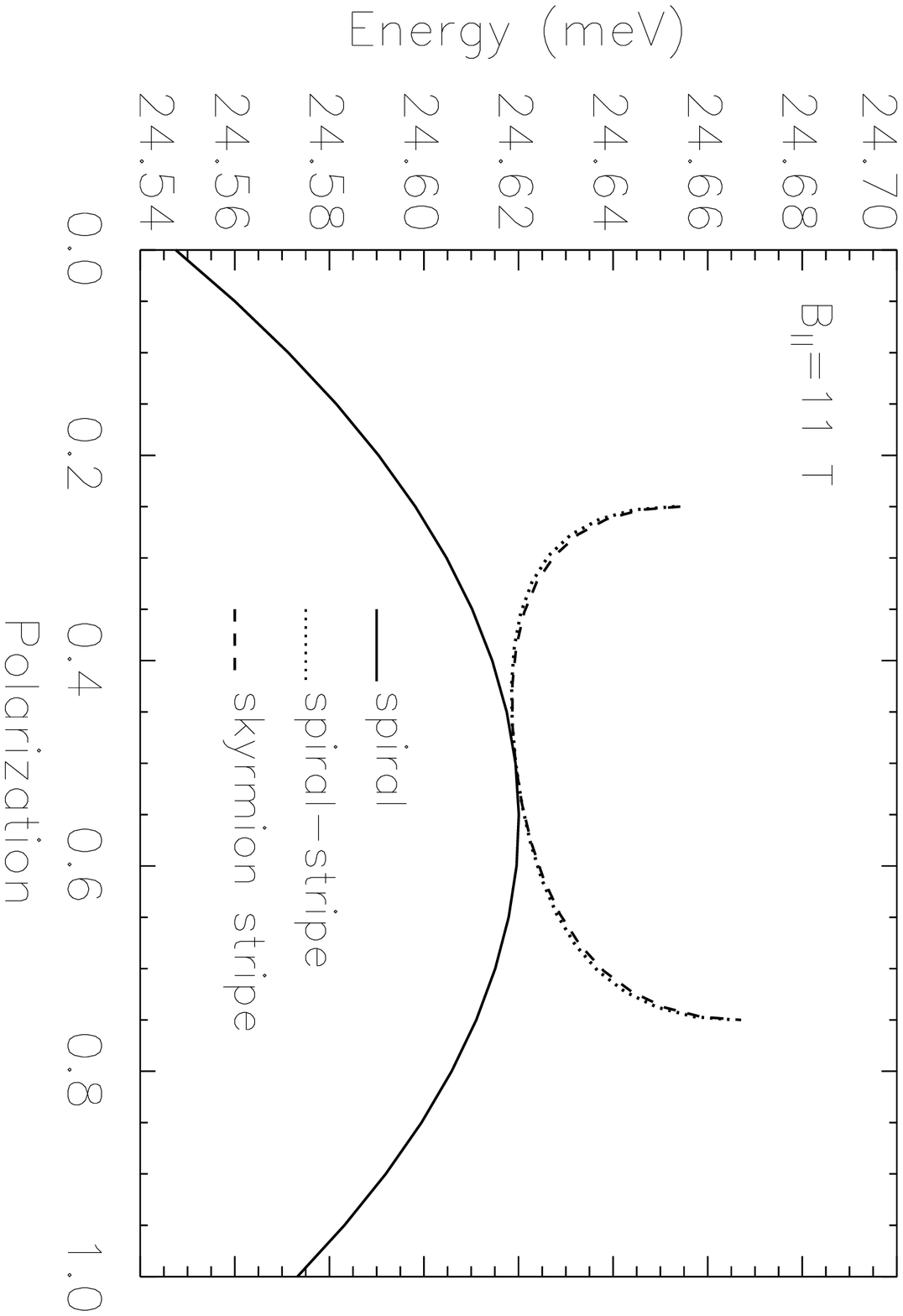}
   }
  \hss}
 }

\caption{
Comparison of energies between three
many-body phases of a wide well system at $\nu=6$ near the
level crossing point ($B^\ast_\|=11.1$ T) : spiral (solid line), spiral
stripe (dotted line) and skyrmion stripe (dashed line).
The system parameters are chosen to be the same as those in Fig.
\ref{energy_levels_figure}.
The polarization is defined by $\Theta_2(0)$ (see Eq. (\ref{def_theta})
and the text),
which is zero for the isospin up state (spin unpolarized)
and is one for the isospin down state (spin fully polarized).
For all three many-body phases, the isospin winding
wavevector $\vec{Q}_\perp=-\vec{Q}_\perp'$ 
has been chosen to be $0.75\, l_0^{-1}$
in $y$ direction (perpendicular to the in-plane field) and zero in
the $x$ direction, which
is the optimal value to minimize the HF energy if $\Theta_2(0)\neq 0, 1$.
(It can be also obtained from the wavevectors of the roton minimum in
the magnetoplasmon excitations as calculated in Ref. [17].)
The phases, $\gamma$ and $\gamma'$ in Eq. (\ref{wavefunction_4l_W2})
are set to be zero.
For the spiral stripe and skyrmion stripe phase,
the stripe phase function, $\psi_k$, is calculated variationally
by using equations in Appendix \ref{psi_k} and parameters defined
in Fig. \ref{fig_psi}. For the convenience of comparison,
here we have fixed $\psi_0=\pi/2$, $\xi=0.5$, $\eta=0$ and
stripe period $a=0.67\, l_0=10^{-6}$ cm, and only $\varphi$ is allowed
to vary from $-\pi/2$ to $+\pi/2$. When $\varphi=0$, the energies
of the stripe phases are the same as the spiral phase.
(Note that the directions of the spiral stripe and the skyrmion stripe
are different: the former one is along $x$ direction while the latter
one is along $y$ direction, see Sections \ref{diff_phases} and \ref{sec_W2}.)
We have checked that using 
other values does not change the figure qualitatively, and also
cannot stabilize any of these many-body phases.
In our calculation shown in this figure, the lowest energy state is always 
the conventional QH state (i.e. either spin unpolarized or spin fully
polarized states at $\Theta_2(0)=0$ or 1 respectively).
However, we find that the energy differences between these states
with the three many-body states (spiral, sprial stripe and skyrmion stripe)
are very small, and the lowest energy of the stripe phase curve (in both
spiral stripe and skyrmion stripe) is at finite $\varphi$, showing
that if only a uniform spiral phase is stablized by some more sophisticated
approximations, the stripe order may also be stablized with 
energy even lower than the uniform spiral phase. 
The stablization of a skyrmion stripe may be 
responsible to the resistance anisotropy observed in Ref. [15]
in the strong in-plane field region (see text).
}
\label{W2_eng_comp}
\end{figure}

For the level crossing at large $B_\|$ region, 
only one level crossing between
$\vec{n}_1=(0,N)$ of spin down and $\vec{n}_2=(0,N+1)$ of spin up
occurs and the next nearest two levels do not cross in the noninteracting
energy spectrum.
From Eqs. (\ref{E_0_W2})-(\ref{E_F_W2}) we find that if we 
fix $\psi_k=\psi_0$ to be uniform and finite ($0<\psi_k<\pi$), the HF 
energy is always minimized at a finite winding wavevector along $y$ 
direction, $\vec{Q}_\perp=\pm (0,Q_y^\ast)$ with $Q_y^\ast\sim 
0.75\, l_0^{-1}$ (the 
magnetic field is along $x$ axis), so the optimum state of broken spin 
symmetry must have a spiral or a collinear spin order (the latter happens when 
$\pm \vec{Q}$ components are present simultaneously). However, in
optimizing the HF energy with respect to $\psi_0$ we find that
the minimum is always at $\psi_0=0$ or $\pi$, so states with 
broken spin symmetry are not favored at the HF level.
In Fig. \ref{W2_eng_comp} we compare the HF energies of 
the isospin spiral, isospin spiral stripe, and isospin
skyrmion stripe phases, calculated using typical realistic
system parameters (for GaAs 2D systems) and by employing a 
more general phase function, $\psi_k$, as shown in Fig. \ref{fig_psi}.  
For the sake of comparison we fix most
parameters in the phase function of $\psi_k$ and let 
$\phi=\psi_1-\psi_2$ to be the only
free parameter for the stripe phases (we have tried different 
ranges of these variational parameters, but the results are qualitatively 
similar to Fig. \ref{W2_eng_comp} and no exotic 
many-body phase is found). In the horizontal axis
of Fig. \ref{W2_eng_comp}, we define
the spin polarization to be $\Theta_2(0)$,
which is proportional to the density of spin triplet excitons excited
from the top filled level to the lowest empty level. Another
coherence parameter, $\Theta_2'(0)$, is chosen to be zero in
Fig. \ref{W2_eng_comp} because the results do not depend qualitatively 
on this choice.
In Fig. \ref{W2_eng_comp} 
we choose $B_\|=11$ T in the calculation, slightly lower than
the level crossing point at $B_\|^\ast=11.1$ Tesla (note that
this is the renormalized level crossing field including
the HF self-energy correction and is therefore lower 
than the noninteracting result).
We find that, within our Hartree-Fock calculation, the energy
of the spiral phase is a convex curve as a function of the polarization,
and therefore 
has {\it no} energy minimum between 
$\Theta_2(0)=0$ (spin unpolarized state) 
and $\Theta_2(0)=1$ (spin polarized state), i.e. either the spin unpolarized
or the spin polarized state is always 
the lowest energy ground state, depending on whether the magnetic field is
smaller or larger than $B_\|^\ast=11.1$ Tesla.
Note that we have chosen the spiral wavevector to be 
$\vec{Q}_\perp^\ast=(0,Q_y^\ast)$, which is the extreme value for the
lowest HF energy for $0<\Theta_2(0)<1$. Unsurprisingly, we find that 
this $Q_y^\ast$ is the same as the wavevector 
obtained from 
the near softening point of the magnetoplasmon excitations \cite{wang02}.
(Therefore we can exclude the simple (incommensurate) 
isospin coherent (stripe) phases from the energy comparison, because 
they do not have spiral order and must have higher energy than the
three cases shown in Fig. \ref{W2_eng_comp}).)
Among the three many-body states of spiral order, 
we find that the spiral phase always has the lowest energy.
Our HF calculation therefore suggests a first order transition between spin
polarized and unpolarized ground states with no intermediate broken
symmetry phases in between (at least within the HF theory). 
We find, however, that the HF energy differences
between simple incompressible states and exotic many-body states, 
such as the spiral phase and the skyrmion stripe phase, are
very small ($<0.1$ meV). We therefore suggest (and speculate) that effects
not included in our analysis, e.g. self-consistent 
calculation for the single electron wavefunction, lower level 
screening, and/or nonparabolic effects of the well 
confinement etc., may now well stabilize the many-body states.
It is also plausible, given the smallness in the HF energy difference,
that higher-order corrections beyond the HF theory could 
stabilize exotic quantum order in this situation.

In Ref. \cite{pan01}, Pan {\it et. al.} observed strong anisotropic
longitudinal resistance when the in-plane field exceeded a 
certain critical value. This is very suggestive of the skyrmion
stripe phase, since the latter has charge modulation in addition to
the spin modulation, and therefore should lead to strong transport
anisotropy.
It is useful to point out that the direction of the charge modulation in 
the skyrmion stripe phase is fixed by the applied parallel magnetic field:
the winding of the transverse components of spin, 
$\langle {\cal I}_{x,y}\rangle$, is set by $\vec{Q}_\perp$ and is
perpendicular to $B_\|$, while the 
modulation of $\langle {\cal I}_{z}\rangle$ is along 
$B_\|$, so we have effectively a one-dimensional charge density wave 
that goes along $B_\|$ (see also the discussion about skyrmion
stripe phase in Section \ref{diff_phases}). Hence the expected 
``low'' resistance direction of the skyrmion stripe phase is perpendicular 
to $B_\|$, which is what was observed in Ref. \cite{pan01}.

An alternative interpretation of the resistance anisotropy has been recently
suggested in Ref. \cite{chalker02}, where Chalker {\it et. al.} argued that
surface disorder will form domains close to the first order phase transition,
that have anisotropic shape due to the presence of the tilted
magnetic field. Naive argument would suggest that these domains differ
only in the spin structure and should not contribute appreciably 
to the transport anisotropy. Analysis presented in this paper suggests,
however, that boundaries between different domains in this case
should be accompanied by the topological spin density, which leads
to change density modulation and may lead to large resistance anisotropy.
In addition there is always the possibility (already mentioned in 
Section \ref{result_W2_1}) that a direct first order phase transition
from the spin unpolarized to the spin polarized phase will give rise 
to domains with different (up or down) spin polarizations in the 
(effectively Ising) ferromagnetic phase. Again, these domains, separated 
by domain walls, would differ only in the spin orientations, and it 
is unclear how this could give rise to the observed resistance anisotropy 
seen in the experiments \cite{pan01,zeitler01}. Also, the domain 
structure should lead to hysteretic behavior in the observed 
resistance, which has not been reported.  We emphasize, however, that
the possibility of a direct first order (Ising type) transition
in the experiment of Ref. \cite{pan01} cannot be ruled out --- in fact,
our HF calculation does indeed predict such a transition (but with very
fragile energetics) as discussed above.

To summarize, within our approximations we do not find any exotic
phases as a true ground state close to the level crossing point
of $W2'$. However, the exotic phases are very close in energy, 
and therefore we speculate 
that the resistance anisotropy observed in Ref. \cite{pan01} may 
arise from the skyrmion stripe phase, which could be stabilized by effects not
included in our theory.

\section{Double well system at $\nu=4N+1$ ($D1$ case)}
\label{sec_D1}

As mentioned in the Introduction, in the double well system at
odd filling factor $\nu=4N+1$, many interesting phenomena
have been explored, such as the interlayer coherence 
(and the commensurate-incommensurate phase transition)
\cite{reviewbook,yang94,moon95,hanna01}, 
unidirectional charge density wave (stripe) state 
\cite{macdonald90_cdw_dqw,brey00,cote02,fertig89_cdw_dqw},
and the in-plane magnetic field induced charge imbalance phase
\cite{radzihovsky01} of the odd filling systems. Following
our earlier work in Ref. \cite{eugene01}, we will use a 
general trial wavefunction, Eq. (\ref{wavefunction_general}), 
to include the isospin stripe order 
and the spiral order simultaneously, to 
obtain a rich quantum phase diagram within a single unified theory
including all the effects mentioned above (which in the past have been 
studied in separate works using different techniques).
In this section, 
we assume the following (reasonable) ordering of energy scales 
for the double well system:
$\omega_\perp\gg \omega_z \gg \Delta_{SAS}$ (see Eq. (\ref{H0_Da})), 
so that the lowest $4N$ filled levels can be assumed to be frozen core 
states with no
coherence effect, and only the top filled level 
has coherence with the lowest
empty level of the same spin polarization (but opposite parity).
Within this approximation it is more convenient to 
follow the standard convention in the literature and consider the
isospin within the layer index basis to calculate the HF energy.

For the convenience of later discussion of the symmetry properties
of the many-body wavefunction, we first write down the many-body state
explicitly in the layer index ($l=\pm$) basis (for notational
simplicity, we suppress the Landau 
level index and the spin index throughout this section):
\begin{eqnarray}
|\Psi_G^{D1}(\psi_k,\vec{Q}_\perp,\gamma)\rangle
&=&\prod_k\left(e^{i(kQ_xl_0^2+\gamma)/2}
\cos(\psi_k/2)c^{D,\dagger}_{+,k-Q_y/2}
+e^{-i(kQ_xl_0^2+\gamma)/2}
\sin(\psi_k/2)c^{D,\dagger}_{-,k+Q_y/2}
\right)|LL\rangle.
\label{wavefunction_D1}
\end{eqnarray}
We then obtain the following expectation
value for the HF energy calculation:
\begin{eqnarray}
&&\langle{\Psi}^{D1}_G|c_{l_1,k_1}^{D,\dagger}
c^{D}_{l_2,k_2}|{\Psi}^{D1}_G\rangle
\nonumber\\
&=&\delta_{l_1,l_2}
\delta_{k_1,k_2}\left[
\cos^2(\psi_{k_1+Q_y/2}/2)\delta_{l_1,+1}+
\sin^2(\psi_{k_1-Q_y/2}/2)/2)\delta_{l_1,-1}\right]
\nonumber\\
&&+\delta_{l_1,-l_2}
\left[\cos(\psi_{k_2-Q_y/2}/2)\sin(\psi_{k_2-Q_y/2}/2)
\delta_{k_1+Q_y/2,k_2-Q_y/2}
\delta_{l_1,+1}e^{-iQ_x(k_2-Q_y/2)l_0^2-i\gamma}
\right.
\nonumber\\
&&\left.+\sin(\psi_{k_1-Q_y/2}/2)
\cos(\psi_{k_1-Q_y/2}/2)
\delta_{k_1-Q_y/2,k_2+Q_y/2}\delta_{l_1,-1}
e^{iQ_x(k_1-Q_y/2)l_0^2+i\gamma}\right].
\end{eqnarray}

\subsection{Hartree-Fock variational energy}

Neglecting the constant energies associated with Landau levels and 
Zeeman splittings, the noninteracting
single electron energy is entirely the tunneling energy:
\begin{eqnarray}
E_0^{D1}&=&-\sum_{k}t
\left[e^{-ikP_yl_0^2}e^{-iQ_xkl_0^2-i\gamma}
+e^{ikP_yl_0^2}e^{iQ_xkl_0^2+i\gamma}\right]
\delta_{P_x,Q_y}
\sin(\psi_k/2)\cos(\psi_k/2)
\nonumber\\
&=&-\Delta_{SAS}\delta_{P_x,Q_y}
\Theta_{3}(P_y+Q_x)\cos(\gamma).
\label{E0_D1}
\end{eqnarray}
The Hartree and the Fock energies can be written as (obtained by using
Eq. (\ref{H1_D})):
\begin{eqnarray}
E^{D1}_{H}&=&
\frac{N_\phi}{2\Omega_\perp}\sum_{q_n}\left[V^{D,++}_{NN,NN}(q_n)
\left(\Theta_1(q_n)^2+\Theta_2(q_n)^2\right)
+2V^{D,+-}_{NN,NN}(q_n)\cos(q_nQ_yl_0^2)
\Theta_1(q_n)\Theta_2(q_n)\right],
\label{E_H_D1}
\end{eqnarray}
and
\begin{eqnarray}
E_{F}^{D1}
&=&\sum_{q_n}\frac{-1}{2\Omega_\perp}\sum_{\vec{q}}\left\{
V^{D,++}_{NN,NN}(\vec{q})\cos(q_nq_yl_0^2)
\left[\Theta_1(q_n)^2+{\Theta_2(q_n)}^2\right]
+2V^{D,+-}_{NN,NN}(\vec{q})\cos((Q_xq_y-Q_yq_x)l_0^2)
\cos(q_nq_yl_0^2){\Theta_{3}(q_n)}^2 \right\},
\label{E_F_D1}
\end{eqnarray}
where we have used $V^{D,++}_{NN,NN}(q_n)=
V^{D,--}_{NN,NN}(q_n)$, and neglected the constant energies 
associated with the frozen core levels.
Combining Eqs. (\ref{E0_D1})-(\ref{E_F_D1}) we obtain
the total HF energy as follows:
\begin{eqnarray}
E^{D1}_{HF}&=&-\Delta_{SAS}\delta_{P_x,Q_y}
\Theta_{3}(P_y+Q_x)\cos(\gamma)
+\frac{1}{2}\sum_{q_n}\left\{\left[E^+_H(q_n)+E^+_F(q_n)\right]\cdot
\left(\Theta_1(q_n)^2+\Theta_2(q_n)^2\right)\right.
\nonumber\\
&&\left.+2E^-_H(q_n;Q_y)
\Theta_1(q_n)\Theta_2(q_n)+
2E^-_F(q_n;\vec{Q}_\perp)\Theta_{3}(q_n)^2
\right\},
\label{E_HF_D1}
\end{eqnarray}
where the definition is obvious by comparing Eq. (\ref{E_HF_D1})
with Eqs. (\ref{E0_D1})-(\ref{E_F_D1}).
The only $\gamma$-dependent term is from the tunneling amplitude, 
reflecting the fact that the
isospin rotational symmetry is broken by electron tunneling.
We will discuss the symmetry properties in details later.
\subsection{Commensurate, incommensurate, and charge imbalance phases}

We first analytically discuss a special class of many-body states
implied by Eq. (\ref{wavefunction_D1}) in the absence of any 
stripe order, i.e.
$\tilde{\psi}_k=\tilde{\psi}_0^\ast$ is a constant.
Taking $q_n=0$ in Eq. (\ref{E_HF_D1}), we obtain the following 
simplified HF energy ($\tilde{\gamma}$ is set to be zero):
\begin{eqnarray}
E^{HF}_0(\vec{Q}_\perp)
&=&-\Theta_{3}(0)\left\{\Delta_{SAS}
\delta_{Q_y,P_x}\delta_{Q_x,-P_y}
+E_\Delta(\vec{Q}_\perp)\Theta_{3}(0)\right\},
\label{E_anal_D1}
\end{eqnarray}
where $E_\Delta(\vec{Q}_\perp)=E^+_H(0)-E^-_H(0;Q_y)+E^+_F(0)-
E^-_F(0;\vec{Q}_\perp)$.
We have used 
$\Theta_1(0)\Theta_2(0)=\Theta_3(0)^2$ and neglected
the irrelevant constant energy.
Following the existing literature, 
we separate the discussion in two parts: systems in an
incommensurate state (i.e. the coherent phase we defined in Section 
\ref{diff_phases} with the extreme value of winding
wavevector,  $\vec{Q}^\ast_\perp=0$), 
and systems in a commensurate state (i.e. the spiral phase defined
in Section \ref{diff_phases} with $\vec{Q}^\ast_\perp
=(-P_y,P_x)$). When the system is in an incommensurate state,
the tunneling amplitude is effectively zero according to Eq. 
(\ref{E_anal_D1}),
and therefore the extreme value of $\tilde{\psi}_0$ 
is determined by the sign of 
$E_\Delta(\vec{Q}^\ast_\perp)$.
For $E_\Delta(\vec{Q}^\ast_\perp) > 0$
(i.e. the Hartree energy dominates the Fock energy),
the minimum energy is at $\Theta^\ast_{3}(0)=1/2$ 
or $\psi^\ast_0=\pi/2$, 
indicating an equal population of electrons in
the two layers. However, 
if $E_\Delta(\vec{Q}^\ast_\perp) < 0$
(i.e. the Fock energy dominates the Hartree energy),
the minimum value of $E^{HF}_0(\vec{Q}^\ast_\perp)$
is at $\Theta^\ast_{3}(0)=0$ (i.e. $\psi^\ast_0=0$ or $\pi$), 
and therefore 
we obtain a fully spontaneous charge 
imbalanced state \cite{radzihovsky01}, where 
all electrons like to accumulate in a single layer rather than distribute
equally in the two layers (this is true, of course, only within the HF
approximation where the correlation energy is totally neglected).
In the commensurate state (i.e. $Q_x=-P_y$ and $Q_y=P_x$) and
$E_\Delta(\vec{Q}_\perp^\ast)<-\Delta_{SAS}$,  
the total energy in Eq. (\ref{E_anal_D1}) is minimized at 
$0<\Theta^\ast_3(0)=|\Delta_{SAS}/2E_\Delta(\vec{Q}^\ast_\perp)|<1/2$,
showing a spontaneous partial charge imbalance phase.
Otherwise, for $E_\Delta(\vec{Q}_\perp^\ast)>-\Delta_{SAS}$, 
the commensurate phase always has equal number of electrons in 
the two layers, i.e. the usual isospin paramagnetic phase.
Therefore, the in-plane magnetic field can cause not only a 
commensurate-incommensurate phase transition, but also a spontaneous
charge imbalance phase when the exchange energy is large. We note that
such exchange driven spontaneous charge imbalance
phases could arise even in the zero-field (i.e. $B_{tot}=0$) non-quantum-Hall
bilayer 2D systems within a restricted HF approximation 
\cite{zerofield_bilayer}, but in the zero-field 
case the corresponding $XY$ ``isospin magnetic'' state has been
shown to be lower in the energy than the Ising-type charge imbalance
phase for the long range Coulomb interaction.

In the present paper, we only consider the
long-range Coulomb interaction, which gives
$E^+_H(0)-E^-_H(0;Q_y)=e^2/\epsilon l_0^2$ ($\epsilon$ is the 
dielectric constant of the system), and therefore the Hartree
electrostatic energy always 
dominates the Fock exchange energy (i.e. $E_\Delta(\vec{Q}_\perp)>0$), 
eliminating any spontaneous charge imbalance between the
two layers. 
On the other hand, as will be shown later, we may obtain 
a commensurate stripe phase with a longer
period ($a\gg l_0$) in the small layer separation region, which is the
asymptotic behavior of the charge imbalance phase recently discussed by
Radzihovsky {\it et. al.} \cite{radzihovsky01}.

\subsection{Numerical results and the stripe phases}

\begin{figure}

 \vbox to 5.8cm {\vss\hbox to 5.cm
 {\hss\
   {\includegraphics{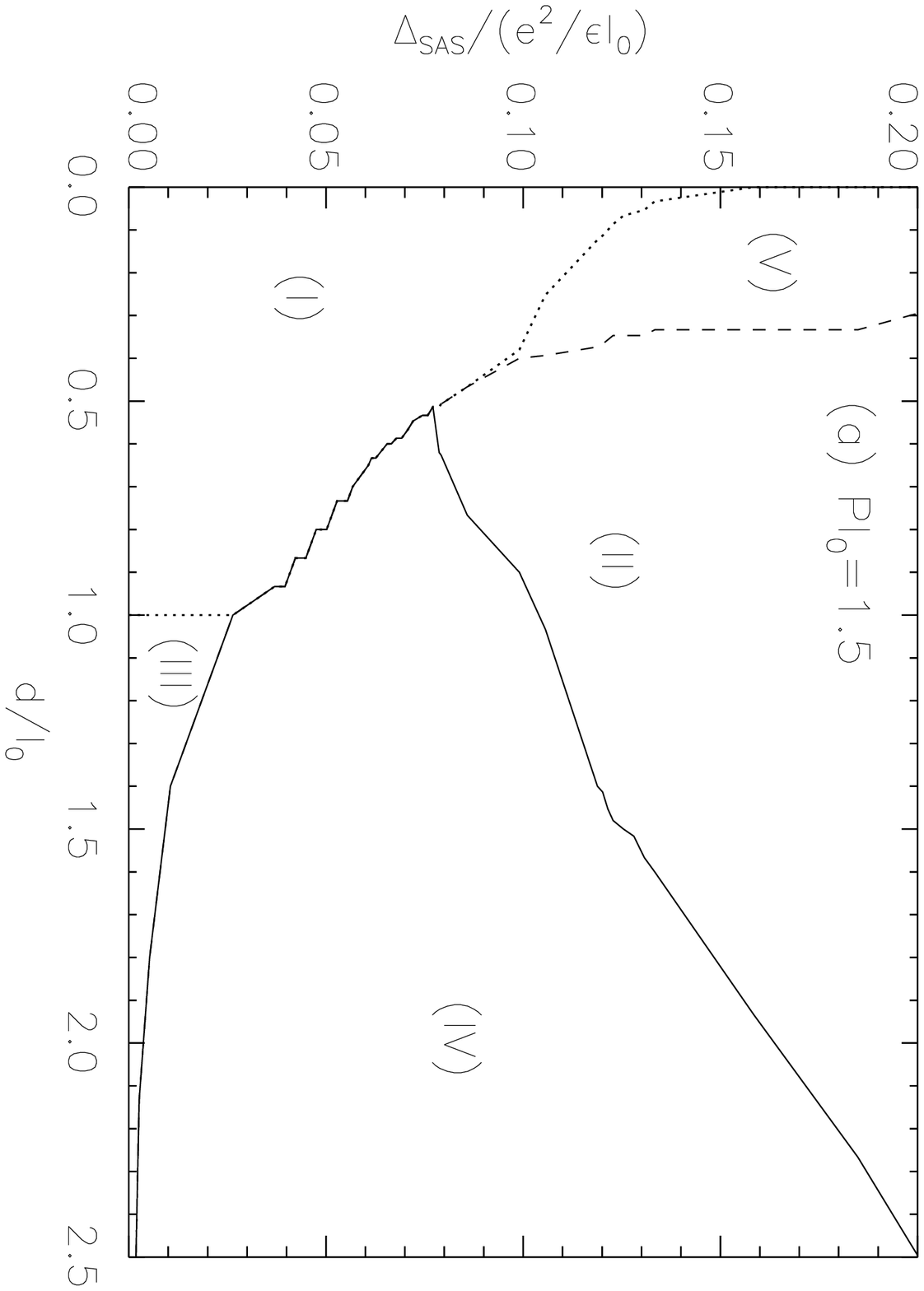}
   }
  \hss}
 }
 \vbox to 5.8cm {\vss\hbox to 5.cm
 {\hss\
   {\includegraphics{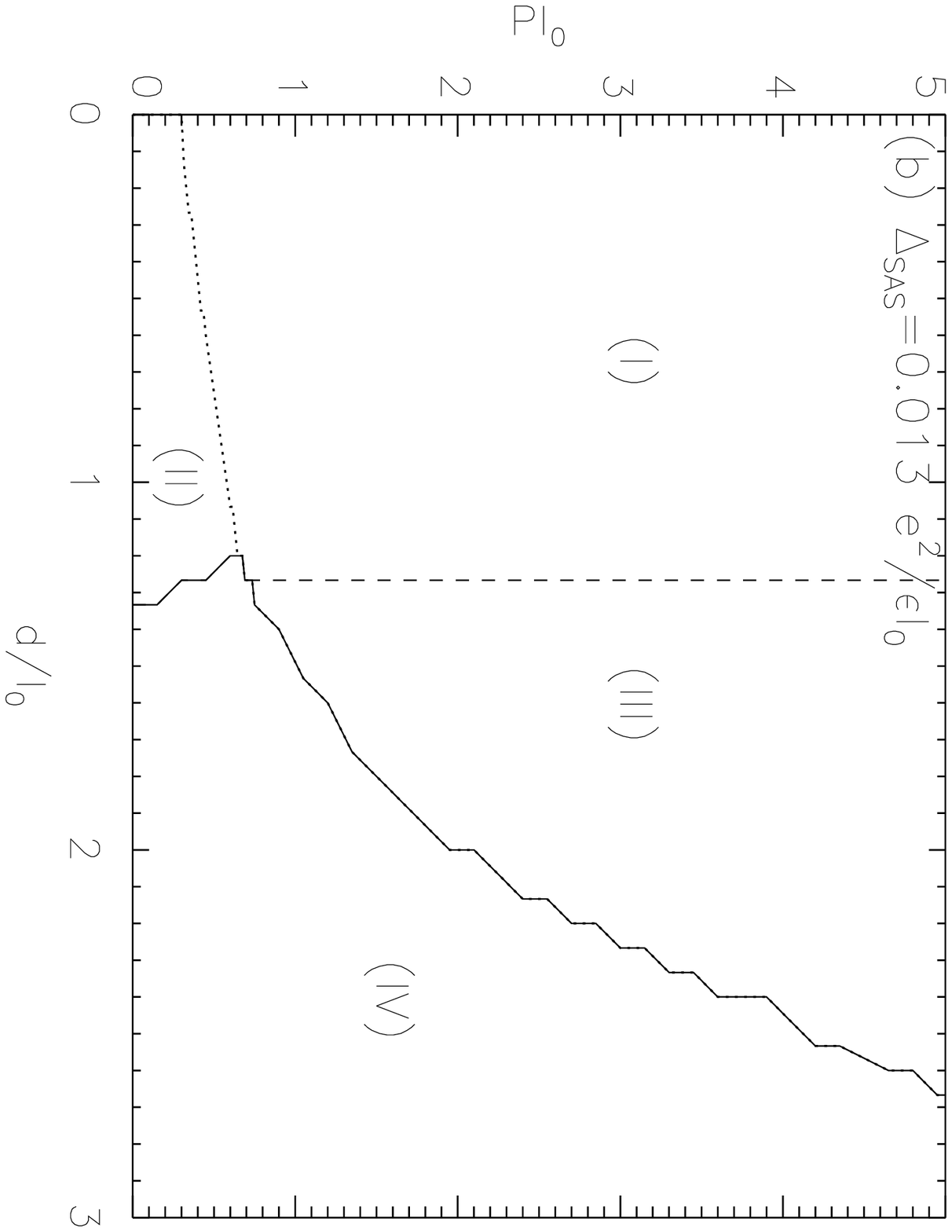}
   }
  \hss}
 }
\caption{Phase diagram of the bilayer system at $\nu=5$
in the presence of parallel magnetic field.
(a) for fixed in-plane magnetic field 
($Pl_0=2 \pi B_\parallel d l_0/\Phi_0=1.5$), and (b) for
fixed tunneling amplitude ($\Delta_{SAS}=0.013 e^2/\epsilon l_0$).
In isospin language defined by layer index basis, 
phase I is an isospin coherent phase, phase II
is an isospin spiral phase, phase III is an isospin stripe
phase, phases IV and V are isospin spiral stripe phases.
Note that phase V has a very long stripe period, and is related to the
charge imbalance phase for the short-ranged interaction (see text).
}
\label{D1_phase_diag}
\end{figure}

Fig. \ref{D1_phase_diag}(a) and (b) are the phase diagrams
we obtain at zero temperature for $\nu=5$ for the in-plane magnetic
field fixed in $x$ direction.
Using the isospin many-body phases defined in Section 
\ref{trial_wavefunction},
phase I is the coherent phase;
phase II is the spiral phase, where the optimal spiral winding
wavevector $\vec{Q}_\perp^\ast=(0,Q^\ast_y)
=(0,P_x)$ is perpendicular to the in-plane field direction;
phase III is the coherent stripe phase,
where the stripe direction can be in arbitrary direction, 
and phases IV and V correspond to the spiral stripe phase,
where $\vec{Q}_\perp^\ast$ is perpendicular to $B_\|$ and the stripe
is aligned in $x$ direction (i.e. $\langle {\cal I}_z\rangle$ modulates
in $y$ direction, parallel to $\vec{Q}$).
To obtain the stripe phase (phases III, IV and V), we numerically minimize
the HF variational energy by using a general stripe
phase function as shown in Fig. \ref{fig_psi} and in 
Appendix \ref{psi_k}.
(The perturbation method developed in Section 
\ref{pert_stripe} also gives similar results, since the stripe 
formation in this system
is a second order transition.)
As mentioned in the beginning of Section \ref{trial_wavefunction}, 
to get the energy of a spiral stripe phase with 
stripe normal direction 
perpendicular to the in-plane field, we can just 
rotate the in-plane field direction from the $x$ to the $y$ axis. 
Therefore the results 
shown in \ref{D1_phase_diag}(a) and (b) are gauge independent.
Note that the stripe period of phase V is very large ($a\gg l_0$),
showing an asymptotic behavior of charge imbalance phase modified
by the long-range Coulomb interaction.
Using the existing terminology of the literature 
\cite{cote02,reviewbook,yang94},
phases I and II are the incommensurate and commensurate phases 
respectively, and 
phases III and IV are the incommensurate and commensurate stripe phases
respectively. We will use this terminology as well as the isospin
phases mentioned above (defined in Section \ref{diff_phases})
for later discussion in this paper.

We note, however, that our HF calculation does not
incorporate the possibility of two decoupled compressible
$\nu=2N+1/2$ states in each layer, which could be 
energetically favored for 
smaller $N$ (lower values of $\nu$, e.e. $\nu=1$). We expect that the phases
discussed in this paper are more likely
to be found for $N>0$, 
when each of the layers becomes susceptible to forming a stripe phase
\cite{review,lilly99,du99}, especially in the presence of a parallel
magnetic field. Particle-hole symmetry implies that
similar states should also occur at filling factors
$4N+3$ by interchanging the role of holes and electrons
in the top filled Landau level.

\subsection{Symmetry properties of the commensurate and incommensurate
states}
\label{symmetry_CI}

Applying Eqs. (\ref{parity_D}) and
(\ref{translation_cx2})-(\ref{translation_cy2}) to the wavefunction,
$|\Psi^{D1}_G(\psi_k^\ast,\vec{Q}_\perp^\ast,\gamma^\ast)\rangle$
in Eq. (\ref{wavefunction_D1}) (where $\psi_k^\ast,\vec{Q}_\perp^\ast$,
and $\gamma^\ast$ denote the extreme values to minimize the HF energy), 
we obtain 
\begin{eqnarray}
\hat{\cal P}|\Psi^{D1}_G(\psi_k^\ast,\vec{Q}_\perp^\ast,
\gamma^\ast)\rangle
&=&\prod_k\left(e^{i(kQ^\ast_xl_0^2+\gamma^\ast)/2}
\cos(\psi_k^\ast/2)c^{D,\dagger}_{-,-k+Q_y^\ast/2}
+e^{-i(kQ_x^\ast l_0^2+\gamma^\ast)/2}
\sin(\psi_k^\ast/2)c^{D,\dagger}_{+,-k-Q_y^\ast/2}
\right)|LL\rangle
\nonumber\\
&=&(-i)^{N_\phi}|\Psi^{D1}_G(\psi_{-k}^\ast+\pi,\vec{Q}_\perp^\ast,
-\gamma^\ast+\pi)\rangle
\label{sym_parity_D1}
\\
\hat{\cal T}_x (R_x)|\Psi_G(\psi_k^\ast,
\vec{Q}_\perp^\ast,\gamma^\ast)\rangle 
&=& \prod_k\left(e^{i(kQ_x^\ast l_0^2+\gamma^\ast-R_xP_y)/2}
\cos(\psi_k^\ast/2)c^{D,\dagger}_{+,k+R_x/l_0^2-Q_y^\ast/2}
\right.
\nonumber\\
&&\hspace{1.5cm}\left.+e^{-i(kQ_x^\ast l_0^2+\gamma-R_xP_y/2)/2}
\sin(\psi_k^\ast/2)c^{D,\dagger}_{-,k+R_x/l_0^2+Q_y^\ast/2}
\right)|LL\rangle
\nonumber\\
&=&|\Psi_G(\psi_{k-R_x/l_0^2}^\ast,
\vec{Q}_\perp^\ast,\gamma^\ast-R_x(Q_x^\ast+P_y))\rangle,
\label{sym_translation_D1x}
\\
\hat{\cal T}_y (R_y)|\Psi_G(\psi_k^\ast,
\vec{Q}_\perp^\ast,\gamma^\ast)\rangle 
&=& \prod_k\left(e^{i(kQ_x^\ast l_0^2+\gamma+R_yP_x)/2}e^{i(k-Q_y^\ast/2)R_y}
\cos(\psi_k^\ast/2)c^{D,\dagger}_{+,k-Q_y^\ast/2}
\right.
\nonumber\\
&&\hspace{1.5cm}\left.+e^{-i(kQ_x^\ast l_0^2+\gamma+R_yP_x/2)/2}
e^{i(k+Q_y^\ast/2)R_y}\sin(\psi_k^\ast/2)c^{D,\dagger}_{-,k+Q_y^\ast/2}
\right)|LL\rangle
\nonumber\\
&=&|\Psi_G(\psi_{k}^\ast,
\vec{Q}_\perp^\ast,\gamma^\ast-R_y(Q_y^\ast-P_x))\rangle,
\label{sym_translation_D1y}
\end{eqnarray} 
where we have used the fact that the guiding center coordinates, $k$,
can be shifted and $\sum_k k=0$ in a completely filled Landau level. 

Now we can study the symmetry properties of the many-body phases
obtained earlier in the HF approximation. In-plane field, $B_\|$, 
is fixed along $x$ axis and hence $P_y=0$.
According to Eqs. (\ref{sym_parity_D1})-(\ref{sym_translation_D1y}),
we find that 
(i) for an incommensurate state ($=$ coherent
phase with neither isospin spiral
nor stripe order: $\psi_k^\ast=\pi/2$, $\vec{Q}^\ast_\perp=0$, and 
$\gamma^\ast=$ arbitrary), $\hat{\cal P}|\Psi^{D1}_G\rangle
=\hat{\cal T}_x|\Psi^{D1}_G\rangle=|\Psi^{D1}_G\rangle$ and 
therefore the parity symmetry and the translational symmetry in $x$
direction are not broken.
However, since $\hat{\cal T}_y (R_y)|\Psi_G(\psi_k^\ast,
\vec{Q}_\perp^\ast,\gamma^\ast)\rangle=|\Psi_G(\psi_k^\ast,
\vec{Q}_\perp^\ast,\gamma^\ast+R_yP_x)\rangle$ is in general not
equal to the original wavefunction, and the HF energy is independent 
of phase $\gamma^\ast$ (the tunneling term is effective absence in the 
incommensurate phase),
the translational symmetry in 
$y$ direction {\it is} therefore broken, with an oscillation period of 
$2\pi/|P_x|$. 
Note that the above translational symmetry breaking in the incommensurate
phase is obtained within the mean field HF approximation, which is {\it not}
the same as the soliton lattice ground state obtained in Ref.
\cite{hanna01} by using an effective field theory to go beyond the mean field
approximation. We should expect to see a gapless Goldstone mode associated
with the broken translational symmetry in the time-dependent Hartree-Fock
approximation.
(ii) For a commensurate phase ($=$ spiral phase:
$\psi_k^\ast=\pi/2$, $\vec{Q}_\perp=(0,P_x)$, and 
$\gamma^\ast=0$), we find that {\it no} symmetry is broken at all,
since no distinct wavefunction is obtained after applying 
$\hat{P}$, $\hat{T}_x$ and $\hat{T}_y$ operators.
In other words, the commensurate (or spiral) phase is basically
the same as an isospin polarized state from the symmetry point of view.
It is not surprising because one can show that such commensurate
(or isospin spiral) state is essentially the same as the noninteracting
ground state, $\prod_k a^\dagger_{n,+,s,k}|LL\rangle$ 
(where $a^\dagger_{n,\alpha,s,k}$ is defined in Eq. (\ref{change_basis})
for the noninteracting energy eigenstate),
of the double well system,
which is certainly an eigenstate of all of these symmetry operators.
(iii) For an incommensurate stripe phase ($=$ coherent stripe:
$\psi_k^\ast=\psi^\ast_{k+a/2\pi l_0^2}$, $\vec{Q}_\perp=0$, and 
$\gamma^\ast=$ arbitrary), we find that the translational
symmetries in both $x$ and $y$ directions are broken.
However, we note that another incommensurate stripe phase, whose isospin 
$z$ component, $\langle {\cal I}_z\rangle$, modulates in $y$ direction
(perpendicular to $B_\|$), 
is also a degenerate ground state at the mean field level,
breaking the translational symmetry 
in $y$ direction {\it only}. The wavefunction of this second 
incommensurate stripe phase cannot be simply described
in the present Landau gauge, $\vec{A}_{[y]}(\vec{r}\,)$, and therefore 
its energy is not shown in the equations presented in this section. 
We may, however, consider an equivalent state by assuming the
in-plane magnetic field to be along $y$ direction with stripe modulation
along $x$ so that the translational symmetry is broken only in the direction
perpendicular to the in-plane field direction in this second type of 
incommensurate stripe phase. Parity symmetry is, however, broken 
in both the incommensurate stripe phases.
(iv) Finally, for a commensurate stripe ($=$ spiral stripe:
$\psi_k^\ast$ modulates in $y$ direction, $\vec{Q}_\perp=(0,P_x)$, and 
$\gamma^\ast=0$), its wavefunction cannot simply be described by 
the present gauge $\vec{A}_{[y]}(\vec{r}\,)$, and we can obtain 
it by effectively rotating the in-plane field direction 
as described above.
By changing the in-plane field direction from $\hat{x}$ to $\hat{y}$,
the spiral wavevector becomes $\vec{Q}_\perp^\ast=(-P_y,0)$, and
Eqs. (\ref{sym_parity_D1})-(\ref{sym_translation_D1y}) tell us 
that both parity symmetry and translational symmetry in $x$ direction 
(now it is perpendicular to the in-plane field) are broken, 
while translational symmetry in $y$ direction (parallel to the 
in-plane field) is preserved. We then 
conclude that the parity symmetry and the translational symmetry
in the direction {\it perpendicular} to the in-plane field (i.e. in $y$ axis) 
are both broken in this commensurate stripe phase in the presence of $B_\|$ 
field (in $x$ direction). 
This direction of broken translational symmetry
is also parallel to the direction of isospin spiral
wavefunction and the stripe oscillation (as defined for the spiral 
stripe phase in Section \ref{symmetry_wavefunction}).

\section{Double well system at $\nu=4N+2$ ($D2$ case)}
\label{sec_D2}

For a double well system at even filling factor, $\nu=4N+2$, both spin and 
layer indices are involved in the level crossing region 
(see Fig. \ref{level_fig}(f)), where the cyclotron 
resonance energy, $\omega_\perp$,
is (realistically) assumed to be much 
larger than the tunneling energy ($\Delta_{SAS}$)
and the Zeeman energy ($\omega_z$) so that
the two top filled levels and the two lowest 
empty levels belong to the same orbital quantum number, well-separated
from all other filled or empty levels, which are considered frozen and 
neglected from our consideration.
As mentioned in Section \ref{Hamiltonian_D}, an appropriate 
basis for the isospinor for this system is the basis of noninteracting
energy eigenstates,
which have definite parity and spin symmetries.
(Our definition of isospin coherent and isospin spiral phases in Section
\ref{diff_phases} then just corresponds to the fully commensurate
and partially incommensurate phases in the literature
\cite{burkov02}. For the convenience of comparison, we will use the 
conventional language for most of our discussion in this section.)
We note that several interesting phenomena associated with the CAF phase
in this system 
have been studied in the literature
in the recent years
\cite{canted_phase,bilayer_sym_break,bilayer_sym_break2,burkov02,caf_exp1,caf_exp2,yang99_effect_D2}.
A spin symmetry broken canted antiferromagnetic phase
can be stabilized in this system in addition to the usual 
symmetric and fully ferromagnetic states in the $B_\|=0$ situation
\cite{canted_phase}. 
In the canted phase electrons in the 
two layers hold the same $z$ component of spin polarization,
while they have opposite spin direction in the $x-y$ plane,
showing a two-dimensional antiferromagnetic order.
When an in-plane magnetic field is applied, such 
a canted phase remaining in a commensurate
state and does not exhibit a commensurate-incommensurate phase transition at
large $B_\|$ as observed in $\nu=4N+1$ case \cite{burkov02}.
However, the possible stripe phase formation in the presence of an in-plane
magnetic field for a larger layer separation has not yet been explored in the
literature. For the sake of completeness,
we will use the trial wavefunction proposed in
Section \ref{trial_wf_4l} and the perturbation method developed in
Section \ref{pert_stripe} to study the possibility of a stripe formation
in the double well system at even filling factors.

For a double well system at $\nu=4N+2$, we have argued that the trial 
wavefunction proposed in Eq. (\ref{wavefunction_4l_D2}), although not
the most possible wavefunction, should be a reasonable approximation
to describe the ground state in the presence of in-plane
magnetic field. Thus we again start from calculating the following 
expectation value by using the trial wavefunction, $|{\Psi}^{D2}_G\rangle$
shown in Eq. (\ref{wavefunction_4l_D2}):
\begin{eqnarray}
&&\langle{\Psi}^{D2}_G|a_{m_1,\alpha_1,\sigma_1,k_1}^\dagger
a^{}_{m_2,\alpha_2,\sigma_2,k_2}|{\Psi}^{D2}_G\rangle
\nonumber\\
&=&\delta_{m_1,m_2}\delta_{\alpha_1,\alpha_2}
\delta_{\sigma_1,\sigma_2}\delta_{k_1,k_2}\left[
\cos^2(\psi_{k_1+Q_y/2}/2)\delta_{m_1,N}
\delta_{\alpha_1,+1}\delta_{\sigma_1,1/2}+
\sin^2(\psi_{k_1-Q_y/2}/2)\delta_{m_1,N}
\delta_{\alpha_1,-1}\delta_{\sigma_1,-1/2}
\right.\nonumber\\
&&+\left.\cos^2(\psi'_{k_1+Q'_y/2}/2)\delta_{m_1,N}
\delta_{\alpha_1,+1}\delta_{\sigma_1,-1/2}+
\sin^2(\psi'_{k_1-Q'_y/2}/2)\delta_{m_1,N}
\delta_{\alpha_1,-1}\delta_{\sigma_1,1/2}+
\theta(N-m_1)\right]
\nonumber\\
&&+\delta_{m_1,N}\delta_{\alpha_1,-\alpha_2}\delta_{\sigma_1,-\sigma_2}
\left[\cos(\psi_{k_2-Q_y/2}/2)\sin(\psi_{k_2-Q_y/2}/2)
e^{-iQ_x(k_2-Q_y/2)l_0^2-i\gamma}\delta_{k_1,k_2-Q_y}\delta_{\alpha_1,+1}
\delta_{\sigma_1,+1/2}
\right. \nonumber\\
&&+\cos(\psi_{k_1-Q_y/2}/2)\sin(\psi_{k_1-Q_y/2}/2)
e^{iQ_x(k_1-Q_y/2)l_0^2+i\gamma}\delta_{k_2,k_1-Q_y}\delta_{\alpha_1,-1}
\delta_{\sigma_1,-1/2}
\nonumber\\
&&+\cos(\psi'_{k_2-Q'_y/2}/2)\sin(\psi'_{k_2-Q'_y/2}/2)
e^{-iQ'_x(k_2-Q'_y/2)l_0^2-i\gamma'}\delta_{k_1,k_2-Q'_y}\delta_{\alpha_1,+1}
\delta_{\sigma_1,-1/2}
\nonumber\\
&&+\left.\cos(\psi'_{k_1-Q'_y/2}/2)\sin(\psi'_{k_1-Q'_y/2}/2)
e^{iQ'_x(k_1-Q'_y/2)l_0^2+i\gamma'}\delta_{k_2,k_1-Q'_y}\delta_{\alpha_1,-1}
\delta_{\sigma_1,1/2}\right],
\end{eqnarray}
which is basically the same as Eq. (\ref{exp_value_W2}) for a wide well
system at $\nu=2N+2$, except for the different quantum numbers associated
with the Landau levels.

\subsection{Hartree-Fock variational energy}

The full Hartree-Fock energies (noninteracting, Hartree, and Fock energies
respectively) 
can be obtained as follows:
\begin{eqnarray}
E_0^{D2}&=&\frac{1}{2}
(-\Delta_{SAS}+\omega_z)(\Theta_1(0)-\Theta_2(0))
+\frac{-1}{2}(\Delta_{SAS}+\omega_z)(\Theta'_1(0)-\Theta'_2(0)),
\label{E0_D2}
\\
E_H^{D2}
&=&\frac{N_\phi}{2\Omega_\perp}\sum_{q_n\neq 0}{V}^I_{NN,NN}(q_n)
\left[\Theta_1(q_n)^2+\Theta_2(q_n)^2+\Theta'_1(q_n)^2+\Theta'_2(q_n)^2
\right.\nonumber\\
&&\left.+2\cos(Q_yq_nl_0^2)\Theta_1(q_n)\Theta_2(q_n)
+2\cos(Q'_yq_nl_0^2)\Theta'_1(q_n)\Theta'_2(q_n)
\right.
\nonumber\\
&&+2\cos((Q_y-Q_y')q_nl_0^2/2)\left(\Theta_1(q_n)\Theta'_1(q_n)
+\Theta_2(q_n)\Theta'_2(q_n)\right)
\nonumber\\
&&\left.+2\cos((Q_y+Q_y')q_nl_0^2/2)\left(\Theta_1(q_n)\Theta'_2(q_n)
+\Theta'_1(q_n)\Theta_2(q_n)\right)\right],
\label{E_H_D2}
\end{eqnarray}
and
\begin{eqnarray}
E_F^{D2}
&=&\frac{-1}{2\Omega_\perp}\sum_{\vec{q}_\perp}{V}_{NN,NN}^I(\vec{q}_\perp)
\sum_{n}\cos(q_nq_yl_0^2)\left[{\Theta_1(q_n)}^2
+{\Theta_2(q_n)}^2+2\cos((Q_xq_y-Q_yq_x)l_0^2){\Theta_{3}(q_n)}^2\right.
\nonumber\\
&&\hspace{1cm}+\left.{\Theta'_1(q_n)}^2+{\Theta'_2(q_n)}^2
+2\cos((Q'_xq_y-Q'_yq_x)l_0^2){\Theta'_{3}(q_n)}^2\right]
\nonumber\\
&&+\frac{-2}{2\Omega_\perp}\sum_{\vec{q}_\perp}{V}_{NN,NN}^o(\vec{q}_\perp)
\sum_{n}\cos(q_nq_yl_0^2)\cos(q_n(Q_y+Q_y')l_0^2)
\left[\Theta_1(q_n)\Theta'_2(q_n)+
\Theta_2(q_n)\Theta'_1(q_n)\right]
\nonumber\\
&&+\frac{-4\delta_{Q_y,-Q_y'}}{2\Omega_\perp}\sum_{\vec{q}_\perp}
{V}_{NN,NN}^o(\vec{q}_\perp)\cos(\gamma+\gamma')
\sum_{n}\Theta_{3}(q_n-Q_x)\Theta'_{3}(q_n+Q_x')\cos((q_xQ_y-q_nq_y)l_0^2),
\label{E_F_D2}
\end{eqnarray}
where the last term of the exchange energy plays the same role as
the last term of Eq. (\ref{E_F_W2}) in the wide well system at $\nu=2N+2$, 
coupling the nearest and the next nearest pairs of levels to 
stabilize a coherent phase.
The minimum of the HF energy is achieved for 
$\gamma+\gamma'=2m\pi$ ($m$ is an integer)
and arbitrary $\gamma-\gamma'$.

\subsection{Nonstripe phases}
\label{CAF}

For the sake of completeness and later discussion,
here we first briefly review some results of the CAF phase 
in a double well system in the presence of in-plane magnetic field 
at even filling factors in our theory.
We note that this system
has also been studied recently in Ref. \cite{burkov02} 
by numerically solving a single
electron HF equation. 
As has been mentioned in Section \ref{4l_D}, in the energy eigenstate
basis we use in Eq. (\ref{change_basis}), the in-plane magnetic field has 
been incorporated in the phase difference between the electrons
in the right layer and in the left layer, so that a fully incommensurate
state \cite{burkov02} has been automatically excluded
in our trial wavefunction (see discussion in Section \ref{4l_D}).
The HF energy of such a fully incommensurate state, however,
can still be calculated if we let tunneling and in-plane magnetic 
field to be zero in the above HF energy expression and let $\vec{P}_\perp
=\vec{Q}_\perp=\vec{Q}_\perp'=0$, since it is well-known 
that the tunneling energy is effectively absent when the 
in-plane field exceeds the critical value for 
the commensurate-to-incommensurate
phase transition. The HF energy of a fully commensurate
state is calculated by setting $\vec{Q}_\perp=-\vec{Q}_\perp'=0$ in the HF
variational energy, and the energy of a partially
commensurate/incommensurate state is obtained by setting 
$\vec{Q}_\perp\neq -\vec{Q}_\perp'$
or $\vec{Q}_\perp=-\vec{Q}_\perp'\neq 0$.
The stripe phase functions, $\psi_k$ and $\psi_k'$, are just 
constant variational parameters for studying the nonstripe phases.

Inspecting the last term of Eq. (\ref{E_F_D2})), it is easy to see
that the minimum energy should be always at $\vec{Q}_\perp=-\vec{Q}_\perp'$ 
to stabilize the CAF phase even in the absence of in-plane field, 
i.e. we need only consider
$\vec{Q}_\perp'=-\vec{Q}_\perp$ for both fully commensurate
($\vec{Q}_\perp=0$) and partially commensurate/incommensurate 
($\vec{Q}_\perp\neq 0$) states. (In the terminology defined in Section
\ref{diff_phases} of this paper, these are just the isospin coherent and 
isospin spiral phases respectively in the noninteracting 
energy eigenstate basis. As stated above, we are using 
the conventional terminology 
for most of our discussion in this section to avoid confusion.)
Therefore, focusing on the nonstripe phase, the HF energy shown above 
can be simplified to be 
\begin{eqnarray}
E^{D2,u}_{HF}(\vec{Q}_\perp)
&=&(\Delta_{SAS}-\omega_z)\Theta_2(0)
+(\Delta_{SAS}+\omega_z)\Theta'_2(0)
\nonumber\\
&&+\frac{1}{2}E_F^I(0,0)\left[\Theta_1(0)^2+\Theta_2(0)^2+
\Theta'_1(0)^2+\Theta'_2(0)^2\right]
+E_F^I(\vec{Q}_\perp,0)\left[\Theta_{3}(0)^2+\Theta'_{3}(0)^2\right]
\nonumber\\
&&+E^o_F(0,0)\left[\Theta_1(0)\Theta'_2(0)
+\Theta_2(0)\Theta'_1(0)\right]
+2E^o_F(\vec{Q}_\perp,0)\Theta_{3}(0)\Theta'_{3}(0),
\label{E_canted}
\end{eqnarray}
where the analytic expression for
$E_{H,F}^{I,o}(\vec{Q}_\perp,0)$ is shown in Appendix
\ref{E_HF_Io}. From Eq. (\ref{E_canted}) we find that, as expected, 
the Hartree energy
does not contribute to the HF energy of a uniform phase, and only
exchange energy (negative value) is relevant.

By comparing the HF energy of the three different phases:
fully incommensurate ($\Delta_{SAS}=0$ and $\vec{P}_\perp=\vec{Q}_\perp=0$),
fully commensurate ($\vec{Q}_\perp=0$), and partially 
commensurate/incommensurate ($\vec{Q}_\perp\neq 0$) phases,
our numerical calculation shows that the ground state is always
the fully commensurate phase (which has the lowest energy), i.e.
$\vec{Q}_\perp=\vec{Q}_\perp'=0$ as an isospin coherent phase in the
noninteracting eigenstate basis. Therefore we do not find the 
commensurate-incommensurate phase transition (as observed in the odd filling
systems) in this even filling factor situation. 
Following the arguments in Section \ref{symmetry} (see particularly
Eqs. (\ref{uniform_parity})-(\ref{uniform_spin})), we find that this fully
commensurate state breaks both 
parity and spin rotational symmetries (but not the translational symmetry).
To understand that these are two separate broken symmetries one can
consider a non-HF state that has uniform $\psi_k$ and $\psi_k'$,
and $(\gamma,\gamma')$ fluctuating between $(-\theta/2,\theta/2)$
and $(\pi-\theta/2,\pi+\theta/2)$ with fixed $\theta$.
Such a state clearly breaks the spin symmetry, but not parity 
(which changes $\gamma$ and $\gamma'$ by $\pi$, see 
Eq. (\ref{uniform_parity})), and may be described as a spin nematic state.

\begin{figure}

 \vbox to 5.8cm {\vss\hbox to 5.cm
 {\hss\
   {\includegraphics{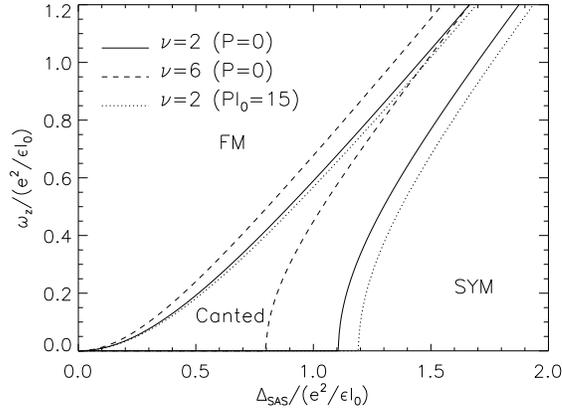}
   }
  \hss}
 }
\caption{Phase diagram of double well systems at $\nu=4N+2$ 
in the presence of an in-plane magnetic field.
The solid and dashed lines are for $\nu=2$ ($N=0$) and $\nu=6$ ($N=1$) without
in-plane magnetic field, while the dotted lines are for $\nu=2$ with
strong in-plane magnetic field ($Pl_0=15$). The layer separation, $d$, is
$0.067 l_0$.
FM is the ferromagnetic state with both filled levels being in the
same spin polarized direction, and SYM is the symmetric state, where both
spin indices are equally occupied in a symmetric orbital Landau level. 
In the middle is the canted antiferromagnetic 
phase, which breaks the spin and parity symmetries 
and is a commensurate state 
when finite in-plane magnetic field is 
applied. 
}
\label{D2_phase_diag}
\end{figure}

The equation for the phase boundary between the fully commensurate
phase (i.e. commensurate CAF phase) 
and the symmetric phase (both filled levels are orbital symmetric, 
$\alpha=+$, but with different spin directions) is
\begin{eqnarray}
\omega_z(\Delta_{SAS})=\sqrt{(\Delta_{SAS}-E_F^o(0,0))^2
-E_F^o(0,0)^2},
\label{boundary_sc}
\end{eqnarray}
where $\omega_z$ is the Zeeman energy. Analogously 
the equation for the phase boundary between the commensurate
CAF phase and the fully spin polarized state is
\begin{eqnarray}
\omega_z^{cf}(\Delta_{SAS})=\sqrt{E_F^o(0,0)^2+\Delta_{SAS}^2}
-E_F^o(0,0).
\label{boundary_fc}
\end{eqnarray}
The calculated phase diagrams of different filling factors and different 
in-plane magnetic fields are shown in Fig. \ref{D2_phase_diag}.

We can also use a similar method to calculate the triplet
magnetoplasmon modes by taking $\Theta_2(0)$ and $\Theta'_2(0)$
to be small in Eq. (\ref{E_canted}), because the noninteracting Hamiltonian has
been diagonalized in the energy eigenstate basis.
However, since such calculations already exist in the literature both
for zero in-plane magnetic field
and in the presence of an
in-plane field \cite{anna}, we do not
show the calculated magnetoplasmon dispersion in this paper.

\subsection{Exploration of the stripe formation}

When considering the possible stripe formation via the periodic functions
of $\psi_k$ and $\psi_k'$ in the trial wavefunction, we need to retain
the Hartree energy, which is canceled in the uniform 
non-stripe phase discussed above. 
Assuming $\vec{Q}_\perp=0$ as in the uniform phase
the Hartree-Fock energy in 
(Eqs. (\ref{E0_D2})-(\ref{E_F_D2})) for the striped case becomes
\begin{eqnarray}
E^{D2,s}_{HF}(0)
&=&(\Delta_{SAS}-\omega_z)\Theta_2(0)
+(\Delta_{SAS}+\omega_z)\Theta'_2(0)
\nonumber\\
&&+\frac{1}{2}\sum_{q_n\neq 0}E^I_H(q_n)\left[
\Theta_1(q_n)+\Theta_2(q_n)+\Theta'_1(q_n)+\Theta'_2(q_n)\right]^2
\nonumber\\
&&-\frac{1}{2}\sum_{q_n}\left\{E_F^I(0,q_n)\left[\Theta_1(q_n)^2
+\Theta_2(q_n)^2+\Theta'_1(q_n)^2+\Theta'_2(q_n)^2
+2\Theta_{3}(q_n)^2+2\Theta'_{3}(q_n)^2\right]\right.
\nonumber\\
&&+\left.2E^o_F(0,q_n)\left[\Theta_1(q_n)\Theta'_2(q_n)+
\Theta_2(q_n)\Theta'_1(q_n)+2\Theta_{3}(q_n)\Theta'_{3}(q_n)\right]\right\},
\label{E_D2_stripe}
\end{eqnarray}
where $q_n=2\pi n/a$ and $a$ is the stripe period.
We define $E^I_H(q_n)=(2\pi l_0^2)^{-1}V^I_{NN,NN}(q_n)$ 
and $E^I_F(q_n)=\Omega_\perp^{-1}
\sum_{\vec{q}_\perp}V^I_{NN,NN}(\vec{q}_\perp)\cos(q_nq_yl_0^2)$, and 
neglect the divergent Hartree energy at $q_n=0$, which is canceled
by the background positive charge (providing the overall
charge neutrality). 
Using the perturbation method developed in Section \ref{pert_stripe},
which is equivalent to studying the mode softening of the Goldstone mode
inside the CAF phase,
we obtain the following perturbation energy matrix 
(see Eq. (\ref{E_pert_4l})):
\begin{eqnarray}
{\bf E}^{HF}_{pert}(\tilde{q})&=&
\left[\begin{array}{c}
(\Delta_{SAS}-\omega_z)\cos(\psi_0^\ast)+
\left[E_F^I(0)-E_F^I(\tilde{q})\right]
-E^o_F(0)\cos(\psi_0^\ast+\psi_0'{}^\ast) \\
-E^o_F(\tilde{q})\cos(\psi_0^\ast+\psi_0'{}^\ast) 
\end{array}
\right.
\nonumber\\
&&\hspace{2cm}
\left.\begin{array}{c}
-E^o_F(\tilde{q})\cos(\psi_0^\ast+\psi_0'{}^\ast)
\\
(\Delta_{SAS}+\omega_z)\cos(\psi'_0{}^\ast)+
\left[E_F^I(0)-E_F^I(\tilde{q})\right]
-E^o_F(0)\cos(\psi_0^\ast+\psi_0'{}^\ast)
\end{array}
\right],
\label{E_D2_stripe_perturb}
\end{eqnarray}
where $\psi_0^\ast$ and $\psi_0'{}^\ast$ are the optimal values obtained
from minimizing the total energy inside the commensurate
canted phase region, and
$\tilde{q}$ is the wavevector of the test small stripe as shown
in Eq. (\ref{perturb_psi}). For simplicity, we have assumed that the stripe
periods of the two stripe phase functions, $\psi_k$
and $\psi_k'$, are the same, $2\pi/\tilde{q}$. The sign
of the eigenvalues of Eq. (\ref{E_D2_stripe_perturb}) then 
determines the existence of stripe formation as discussed in 
Section \ref{pert_stripe}.
Our numerical calculation shows that both of the eigenvalues of
${\bf E}^{HF}_{pert}(\vec{q})$
in Eq. (\ref{E_D2_stripe_perturb}) are always positive within the 
canted phase region, showing that {\it no} stripe phase 
should occurs with an energy lower
than the commensurate CAF phase. We have also studied the possibility of
first order phase transition to a stripe phase, which cannot be included in
the perturbation theory, by directly comparing various ground state energies.
We still find that 
no stripe phase can be stabilized energetically.
Therefore, unlike the rich phase diagram shown in the odd filling factor
double well system
at $\nu=4N+1$, the system at even filling $\nu=4N+2$ in the presence of in-plane field
has neither the commensurate-to-incommensurate phase transition 
in the large $B_\|$ region (see discussion below Eq. (\ref{E_canted})) nor
the uniform-to-stripe phase transition in the large layer separation region.
The former result is also consistent with the recent result obtained in an
effective field theory \cite{yang99_effect_D2}. 
We discuss this result further in Section \ref{discussion}.

\section{Discussion}
\label{discussion}

In this section, we discuss and compare the quantum phases 
obtained in different systems studied in this paper (as well as those already
existing in the literature).

\subsection{Directions of isospin stripe and isospin spiral orders}
\label{order_direction}

In Table \ref{directions} we summarize the isospin spiral and isospin 
stripe directions obtained 
in the three different cases, $W1'$, $W2'$, and $D1$, 
where at least a spiral or a stripe order exists in our HF calculation.
For the convenience of discussion, we will define the isospin components 
to be in the layer index basis for the double well systems at $\nu=4N+1$, 
i.e. the isospin spiral/coherent (stripe) phases in $D1$ case
correspond to the conventional commensurate/incommensurate 
(stripe) states respectively. We do not include the commensurate
CAF phase in $D2$ system, because it also has a spiral order following the 
in-plane field in the layer index basis and therefore is similar to
the commensurate state (isospin spiral phase) of $D1$ system.

Note that although the definition of isospin components are different
for these systems (see Table \ref{system_notation}), all the
isospin spiral orders select 
a wavevector, $\vec{Q}_\perp$ (but 
$\pm \vec{Q}_\perp$ are
degenerate for the wide well system), to be perpendicular to the 
in-plane magnetic field.
The mechanisms for the spiral winding in wide wells 
and in double wells are very different. 
In a wide well system with a strong in-plane magnetic field, the electron
wavefunction is distorted by the anisotropic field and therefore
the electron-hole binding energy is the strongest if
the isospin is winding along the direction perpendicular to the in-plane 
field \cite{wang02}. More precisely, this is true only in the presence
of a very strong perpendicular magnetic field or in wells
which are not too wide, 
so that the effective cyclotron resonance energy, $\omega_\perp$, 
is comparable to or larger than 
the confinement energy, $\omega_0$. 
The direction of the spiral winding 
can change to the in-plane field direction if the perpendicular 
field is so weak that $\omega_\perp\ll\omega_0$.
(We will show this result explicitly in below).
This effect of spiral winding locking is less important in the
weak $B_\|$ region, leading to the isotropic phases observed 
in the intersubband level crossing region ($W1$). 
In a double well system with zero 
well width, however, the in-plane magnetic field does not affect the
single electron wavefunction, but only 
affects the tunneling amplitude through the
Aharonov-Bohm phase factor, which selects a specific spiral
wavevector, $\vec{Q}_\perp=(0,P_x)=(0,edB_\|/c)$, 
perpendicular to the in-plane magnetic field. 
When $B_\|$ is weak, the isospin spiral order follows the wavevector
of the tunneling amplitude, since it minimizes the tunneling
energy and does not cost much in the Coulomb exchange energy.
When $B_\|$ exceeds a certain critical value, the energetic cost of winding
from the point of view of exchange energy becomes prohibitively high, and the
system goes into the incommensurate phase with no isospin winding. 
Therefore the origin of isospin spiral order in a wide well
system and a double well system is very different, although both are
perpendicular to the in-plane magnetic field.

Now we discuss the mechanism, which determines the directions
of the isospin stripe orders (if it exists) in these systems.
From Table \ref{directions} we note that the stripe modulation of $I_z$
(i.e. the normal vector, $\hat{n}$) of the stripe is along $x$ axis,
parallel to the $B_\|$ field for
a wide well system at both even and odd filling factors
(for even filling system, $W2'$, we cannot stabilize a many-body
phase within the HF approximation, but speculate, based on very small
calculated HF energy differences, that
the resistance anisotropy observed in the 
experiment \cite{pan01} could result from a skyrmion stripe phase
near the level crossing region, which could perhaps be stabilized by going
beyond our approximation scheme (see Section \ref{sec_W2}). 
But the stripe normal vector is along $y$ axis,
perpendicular to the $B_\|$ field, for a double well system at $\nu=4N+1$
(no stripes are found in a double well system at $\nu=4N+2$).
The different stripe directions in the two systems could be understood
as a result of competition between two effects: one is the anisotropy
energy (i.e. the total energy difference between a stripe perpendicular to
the in-plane field and a stripe parallel to the in-plane field,
see \cite{jungwirth99}) induced by the in-plane magnetic field via
the anisotropic distortion of electron wavefunction, 
and the other one is the exchange
interaction between the spiral order (oscillation
of $\langle I_x\rangle$ and $\langle I_y\rangle$) and the stripe order
(oscillation of $\langle I_z \rangle$). This effect reflects on 
the term, $E^{W1}_{F6}(q_n,0\,;Q_x,Q_y)$ of Eq. (\ref{E_F^W1}) 
for a wide well at $\nu=2N+1$ 
and on the last term of Eq. (\ref{E_F_D1}) for a double well at $\nu=4N+1$
(the similar terms in the even filling systems can be also found 
in Eq. (\ref{E_F_W2}) and Eq. (\ref{E_F_D2}) for
wide well and double well systems respective: 
proportional to $\Theta_3(q_n)^2$ and $\Theta'_3(q_n)^2$).
Such interaction between the spiral order 
and the stripe order 
prefers to keep the wavevectors of these two oscillations 
parallel with each
other in order to optimize the exchange energy. 
On the other hand, the anisotropy energy induced by the 
anisotropic electron wavefunction in the
finite width well prefers to form a stripe aligned perpendicular to
the in-plane field direction, i.e. its normal direction is parallel to
$\hat{x}$ axis \cite{footnote_stripedirection}.

If the spiral-stripe exchange energy dominates the anisotropic energy
(like in the double well system, where the zero-well-width electron 
wavefunction is isotropic and hence the anisotropy energy is zero),
the stripe
order prefers to select a normal wavevector parallel to the wavevector 
of the spiral order, resulting in a spiral 
stripe phase. The stripe direction in this case is governed
by the spiral order, which is perpendicular
to the in-plane field in the commensurate phase as we find in the numerical
results shown in Section \ref{sec_D1} and in Fig. \ref{D1_phase_diag}(a)
(phase (IV) and (V)). When the in-plane field
is so strong that the double well system undergoes a first order phase 
transition to an incommensurate phase, the spiral order disappears and
therefore the stripe direction is not locked by the in-plane field,
being in an arbitrary direction in the two-dimensional plane
(phase (III) in Fig. \ref{D1_phase_diag}) \cite{brey00}.
On the other hand, when we consider a single 
finite width well in the presence 
of in-plane field, the anisotropy energy is finite and competes with the
spiral-stripe exchange energy.
If the anisotropy energy dominates, the spiral-stripe exchange energy
can {\it not} force the stripe and spiral orders to be parallel with each other
(and hence perpendicular to the in-plane field). Therefore 
the density modulation of 
the stripe phase prefers to stay in the direction of in-plane
field \cite{footnote_stripedirection},
having a direction perpendicular to the spiral order, 
resulting in a skyrmion stripe phase.
Therefore the mechanism we discussed so far can explain all 
of the spiral and stripe directions
obtained individually in the previous sections within the HF approximation.

As a final remark, we note that the HF energy difference between
the spiral stripe phase and the skyrmion stripe phase are very small
compared to the other energy scales in our numerical calculation, both
in wide well systems and in double well systems. Therefore we expect
that various effects (e.g. disorder and impurity scattering) ignored
in our calculations could have strong influence in determining the eventual
ground state of the system, providing an experimental stripe direction
different from that obtained in our idea HF theory.
The actual finite width
effects of a double quantum well system in a realistic experiment
could also lead to the stabilization of the
skyrmion stripe (rather than a spiral stripe) phase, 
especially when the spiral-stripe coupling
is weakened by the finite temperature and/or finite disorder effects.  

\subsection{Role of spin degree of freedom in level crossing}
\label{comparison}

From the results presented in Sections \ref{sec_W1} and \ref{sec_W2} for
wide well systems, we find that systems of odd filling factors 
($W1$ and $W1'$) have more interesting coherent phases with exotic 
quantum order than the 
systems of even filling factor ($W2$ and $W2'$). 
As summarized in Table \ref{system_notation}, the HF trial
wavefunction, Eq. (\ref{wavefunction_general}), stabilizes
an isospin coherent phase in $W1$ , and an isospin skyrmion stripe phase
in $W1'$. For systems at even filling factors, strictly speaking,
no novel quantum phases are obtained in the wide well systems within our HF 
approximation (although sometimes the exotic phases are very quantum closeby
in energy). 
Since the orbital wavefunctions of the two crossing levels 
in the even filling systems are
the same as those in the odd filling systems,
it is natural to attribute the important difference
between the two systems to the additional
spin degree of freedom of the two crossing (or degenerate) levels,
that is present in the $W1/W1'$ cases.
More precisely, we can compare the formulae 
of the HF variational energy of an odd filling system shown 
in Section \ref{sec_W1}
with the formulae of HF energy of even filling system 
in Section \ref{sec_W2}, where  
we can simply take $\psi_k'=0$
in Eq. (\ref{wavefunction_4l_W2}) and Eq. (\ref{exp_value_W2}), and 
consider the trial wavefunction constructed by the two crossing levels
only (instead of the four degenerate levels),
similar to Eq. (\ref{wavefunction_general}). In such situations,
the real spin quantum number of the two isospin states are of opposite sign,
different from the level crossing in the odd filling system.
The resulting HF variational energy (denoted by underlines) then becomes
(compared to Eqs. (\ref{E_0^W1})-(\ref{E_F^W1}))
\begin{eqnarray}
\underline{E}^{W2}_0 &=&
E^{0,W}_{\vec{n}_1,\downarrow}\Theta_1(0)+
E^{0,W}_{\vec{n}_2,\uparrow}\Theta_2(0),
\label{E_0_W2_2}
\\
\underline{E}^{W2}_H&=&\frac{N_\phi}{2\Omega_\perp}\sum_{q_n}\left[
\tilde{V}^W_{\vec{n}_1\vec{n}_1,\vec{n}_1\vec{n}_1}(q_n)
\Theta_1(q_n)^2+
\tilde{V}^W_{\vec{n}_2\vec{n}_2,\vec{n}_2\vec{n}_2}(q_n)
\Theta_2(q_n)^2
+2\tilde{V}^W_{\vec{n}_1\vec{n}_1,\vec{n}_2\vec{n}_2}(q_n)
\Theta_1(q_n)\Theta_2(q_n)
\cos(q_nQ_yl_0^2)\right],
\label{E_H_W2_2}
\\
\underline{E}^{W2}_{F}\nonumber
&=&\frac{-1}{2\Omega}\sum_{\vec{q}_\perp}
\left[\tilde{V}^W_{\vec{n}_1\vec{n}_1,\vec{n}_1\vec{n}_1}(\vec{q}_\perp)
\sum_{q_n}\cos(q_nq_yl_0^2)
\Theta_1(q_n)^2
+\tilde{V}^W_{\vec{n}_2\vec{n}_2,\vec{n}_2\vec{n}_2}(\vec{q})
\sum_{q_n}\cos(q_nq_yl_0^2)
{\Theta_2(q_n)}^2\right]
\nonumber\\
&&+\frac{-2}{2\Omega_\perp}\sum_{\vec{q}_\perp}
\tilde{V}^W_{\vec{n}_1\vec{n}_1,\vec{n}_2\vec{n}_2}(\vec{q}_\perp)
\cos((Q_xq_y-Q_yq_x)l_0^2)
\sum_{q_n}\cos(q_nq_yl_0^2)\Theta_{3}(q_n)^2,
\label{E_F_W2_2}
\end{eqnarray}
where we have neglected those uniform terms (linearly proportional to
$\Theta_i(0)$ for $i=1,2,3$) for simplicity.
Comparing Eqs. (\ref{E_0_W2_2})-(\ref{E_F_W2_2}) with the HF energy of 
an odd filling system in Eqs. (\ref{E_0^W1})-(\ref{E_F^W1}), we find
that the odd filling factor systems have one additional term in 
both Hartree and Fock energies
(i.e. the $E^{W1}_{H6}$ and $E^{W1}_{F3}$ terms) 
after neglecting the uniform terms and those singular terms proportional to
$\delta_{Q_y,0}$ shown in 
Eqs. (\ref{E_0^W1})-(\ref{E_F^W1})). These two additional terms
result from the direct and the exchange contractions of 
the correlation, $\langle c^\dagger_{\vec{n}_1,\sigma}
c^{}_{\vec{n}_2,\sigma}c^\dagger_{\vec{n}_2,\sigma'}c^{}_{\vec{n}_1,\sigma'}
\rangle$, which is absent at even filling factors, because the two
relevant levels, $\vec{n}_1$ and $\vec{n}_2$, have opposite spin direction.
Since such additional exchange energy is larger than 
the additional Hartree energy, a coherent phase near the 
level crossing point can be stabilized more easily in an odd filling system
than in an even filling system.
This mechanism explains the results we obtain in the wide well system
within the HF approximation, 
although it does not exclude the possibility of having a many-body phase
in an even filling system in more sophisticated theories.

Such effects of spin degree of freedom can also be observed from the
results of the double well systems ($D1$ and $D2$). When considering only
the two crossing levels in $\nu=4N+2$ case, we will not obtain any many-body
state for the same reason as mentioned above. The coherent phase
(i.e. commensurate canted antiferromagnetic phase) obtained in the literature
\cite{canted_phase,bilayer_sym_break,burkov02} and in this paper
strictly relies on the incorporation of all the four 
degenerate levels (see Sections
\ref{trial_wf_4l} and \ref{sec_D2}). 
On the other hand, 
at least four different many-body 
phases (see discussion in Section \ref{sec_D1} and
Fig. \ref{D1_phase_diag}) are obtained as ground states
at $\nu=4N+1$ within the HF approximation. 
In the odd filling case, electrons in the top level 
equally occupy the two layers, and therefore 
when the layer separation is larger than the order of 
$l_0$, interlayer coherence becomes relatively weaker and a stripe 
formation may occur in order to optimize the intralayer exchange energy in
each layer, reflecting the charge density wave instability studied
in the high half-odd-integer single layer systems 
\cite{lilly99,du99}. Such a simple 
mechanism, however, does not seen to 
apply to $\nu=4N+2$ case, because every flux
quantum is occupied by an electron in each layer --- any nonuniform
density modulation, as in a stripe phase, will have to pay a large 
direct energy for the double occupancy of electrons of opposite spins in 
each layer in each flux quantum.
Since the commensurate canted
phase can successfully lower the exchange energy by canting both spin
and isospin degrees of freedom, a stripe phase becomes unlikely
in the even filling factor $\nu=4N+2$ bilayer case (at least in the absence
of any external bias voltage). When an external gate voltage
is applied in the growth direction of the two layers, the two
layers become partially filled, and  
there might be some interesting stripe phase in this 
situation, at least for the following extreme case: when 
the gate voltage between the two layers is so large that the filling factor
of the left layer is near $2N+1/2$ and of the right layer is  
near $2N+3/2$, the two layers can form stripe phases individually
similar to those well-known single layer systems at high half-odd-integer 
filling factors ($N\ge 1$) \cite{review} (especially if the layer separation 
is so large that the tunneling energy and interlayer interaction are
negligible). 

\subsection{Comparison with the exchange induced spin density wave
in a very wide well}
\label{sdw}
 
It is very interesting and instructive to compare our results of stripe phases 
with the spin density instabilities proposed first by Brey and Halperin
(BH) \cite{brey89_sdw} in a wide parabolic well system and
by Das Sarma and Tamborenea
\cite{dassarma94_sdw_dqw} in a zero-field double quantum well system. 
In Ref. \cite{brey89_sdw}, it is found that when an 
in-plane magnetic field is applied (without any perpendicular magnetic 
field) to a very wide parabolic well (width is $\sim 4000$ \AA), 
Hartree energy between electrons inside the well modifies the electron
density profile to be almost a uniform slab, making the parabolic well 
a good approximation of the 3D
jellium model where electrons move in a constant positive charge background.
Calculation of spin density correlation function
shows that a divergent singularity occurs at a certain wavevector 
in $x$ direction (parallel to the in-plane field), 
$Q^{BH}_x=k_{F,\uparrow}+k_{F,\downarrow}$, 
where $k_{F,\sigma}$ is the Fermi wavevector of spin
$\sigma$ electrons. They proposed that such spin density wave instability 
is similar to those originally proposed by Celli and Mermin \cite{celli65_3d}
in three-dimensional systems and may cause a resistance anisotropy
in the $x-y$ plane of the well. This result \cite{brey89_sdw} 
seems to contradict
the spiral direction suggested by the magneto-roton minimum of the
magnetoplasmon mode \cite{wang02} as well as our HF calculation in this paper
for a wide parabolic well system at the even filling factor, where the
softening of the spin triplet magnetoplasmon mode 
(i.e. the divergence of a spin 
density correlation function) is in $y$ direction, perpendicular to
the in-plane magnetic field.
Besides, the wavevector of the BH spin density wave mode 
softening, $Q^{BH}_x$, depends
mainly on the electron density and is almost independent of the magnetic field
strength, while the wavevector of the mode softening in our calculation 
crucially depends on the in-plane magnetic field strength and is almost
independent of the electron total density \cite{wang02}.
Here we analyze the superficial contradiction between our results
in this paper and the BH results, providing
a deeper understanding of these two calculations. We find
that both results
are correct, and their different characteristic spin density wave 
vectors arise entirely from considerations of different limits of system
parameters in the two situations.

First we should clarify the energy scales of the wide well system 
we are going to discuss. For the quantum Hall situation of 
our interest in this paper, the perpendicular 
magnetic field is always finite ($B_\perp\sim 3$ Tesla, i.e. $\omega_\perp
\sim 5$ meV) to ensure the existence of a discrete orbital Landau
level spectrum. The well width is chosen to have
a comparable ($\omega_0\sim\omega_\perp$) bare confinement energy 
$\omega_0\sim 7$ meV. 
The (iso)spin density wave instability 
(for both odd and even filling factor) in this system occurs in a very high
in-plane magnetic field region ($B_\|\sim 10-30$ T, or $\omega_\|
\sim 17-50$ meV). Therefore we can obtain the following 
approximate order of energy scales 
for the system of our interest in the current paper
(using the even filling factor system as an example)
\begin{eqnarray}
\omega_\perp\sim\omega_0\ll \omega_\|, \mbox{   and   } 
N_\uparrow\sim N_\downarrow \sim {\cal O}(1),
\label{sdw_limit1}
\end{eqnarray}
where $N_\sigma$ is the Landau level index of the highest filled level
of spin $\sigma$, and the total 
filling factor $\nu=N_\uparrow+N_\downarrow+2$ is of order of unity,
A small parameter, $\epsilon\equiv 
Max(\omega_\perp,\omega_0)/\omega_\|\ll 1$ can therefore be defined. 
In this limit, we have $\omega_1\to\omega_\|$ and 
$\omega_2\to\epsilon^2\omega_\|$ for the two noninteracting Landau energy
separation (see Eq. (\ref{omega12})). Besides, the two effective magnetic 
lengths, $l_{1,2}\equiv\sqrt{1/m^\ast\omega_{1,2}}$, defined in 
Eq. (\ref{wf_i}) (or see Ref. \cite{wang02}) become
$l_0\epsilon^{1/2}$ and $l_0\epsilon^{-1/2}$ respectively. 
On the other hand, for the 
system discussed by BH \cite{brey89_sdw}, no perpendicular magnetic
field is applied and the predicted spin density wave instability occurs at an
intermediate strength of $B_\|$ ($\sim 1$ Tesla, i.e. $\omega_\|\sim 1.7$ meV) 
for a very wide well (for width about 4000 \AA, the parabolic confinement
energy $\omega_0<0.05$ meV). Therefore we can obtain the 
following approximate order of energy scales for the BH system:
\begin{eqnarray}
\omega_\perp\ll\omega_0\ll \omega_\|, \mbox{   and   }
N_\uparrow\sim N_{tot,\uparrow},\ N_\downarrow \sim N_{tot,\downarrow},
\label{sdw_limit2}
\end{eqnarray}
where $N_{tot,\uparrow(\downarrow)}$ is the total electron number of spin
up(down) (in the direction parallel to $B_\|$), and
hence the Landau level degeneracy, $N_\phi$, is only of the order of unity
in this weak $B_\perp$ limit. Therefore 
there are two small parameters we can construct
for the BH system: $\epsilon\equiv Max(\omega_\perp,\omega_0)/\omega_\|
=\omega_0/\omega_\|\ll 1$ as defined earlier
and $\alpha\equiv(\omega_\perp/\omega_0)^2
\ll 1$, which is of order unity in the system of our 
interest in this paper.
In this limit (Eq. (\ref{sdw_limit2})), $\omega_{1,2}$ 
are still the same as above (i.e. close
to $\omega_\|$ and $\epsilon^2\omega_\|$ respectively), while $l_1\to
l_0\epsilon^{1/2}\alpha$ and $l_2\to l_0\epsilon^{-1/2}\alpha$.
Note that although both $\epsilon$ and $\alpha$ are small numbers for 
the BH system, an additional constraint, $\alpha\ll\epsilon$, exists also,
to ensure the fact that no perpendicular magnetic field is applied. 
We will show below that this additional constraint 
(i.e. $\alpha\ll\epsilon$) and
the resulting small Landau level degeneracy ($N_\phi\sim{\cal O}(1)$) are
the key points needed to understand BH's result of 
SDW instability from the perspective of our theory.

For the convenience of comparison, 
we still consider the two level degeneracy on the top of the filled Landau
levels and use the "core state" approximation for a qualitative discussion,
although it is certainly a bad approximation in the limit of
Eq. (\ref{sdw_limit2}), where
the Landau level energy separation goes to zero.
To simplify our analysis, we calculate $E^{W2}_{X}(\vec{Q}_\perp)$ shown
in Eq. (\ref{E_W2_X}) only, which corresponds to the electron-hole
binding energy via ladder diagrams and gives rise to the roton
minimum of the magnetoplasmon dispersion \cite{wang02},
and use a zero-range contact electron-electron interaction.
We can obtain the following
approximate $E^{W2}_X(\vec{Q}_\perp)$ by using the analytical
expression of $A(\vec{q}\,)$ in Eq. (\ref{A_explicit}) and
taking the $\epsilon\to 0$ limit:
\begin{eqnarray}
E^{W2}_X(\vec{Q}_\perp)&=&\frac{-V_0}{\Omega}\sum_{\vec{q}}
A^W_{\vec{n}_1\vec{n}_1}(-\vec{q})A^W_{\vec{n}_2\vec{n}_2}(\vec{q})
\cos((q_xQ_y-q_yQ_x)l_0^2)
\nonumber\\
&\sim&\frac{-V_0}{\Omega}\sum_{\vec{q}}
\exp\left[-\frac{\epsilon^2\alpha(q_yl_0)^2+
(\epsilon\sqrt{\alpha}q_xl_0-q_zl_0)^2\epsilon^2\alpha^4}
{2\epsilon\alpha^2}\right]
\exp\left[-\frac{(q_yl_0)^2+
(q_xl_0+\epsilon\sqrt{\alpha}q_zl_0)^2\alpha^4/\epsilon^2}
{2\alpha^2/\epsilon}\right]
\nonumber\\
&&\times L^0_{N_\uparrow+1}\left(\frac{(q_yl_0)^2+
(q_xl_0+\epsilon\sqrt{\alpha}q_zl_0)^2\alpha^4/\epsilon^2}
{2\alpha^2/\epsilon}\right)
L^0_{N_\downarrow}\left(\frac{(q_yl_0)^2+
(q_xl_0+\epsilon\sqrt{\alpha}q_zl_0)^2\alpha^4/\epsilon^2}
{2\alpha^2/\epsilon}\right)
\cos((q_xQ_y-q_yQ_x)l_0^2),
\label{E_X_approx}
\end{eqnarray}
where we have let $\vec{n}_1=(0,N_\downarrow)$ and 
$\vec{n}_2=(0,N_\uparrow+1)$
for the orbital level index of the top filled and the lowest empty levels
respectively.

For the quantum Hall system of our interest in this paper, $\alpha$ is 
of the order of unity (see Eq. (\ref{sdw_limit1})), and then the above 
equation can be simplified further (denoted by $E^{W2,1}_X(\vec{Q}_\perp)$) 
by keeping only the leading term in $\epsilon$:
\begin{eqnarray}
E^{W2,1}_X(\vec{Q}_\perp)&\sim&\frac{-V_0}{\Omega}\sum_{\vec{q}}
\exp\left[-\frac{(q_zl_0)^2\epsilon\alpha^2}{2}\right]
\exp\left[-\frac{(q_xl_0)^2\alpha^2}{2\epsilon}\right]
L^0_{N_\uparrow+1}\left(\frac{(q_xl_0)^2\alpha^2}{2\epsilon}\right)
L^0_{N_\downarrow}\left(\frac{(q_xl_0)^2\alpha^2}{2\epsilon}\right)
\cos((q_xQ_y-q_yQ_x)l_0^2)
\nonumber\\
&\propto&-\frac{V_0\delta_{Q_x,0}}{2\pi}\int dq_x
\exp\left[-\frac{q_x^2l_2^2}{2}\right]
\left(\frac{q_x^2l_2^2}{2}\right)^{2N+1}
\cos(q_xQ_yl_0^2)
\label{E_x_approx_1}
\end{eqnarray}
where we have used $L_n^0(x)\propto x^n$ for $x\gg 1$, $N_\uparrow=
N_\downarrow=N=(\nu-2)/2$ for even filling system, and $l_2/l_0\sim 
\alpha/\sqrt{\epsilon}$.
It is very easy to see from Eq. (\ref{E_x_approx_1}) that 
$E^{W2,1}_X(\vec{Q}_\perp)$
is nonzero {\it only} at $Q_x=0$ and a finite $Q_y$. The extreme value of $Q_y$
to optimize the exchange energy is given by the length scale generated
by the competition between the exponential function and the power-law
function, proportional to $\sqrt{N}\,l_2^{-1}$. 
This is consistent with
our numerical results obtained either from the HF 
variational calculation in this paper or from the collective 
mode calculation in our earlier work \cite{wang02}.

On the other hand, if we take $\alpha\to 0$ first in Eq. (\ref{E_X_approx})
as appropriate for the situation considered by BH, we obtain the electron-hole
binding energy (denoted by $E^{W2,2}_X(\vec{Q}_\perp)$) to be
\begin{eqnarray}
E^{W2,2}_X(\vec{Q}_\perp)
&\sim&\frac{-V_0}{\Omega}\sum_{\vec{q}}
\exp\left[-\frac{\epsilon(q_yl_0)^2}{2\alpha}\right]
\exp\left[-\frac{\epsilon (q_yl_0)^2}{2\alpha^2}\right]
L^0_{N_\uparrow+1}\left(\frac{\epsilon (q_yl_0)^2}{2\alpha^2}\right)
L^0_{N_\downarrow}\left(\frac{\epsilon (q_yl_0)^2}{2\alpha^2}\right)
\cos((q_xQ_y-q_yQ_x)l_0^2)
\nonumber\\
&\sim&B-\delta_{Q_y,0}C
\int_{l_2/l_0^2}^{+\infty} 
\frac{dq_yl_2}{q_yl^2_0}
\cos(\sqrt{2N_\uparrow}\,q_y l^2_0/l_2)
\cos(\sqrt{2N_\downarrow}\,q_y l^2_0/l_2)
\cos(q_yQ_xl_0^2),
\label{E_X_approx2}
\end{eqnarray}
where we have used the asymptotic formulae, $L_{n}^0(x)\propto e^{x/2}
x^{-1/4}\cos(2\sqrt{n x})$ for large $n$ \cite{ryzhik}, and have let the
constant contribution from integration of small $q_y$ region to be denoted by
$B$; $C$ is a constant factor. 
It is easy to see that the maximum
electron-hole (exciton) binding energy occurs at $Q_y=0$ and $Q_x\sim
\pm (k_{F,\uparrow}+k_{F,\downarrow})$, where the Fermi wavevector
is determined by the Fermi energy, $k_{F,\sigma}=
\sqrt{2 m^\ast (E_{F,\sigma}-E_{k=0})}\sim
\sqrt{2 m^\ast(E^{0,W}_{(0,N_\sigma),\sigma}-E^{0,W}_{(0,0),\sigma})}
=\sqrt{2 m^\ast N_{\sigma}\omega_2}
=\sqrt{2N_{\sigma}}\,l_2^{-1}$, for spin $\sigma$ subband ($E_k=k^2/2m^\ast$
is the usual free particle zero-field energy dispersion 
with effective mass $m^\ast$). 
Therefore Eq. (\ref{E_X_approx2}) is consistent with the result obtained
by BH, showing a SDW instability along the in-plane field direction ($x$)
with a wavevector, $Q_x=\pm(k_{F,\uparrow}+k_{F,\downarrow})$. 
Note that the maximum $E^{W2,2}_X(Q_x,0)$ does not occur at 
$Q_x=\pm(k_{F,\uparrow}-k_{F,\downarrow})$ because the decaying function,
$q_y^{-1}$ in Eq. (\ref{E_X_approx2}) has higher 
contribution from the small
$q_y$ region, and hence a larger value of $|Q_x|$ (i.e. 
$|k_{F,\uparrow}+k_{F,\downarrow}|$) should a give stronger electron-hole
binding energy than a smaller value of $|Q_x|$ (i.e. 
$|k_{F,\uparrow}-k_{F,\downarrow}|$). This effect is enhanced for the
long-ranged Coulomb interaction, 
which has a faster decaying function in $q_y$. Therefore our
simple analysis above shows that the apparent inconsistency between
of the SDW spiral winding direction studied in our work \cite{wang02}
and in the BH work \cite{brey89_sdw} is just due to the different limits of
interest in the two works: A quantum Hall situation with finite $B_\perp$
in our case and the ``zero-field'' $B_\perp=0$ situation in Ref. 
\cite{brey89_sdw}. 

We now summarize the above discussion on the spiral 
(spin-density-wave) order of a wide well system in the presence of
in-plane magnetic field (along $x$ axis): when a strong perpendicular 
magnetic field is applied so that only few Landau 
levels are occupied and the in-plane field is tuned to be close to
a level crossing (or a level near degeneracy) point, the (iso)spin spiral
winding prefers a specific wavevector in a direction ($y$) 
perpendicular to the in-plane field, while the (iso)spin 
$\langle I_z\rangle$
oscillates in $x$ direction;
when the perpendicular magnetic field is reduced to a very small value,
the electron wavefunction near Fermi energy becomes a plane wave and 
strongly modifies the interaction matrix elements
through the form function ---
the spin-flip exchange energy then stabilizes 
the spin density wave at a wavevector $Q^\ast_x\sim 
\pm (k_{K,\uparrow}+k_{F,\downarrow})$.
The spin (or isospin) density wave state actually 
results from two different mechanisms in the two different limits
of the same system. These results are independent
of the gauge choice in the theory.

\subsection{Effective Hamiltonian of wide well 
at $\nu=2N+1$ around intersubband level crossing point ($W1$ case)}
\label{effective_theory}

In the previous sections, we have been using the concept of ``isospin''
to discuss the ground state wavefunction properties in both
the wide well systems and the double well systems.
Without losing generality, one can also apply the isospin concept 
to effectively 
represent the Hamiltonians of these systems so that some aspects of this 
physical properties (e.g. collective excitations or finite 
temperature effects) 
may be studied in a more comprehensive way. 
Here we will give the effective Hamiltonian of a wide well system 
at $\nu=2N+1$ near the intersubband level crossing point 
in the isospin representation after projecting
the whole system into the two crossing
levels we consider in this paper. According to our HF calculation results
presented in Section \ref{sec_W1}, only the uniform coherent phase is stabilized
in this intersubband level crossing region, and therefore we will not
consider the stripe order (modulation of isospin $z$ component)
here for simplicity.

For convenience of comparison, we redefine the isospin component
to be (c.f. Eqs. (\ref{rho_I}) and (\ref{S_def})):
\begin{eqnarray}
\tilde{\cal {I}}^{\,W1}_\alpha(\vec{r}_\perp)&=&\frac{1}{2\Omega_\perp}
\sum_{\vec{q}_\perp}e^{i\vec{q}_\perp\cdot\vec{r}_\perp}
\rho^{W1}_{I_1I_2}(\vec{q}_\perp)\,\sigma^\alpha_{I_1I_2},
\label{newisospin_I}
\end{eqnarray}
where $\alpha=x,y,z$ and $\sigma^\alpha_{I_1I_2}$ is the Pauli matrix element.
Therefore the low energy effective Hamiltonian for
a wide well system at $\nu=2N+1$ can be obtained from Eq. (\ref{H_W_rho})
by using Eq. (\ref{newisospin_I}) and considering the long wavelength limit:
\begin{eqnarray}
{\cal H}^{W1}_{eff}
&=&\int d\vec{r}_\perp\left\{
(h+g^{}_{z0})\,\tilde{\cal I}^{W1}_z+g_{zz}^{}
\left(\tilde{\cal I}^{W1}_z\right)^2
+\left[g_{\perp}^{}+g_{X}^{}\cos(2\gamma)+
\tilde{g}_{\alpha\beta}(\gamma)
(\nabla_\alpha\gamma)(\nabla_\beta\gamma)\right]
\left(\tilde{\cal I}^{W1}_\perp\right)^2\right\}
\label{H_eff}
\end{eqnarray}
where we have used $\tilde{\cal I}^{W1}_{x}=\tilde{\cal I}^{W1}_\perp
\cos\gamma$ and $\tilde{\cal I}^{W1}_{y}=\tilde{\cal I}^{W1}_\perp
\sin\gamma$ to address the phase dependence in the isospin $x-y$ components.
$\nabla_\alpha$ is the spatial derivative of phase $\gamma$ 
in $\alpha=x,y$ direction.
The coefficients shown above are respectively
\begin{eqnarray}
h&=&E^{0,W}_{\vec{n}_1,\uparrow}-E^{0,W}_{\vec{n}_2,\uparrow},
\\
g^{}_{z0}&=&\frac{1}{2}\left[
{\cal V}^W_{\Uparrow\Uparrow,\Uparrow\Uparrow}(0)
-{\cal V}^W_{\Downarrow\Downarrow,\Downarrow\Downarrow}(0)\right],
\\
g^{}_{zz}&=&\frac{1}{2}\left[
{\cal V}^W_{\Uparrow\Uparrow,\Uparrow\Uparrow}(0)
+{\cal V}^W_{\Downarrow\Downarrow,\Downarrow\Downarrow}(0)
-2{\cal V}^W_{\Uparrow\Uparrow,\Downarrow\Downarrow}(0)\right],
\\
g^{}_{\perp}&=&
2{\cal V}^W_{\Uparrow\Downarrow,\Downarrow\Uparrow}(0),
\\
g^{}_{X}&=&{\cal V}^W_{\Uparrow\Downarrow,\Uparrow\Downarrow}(0),
\label{g_X}
\\
\tilde{g}_{\alpha\beta}(\gamma) &=& 
\delta_{\alpha\beta}\partial_\alpha\partial_\beta{\cal V}^W
_{\Uparrow\Downarrow,\Downarrow\Uparrow}(0)
-\left|\partial_\alpha\partial_\beta{\cal V}^W
_{\Uparrow\Downarrow,\Uparrow\Downarrow}(0)\right|
\cos\left(2\gamma+\frac{\pi}{2}(1-\delta_{\alpha\beta})\right),
\label{g_ab}
\end{eqnarray}
where we have used the fact that ${\cal V}^W
_{\Uparrow\Downarrow,\Uparrow\Downarrow}(0)$ and
$\partial^2_{x(y)}{\cal V}^W
_{\Uparrow\Downarrow,\Uparrow\Downarrow}(0)$ ($\partial_\alpha$
is the derivative in the $\alpha=x(y)$ component of 
wavevector, $\vec{q}_\perp$) are pure real, and
$\partial_{x}\partial_y {\cal V}^W
_{\Uparrow\Downarrow,\Uparrow\Downarrow}(0)$
is pure imaginary giving
an additional phase $\pi/2$. The effective interaction matrix
element, ${\cal V}^W_{I_1I_2,I_3I_4}(\vec{q}_\perp)$, can be obtained
within the Hartree-Fock theory:
\begin{eqnarray}
{\cal V}^W_{I_1I_2,I_3I_4}(\vec{q}_\perp)&\equiv&
\frac{1}{2\pi l_0^2}\tilde{V}^W_{I_1I_2,I_3I_4}(\vec{q}_\perp)
-\frac{1}{\Omega_\perp}
\sum_{\vec{p}_\perp}\tilde{V}^W_{I_1I_4,I_3I_2}(\vec{p}_\perp)
\,e^{i(q_xp_y-p_{x}q_{y})l_0^2}.
\label{cal_V}
\end{eqnarray}
From the effective Hamiltonian in Eq. (\ref{H_eff}), we can see that 
the effective ``Zeeman'' energy, $h+g_{z0}$, is the same as the single 
electron energy gap between the two crossing levels (including the
HF correction). Therefore when such 
``Zeeman''  energy is dominant in Eq. (\ref{H_eff})
(i.e. the two levels are far from crossing region), the isospin direction
must be polarized at $\tilde{\cal I}^{W1}_z=\pm 1/2$ and hence 
the ground state has no transverse component, i.e. 
$\tilde{\cal I}^{W1}_\perp=0$. 
These are the conventional integer QH states with 
electrons of the top filled Landau level being
at level $\vec{n}=(1,0)$ (for isospin up) and at $\vec{n}=(0,N)$ (for isospin
down) respectively. In the intermediate region between the two states
(see also Fig. \ref{W1_cross}(b)), where
$|h+g_{z0}|$ is small compared with the other energy scales, the transverse
isospin component, $\tilde{\cal I}^{W1}_\perp$ may become finite
to minimize the energy so that
the isospin polarization becomes canted. The prefactor, $\cos(2\gamma)$, 
of the $g^{}_X$ term in Eq. (\ref{H_eff}) selects $\gamma=0,\pi$ to 
be the two degenerate points of the lowest energy, showing that 
only $\hat{\cal I}^{W1}_x$ is finite and there is no isospin in $y$ direction.
This two-point degeneracy results from the fact that only the parity 
symmetry is broken in the intermediate region.
The coefficient of the gradient term, $\tilde{g}_{\alpha\beta}(\gamma)$, 
is positive in such an intersubband level crossing case so that the
optimal value of phase $\gamma$ is always a constant in space, confirming
the fact that no spiral order exists in the coherent phase.

For the convenience of comparison, here we also
show the effective Hamiltonian of the double well system at $\nu=4N+1$ 
using the same definition of isospins 
in the layer index basis ($B_\|$ is along
$x$ axis):
\begin{eqnarray}
{\cal H}^{D1}_{eff}&=&\int d\vec{r}_\perp\left\{
-\Delta_{SAS}\cos(\gamma-P_xy)\tilde{\cal I}^{D1}_\perp+
\left[{\cal V}^D_{\Uparrow\Uparrow,\Uparrow\Uparrow}(0)
-{\cal V}^D_{\Uparrow\Uparrow,\Downarrow\Downarrow}(0)
\right]\left(\tilde{\cal I}^{D1}_z\right)^2
\right.
\nonumber\\
&&\left.+\left[2{\cal V}^D_{\Uparrow\Downarrow,\Downarrow\Uparrow}(0)
+\partial_\alpha^2{\cal V}^D_{\Uparrow\Downarrow,\Downarrow\Uparrow}(0)
(\nabla_\alpha\gamma)^2
\right]\left(\tilde{\cal I}^{D1}_\perp\right)^2
\right\},
\label{H_D1_eff}
\end{eqnarray}
where ${\cal V}^D_{I_1I_1,I_2I_2}(\vec{q}_\perp)$ is similarly defined
as Eq. (\ref{cal_V}) now for the double well system. 
Without losing generality, the effective 
Hamiltonian of the wide well system 
in Eq. (\ref{H_eff}) is similar to that of the double well system
in Eq. (\ref{H_D1_eff}) \cite{yang94}, except that in the former case
the in-plane field is included implicitly in the interaction matrix element,
$\tilde{V}^{W}_{I_1I_2,I_3I_4}(\vec{q}_\perp)$ and
the isospin stiffnesses in $x$ and $y$ directions
depend on the isospin direction.
The effective ``tunneling'' amplitude, $g_X^{}$, in the wide well system
originates from the Coulomb exchange energy and as such is not an
independent physical quantity (in contrast to) the tunneling 
energy in the double well system, which is an independent physical parameter.
When considering the intersubband level crossing in the small
in-plane magnetic field range (i.e. $W1$ case), 
$\tilde{g}_{\alpha\beta}(\gamma)$ is 
always positive as mentioned above, so that the
phase is fixed to be 0 or $\pi$ similar to the commensurate phase in the
double well system at $\nu=4N+1$, where only $\gamma=0$ is chosen
by the tunneling energy, $\Delta_{SAS}$. 
However, if we apply Eq. (\ref{H_eff}) to the {\it intrasubband} level
near degeneracy case in the strong in-plane field region (i.e. $W1'$ case), 
$\tilde{g}_{\alpha\beta}(\gamma)$ may become negative, so that the
total energy can be minimized at some finite isospin winding in
the isospin $x-y$ plane (say, $\gamma=P_x y$ becomes a function of
spatial coordinate). This result leads to the finite spiral order
of the isospin skyrmion stripe phase as discussed in Section 
\ref{sec_D1} in the large in-plane field region. 
In such a case, the effective ``tunneling'' amplitude, $g^{}_X$,
becomes zero since the space average of $\cos(2\gamma)=\cos(2P_x y)$
vanishes. Phenomenologically this situation is 
similar to the incommensurate state of the double
well system, where the tunneling energy is dominated by the Coulomb 
exchange energy so that the phase is not fixed by the in-plane field.
But in such intrasubband level near degeneracy case,
we will also have to consider the stripe order (gradient
term of $\tilde{\cal I}^{W1}_z$) to obtain the a full effective Hamiltonian
in describing the skyrmion stripe phase obtained in the HF calculation.
Finally we note that in general, the interlevel(layer)
scattering amplitude (proportional to 
${\cal V}^{W(D)}_{\Uparrow\Downarrow,\Uparrow\Downarrow}(\vec{q}_\perp)$) 
is finite in the wide well system, while it is zero 
in the double well system due to the layer separation.
This gives the finite values of $g^{}_X$ and the second 
term of $\tilde{g}^{}_{\alpha\beta}(\gamma)$, according to Eqs. 
(\ref{g_X})-(\ref{g_ab}), and leads to the major difference 
between a wide well system and a double well system. 

From the experimental point of view, however,
a smooth crossover from a wide well system to a double well system can be
observed in the same semiconductor system by tuning some parameters, e.g.
electron density \cite{WWW-DQW} due to the Coulomb screening 
effects on the
confinement potential. It is therefore very interesting
to investigate such crossover from monolayer to bilayer systems
from a more fundamental theory --- 
they are usually assumed to be two different systems when
studied in the literature (the inclusion of a modified form function
to take into account the finite width effect in the double
well system does not help in this comparison). 
The effective Hamiltonians shown above could be a good starting point 
for this problem, but we will not discuss it
further here since it must incorporate one-electron self-consistent potential
\cite{dempsey93,dassarma94_sdw_dqw} in the confinement potential
which is beyond the scope of our approximation scheme. We must emphasize
that it is useful to remember that the bilayer and the wide well systems 
may be continuously tuned into each other by changing the system carrier
density.

\subsection{Instability of the stripe phases and the domain wall formation}
\label{bubble}

In a single well system at partial filling factors,
another interesting kind of charge density wave ground state, the so-called
"bubble phase", may occur when the filling factor of the top level 
is away from the exact half
odd integer values \cite{review}. It is believed that when the filling factor
of the top Landau level is below about 0.4, electrons in the top level
may lower their potential energy by accumulating to become bubbles,
rather than stripes, in the two dimensional plane \cite{bubble-stripe}.
The equivalent "bubbles of holes" may occur
when the filling factor of the top level is between 0.6 and 1. This phase
can be understood as the edge state instability of the stripe 
phase due to the 
backward scattering between the two nearest edges, leading to the
breaking up of the stripes
into segments. These segments of stripes then
rearrange their shapes and positions to form bubbles in a lattice
to lower their total
energy. (The strong field Wigner crystal phase can be viewed 
as an extreme limit of the bubble phase --- having
only one electron in each ``bubble''.)
A possible experimental
evidence for such an interesting phase is the reentrant integer quantum Hall
effect observed recently \cite{review,eisenstein02}. Similar nonstripe density
modulation may also occur in the integral quantum Hall systems 
discussed in this paper. Near the level degeneracy region, the edge state
instability may be strong enough to break the stripe phase
and form several "bubble-like"
domain walls inside which the isospin $\langle I_z\rangle$ is different
from that outside the domain wall. Some spiral winding
structure may appear at the surface of the domain wall to
reduce the Hartree energy (see Fig. \ref{domain}).
Such domain wall structures could be stabilized by the presence
of surface disorder as suggested in Ref. \cite{chalker02}, and may also lead
to the observed resistance anisotropy \cite{pan01,zeitler01}.
However, this domain wall phase cannot be included in the 
trial wavefunctions we propose in this paper
\cite{exact_domain}, and we speculate that 
they may be important in understanding some other experimental results,
such as the hysteretic behavior
in the magnetoresistance experiments \cite{hysteretic,jungwirth01}.

\subsection{The $\nu=1$ bilayer system}
\label{nu=1_bilayer}

There has been a great deal of recent experimental and theoretical interest
in the (weak-tunneling) bilayer double well system at $\nu=1$, i.e. $N=0$
case, which has essentially been left out of our consideration in this 
paper where we have mostly restricted ourselves to the case $N\ge 1$,
with $\nu=4N+1$ ($D1$) or $4N+2$ ($D2$), level crossing bilayer situation.
(The $\nu=1$ bilayer case can be thought of as an isospin level
crossing situation for $N=0$ in our notation.)
Here we provide some brief comments on the $\nu=1$ bilayer situation in 
the context of our theoretical results presented in this paper. The specific 
issue we addresses is the nature of the phase transition in $\nu=1$ 
bilayer system as the layer separation ($d$) is increased in the 
zero (or weak) tunneling situation. It is commonly believed, not however based
on any really compelling evidence, that the $\nu=1$ bilayer system 
undergoes a first order transition from an incompressible interlayer-coherent
quantum hall state (presumably a Halperin (1,1,1) state \cite{halperin83})
to a compressible state (presumably two decoupled $\nu=1/2$ single-layer
states) as $d$ increases above a critical value (which depends
on the tunneling strength). We want to suggest here another distinct
possibility based on our results presented in this paper. It is in principle,
possible for the transition to be a second order quantum phase transition 
(rather than a direct first order transition) to a many-body state with 
exotic quantum order, such as an isospin stripe phase discussed in this 
paper. Such an exotic phase will still be incompressible (for low value 
of disorder), but perhaps with a much smaller value of the 
incompressibility gap. (The system will make a transition to the compressible
state of two decoupled $\nu=1/2$ layers at some still larger nonuniversal
value of $d$.) We believe that there is already some evidence supporting our
suggested ``double-transition'' scenario (a second order transition to a 
weakly incompressible stripe phase at $d=d_{c1}$, followed by a first 
order transition to the compressible decoupled $\nu=1/2$ layers at 
$d=d_{c2}>d_{c1}$). In particular, the early calculation of Fertig
\cite{fertig89_cdw_dqw} finding a finite wavevector magnetoplasmon mode
softening in bilayer TDHF theory indicates the obvious possibility of 
an isospin stripe formation at the characteristic wavevector of mode 
softening. The second order quantum phase transition associated with 
this mode softening leads to an incompressible state (at low disorder) 
which has a finite quasiparticle gap (perhaps reduced from that in the 
uniform (1,1,1) phase). At some higher value of disorder the stripes 
would eventually be pinned independently in each layer, leading to an 
incompressible-to-compressible phase transition as a function of increasing 
disorder. (This is also consistent with experiment where all the 
interesting results are typically obtained in extremely high mobility 
bilayer samples.) The level crossing HF technique used in our current 
work is unfortunately unsuitable to investigate the $\nu=1$ bilayer case, 
which has to be studied by the direct numerical diagonalization technique. 
It is therefore encouraging that there is recent numerical evidence
\cite{park} in apparent support of the scenario as proposed here.
The recent experimental results of inelastic light scattering 
\cite{bialyer_RRS}, showing that the magneto-roton excitation is 
softened at a layer separation very close to the 
incompressible-to-compressible phase boundary, is also consistent
with our scenario that there is an intermediate second order phase transition 
to a stripe phase before the system
undergoes the first order transition to a compressible state.
The magnetoresistance anisotropy measurement 
around the critical layer separation
of the observed magneto-roton excitation softening (which is 
also very close to the 
incompressible-to-compressible phase transition boundary)
should be a good test of such charge density wave order in the
$\nu=1$ incompressible quantum Hall system. 

\section{Summary}
\label{summary}

In this paper, we use the mean-field HF approximation to systematically study 
the possible spontaneous
breaking of parity, spin, and translational symmetries 
in different integer quantum Hall systems in the presence of 
an in-plane magnetic field, concentrating on the level crossing situation in
wide parabolic well and bilayer double well systems for 
both even and odd filling factors.
We propose a general class of variational wavefunctions to include the isospin
spiral and isospin stripe orders simultaneously, and 
discuss the symmetry breaking as well as the exotic quantum order properties
of various many-body phases generated by our wavefunctions. 
Comparing the HF energies
of these many-body phases, we find several of them
can be stabilized near the level crossing or level near degeneracy regions
of different systems, breaking certain system symmetries 
as listed in the last two columns of Table \ref{system_notation}:
(i) for a wide well system at $\nu=2N+1$, we find an isospin coherent phase
in the intersubband level crossing region (small $B_\|$), breaking
the parity symmetry only, and an isospin skyrmion stripe phase
in the intrasubband level near degeneracy region (large $B_\|$), breaking
the parity and the translational symmetries in both $x$ and $y$ directions
simultaneously and having finite topological isospin density,
which leads to appreciable charge oscillations in the direction parallel
to the in-plane field;
(ii) for a wide well system at $\nu=2N+2$, we find direct
first order phase transitions between simple (un)polarized QH states in both
intersubband and intrasubband level crossing regions, but
we suggest that the resistance anisotropy observed recently by Pan {\it
et. al.} \cite{pan01} may possibly be explained by a skyrmion stripe phase
by going beyond the HF approximation;
(iii) for a double well system at $\nu=4N+1$, we stabilize the coherent,
spiral, coherent stripe, and spiral stripe phases in different parameter
regions of the phase diagram (see Fig. \ref{D1_phase_diag}), and 
critically discuss the broken symmetries and the exotic quantum order
of these many-body phases in Section
\ref{symmetry_CI};
(iv) for a double well system at $4N+2$, only a coherent phase 
($=$ commensurate CAF phase in the literature) is stabilized, breaking 
parity and spin rotational symmetries simultaneously.
We also compare our HF results for these different systems in details,
and discuss the influence of wavefunction anisotropy, 
spiral-stripe coupling, spin degree of freedom and finite well width
in Section \ref{discussion}, manifesting the nontrivial effects of 
Coulomb interactions in the multi-component quantum Hall systems.

\section{Acknowledgment}
\label{acknowledgement}

This work is supported by the NSF grant DMR 99-81283, DMR 01-32874,
US-ONR, DARPA, and ARDA. We thank B. I. Halperin for 
helpful discussion and critical reading of the manuscript.
We also acknowledge useful discussions with
C. Kallin, S. Kivelson, A. Lopatnikova, A. MacDonald,
C. Nayak, K. Park, D. Podolsky, L. Radzihovsky, and S. Sachdev.

\appendix
\section{The variational wavefunction for stripes parallel to the 
in-plane field}
\label{diff_direction}

In this section we derive a variational trial wavefunction, which 
describes a stripe phase parallel to the in-plane field in a wide 
well system. As mentioned in Section \ref{trial_wavefunction}, we choose 
the Landau gauge, 
$\vec{A}_{[x]}(\vec{r}\,)=(-B_\perp y, -B_\|z, 0)$, in which 
particle momentum is conserved along $x$ axis.
We will use the notation "$\bar{O}$" to denote 
quantities, $O$, in this gauge in order to avoid confusion. The noninteracting
Hamiltonian, Eq. (\ref{H0_W}), becomes 
\begin{eqnarray}
\bar{H}^W_0&=&
\frac{1}{2m^\ast}\left(p_x-\frac{eB_\perp y}{c}\right)^2+
\frac{1}{2m^\ast}\left(p_y-\frac{eB_\| z}{c}\right)^2+
\frac{p_z^2}{2m^\ast}+\frac{1}{2}m^* \omega_{0}^2 z^2 -\omega_z S_z.
\label{H0_W_2}
\end{eqnarray}
Setting $p_x=k$ to be a good quantum number, we can obtain the single electron
wavefunction similar to the form of Eq. (\ref{wavefunctions_W_y}):
\begin{eqnarray}
\bar{\phi}^W_{\vec{n},s,k}(\vec{r}\,)&=&
\frac{e^{ikx}}{\sqrt{L_x}}\bar{\Phi}^W_{\vec{n}}(y-l_0^2k,z),
\end{eqnarray}
where $\bar{\Phi}_{\vec{n}}(y,z)$ satisfies the following wave equation:
\begin{eqnarray}
\left[\frac{1}{2}m^\ast\omega^2_\perp y^2+
\frac{1}{2m^\ast}\left(p_y-\frac{eB_\| z}{c}\right)^2+
\frac{p_z^2}{2m^\ast}+\frac{1}{2}m^* \omega_{0}^2 z^2 -\omega_z S_z\right]
\bar{\Phi}^W_{\vec{n}}(y,z)=E^{0,W}_{\vec{n},s}\bar{\Phi}^W_{\vec{n}}(y,z).
\label{H0_W_Phi}
\end{eqnarray}
If we define an auxiliary function, 
$\bar{\Phi}_{\vec{n}}^W{}'(p\,l_0^2,z)$ to be
\begin{eqnarray}
\bar{\Phi}_{\vec{n}}^W{}'(p\,l_0^2,z)=\frac{1}{\sqrt{2\pi}\,l_0}
\int dy\,\bar{\Phi}^W_{\vec{n}}
(y,z)\,e^{-ipy},
\label{FT}
\end{eqnarray}
then it is easy to show that $\bar{\Phi}_{\vec{n}}^W{}'(p\,l_0^2,z)$ 
satisfies exactly the same 
equation as $\Phi^W_{\vec{n}}(x,z)$ in Eq. (\ref{H0_W_1}) by redefining
$x=x+kl_0^2$. Therefore
we could write down the complete solution 
of Eq. (\ref{H0_W_Phi}) to be
\begin{eqnarray}
\bar{\Phi}^W_{\vec{n}}(y,z)=\frac{l_0}{\sqrt{2\pi}}\int dp\,e^{ipy}
{\psi}^{(1)}_{n_1}(pl_0^2\cos\theta-z\sin\theta)
\cdot{\psi}^{(2)}_{n_2}(pl_0^2\sin\theta+z\cos\theta),
\end{eqnarray}
where $\psi^{(i)}_{n}(x)$ has been defined in Eq. (\ref{wf_i}).
Note that the energy quantum number, $\vec{n}$, is the same as before because
the energy eigenvalue is independent of the gauge we choose.

Similar to our analysis in Section \ref{trial_wf_2l}, we can construct
the following trial wavefunction for the possible many-body ground state
near the degeneracy point of the two crossing levels (or of the two
nearly degenerate levels):
\begin{eqnarray}
|\bar{\Psi}_G\rangle &=& \sum_k\tilde{\bar{c}}^\dagger_{1,k}|LL\rangle
\nonumber\\
\left[\begin{array}{c}
       \tilde{\bar{c}}_{1,k}^\dagger \\ \tilde{\bar{c}}_{2,k}^\dagger
      \end{array}\right]
&=&\left[\begin{array}{cc}
       e^{-ik\bar{Q}_yl_0^2/2-i\bar{\gamma}/2}\cos(\bar{\psi}_k/2)& 
       e^{ik\bar{Q}_yl_0^2/2+i\bar{\gamma}/2}\sin(\bar{\psi}_k/2) \\
      -e^{-ik\bar{Q}_yl_0^2/2-i\bar{\gamma}/2}\sin(\bar{\psi}_k/2)& 
       e^{ik\bar{Q}_yl_0^2/2+i\bar{\gamma}/2}\cos(\bar{\psi}_k/2)
       \end{array}\right]
\cdot
\left[\begin{array}{c}
       \bar{c}_{\Uparrow,k-\bar{Q}_x/2}^\dagger \\
       \bar{c}_{\Downarrow,k+\bar{Q}_x/2}^\dagger
      \end{array}\right].
\label{general_trial_wf_x}
\end{eqnarray}
Then the isospin components can be obtained easily 
in the new ground state wavefunction:
\begin{eqnarray}
\langle \bar{\cal I}_z(\vec{q}_\perp)\rangle&=&
\frac{1}{2}N_\phi \delta_{q_x,0}\left[
\bar{A}^W_{\Uparrow\Uparrow}(\vec{q}_\perp,0)
\,e^{iq_y\bar{Q}_xl_0^2/2}\Theta_1(q_y)-
\bar{A}^W_{\Downarrow\Downarrow}(\vec{q}_\perp,0)\,e^{-iq_y\bar{Q}_xl_0^2/2}
\Theta_2(q_y)\right]
\label{I_z_x}
\end{eqnarray}
and
\begin{eqnarray}
\langle \bar{\cal I}_+(\vec{q}_\perp)\rangle&=&
N_\phi\delta_{q_x,\bar{Q}_x}\bar{A}^W_{\Uparrow\Downarrow}(\vec{q}_\perp,0)
\Theta_{3}(q_y-\bar{Q}_y)\,e^{i\bar{\gamma}}.
\label{I+_x}
\end{eqnarray}
Therefore the stripe phase constructed in this method
is along $x$ axis and the spiral direction is determined by 
$\vec{\bar{Q}}_\perp$.

Finally it is instructive to point out that the form function obtained
from the noninteracting electron wavefunction in Eq. (\ref{A_def_W}) is 
{\it invariant} under such gauge transformation. Defining 
the form function in the new gauge similar to Eq. (\ref{A_def_W}), we 
have
\begin{eqnarray}
\bar{A}^W_{\vec{n}_1\vec{n}_2}(\vec{q}\,)&=&
\int dy\int dz\,e^{-iq_yy}e^{-iq_zz}
\bar{\Phi}^W_{\vec{n}_1}(y-l_0^2q_x/2,z)
\bar{\Phi}^W_{\vec{n}_2}(y+l_0^2q_x/2,z) \nonumber\\
&=&\frac{l_0^2}{2\pi}\int dy\int dz\,e^{-iq_yy}e^{-iq_zz}
\int dp\,e^{ip(y-l_0^2q_x/2)}\,\bar{\Phi}_{\vec{n}_1}^W{}'(p\,l_0^2,z)
\int dp'\,e^{ip'(y+l_0^2q_x/2)}\,\bar{\Phi}_{\vec{n}_2}^W{}'(p'l_0^2,z)
\nonumber\\
&=&\int dx\int dz\,e^{-iq_xx}e^{-iq_zz}
{\Phi}^W_{\vec{n}_1}(x-l_0^2q_y/2,z)
{\Phi}^W_{\vec{n}_2}(x+l_0^2q_y/2,z)=A^W_{\vec{n}_1\vec{n}_2}(\vec{q}\,),
\end{eqnarray}
where we have changed the integration variables: $p=q_y/2+x/l_0^2$ and
$p'=q_y/2-x/l_0^2$ in the last equation.
Therefore the interaction matrix element, $\bar{V}^W_{\vec{n}_1\vec{n}_2,
\vec{n}_3\vec{n}_4}(\vec{q}\,)$ does not change in the new gauge, and
the HF variational energies for the wide well system in the new gauge,
$\vec{A}_{[x]}(\vec{r}\,)$, can be obtained by doing the following
simple transformation in Eqs. (\ref{E_H^W1}), (\ref{E_F^W1}),
(\ref{E_H_W2}), and (\ref{E_F_W2}):
$\tilde{V}^W_{\vec{n}_1\vec{n}_2,\vec{n}_3\vec{n}_4}(q_n,0)
\longrightarrow \tilde{V}^W_{\vec{n}_1\vec{n}_2,\vec{n}_3\vec{n}_4}(0,q_n)$,
$\tilde{V}^W_{\vec{n}_1\vec{n}_2,\vec{n}_3\vec{n}_4}(q_n,Q_y)
\longrightarrow \tilde{V}^W_{\vec{n}_1\vec{n}_2,\vec{n}_3\vec{n}_4}(Q_y,q_n)$,
$\Theta_{3}(q_n-Q_x)\longrightarrow \Theta_{3}(q_n-Q_y)$,
$\cos(q_nq_yl_0^2)\longrightarrow \cos(q_nq_xl_0^2)$,
$\cos(q_n(q_y+Q_y)l_0^2)\longrightarrow \cos(q_n(q_x+Q_x)l_0^2)$,
$\cos(q_nQ_yl_0^2)\longrightarrow\cos(q_nQ_xl_0^2)$,
$\cos(q_n(Q_y\pm Q_y')l_0^2)\longrightarrow\cos(q_n(Q_x\pm Q_x')l_0^2)$,
$\Theta_{3}(q_n\pm Q_x{}^({}'{}^))\longrightarrow
\Theta_{3}(q_n\pm Q_y{}^({}'{}^))$, and
$\cos((q_xQ_y-q_nq_y)l_0^2)\longrightarrow\cos((q_yQ_x-q_nq_x)l_0^2)$.
Note that all other physical features of stripe phases (for example, 
the charge oscillation induced by the 
topological spin density of a skyrmion stripe phase in 
Appendix \ref{topo_charge} and the perturbation theory for the stripe formation
in Section \ref{pert_stripe}, etc.) developed in the main text for the stripe
aligned along $y$ axis can also directly apply to the stripe phase
along $x$ axis in a similar way.
The stripe phases constructed
in a four level coherent trial wavefunction as shown in 
Section \ref{trial_wf_4l} for even filling systems can also be obtained
similarly.

\section{Analytical expression of the form function}
\label{A_function}

The explicit formula for the form function,
$A^{W}_{\vec{n}_\alpha\vec{n}_\beta}(\vec{q}\,)$, 
defined in Eq. (\ref{A_def_W})
can be evaluated analytically via a special function \cite{wang02}.
Here we just show the analytical results 
(for convenience, we define
$\vec{n}_\alpha=(n_\alpha,n_\alpha')$
and $\vec{n}_\beta=(n_\beta,n_\beta')$ to be the Landau level indices
of a parabolic wide well system):
\begin{eqnarray}
&&A^{W}_{\vec{n}_\alpha\vec{n}_\beta}(\vec{q})=
\sqrt{\frac{n_{\alpha\beta,min}!}{n_{\alpha\beta,max}!}\cdot
\frac{n'_{\alpha\beta,min}!}{n'_{\alpha\beta,max}!}}
\nonumber\\
&&\times
\exp\left[-\frac{\cos^2\theta(q_yl_0)^2+
(\cos\theta q_xl_0-\sin\theta q_zl_0)^2\lambda_1^2}{4\lambda_1}\right]
\exp\left[-\frac{\sin^2\theta(q_yl_0)^2+
(\sin\theta q_xl_0+\cos\theta q_zl_0)^2\lambda_2^2}{4\lambda_2}\right]
\nonumber\\
&&\times
\left(\frac{\mp\cos\theta(q_yl_0)-i
(\cos\theta q_xl_0-\sin\theta q_zl_0)\lambda_1}{\sqrt{2\lambda_1}}\right)
^{m_{\alpha\beta}}
\left(\frac{\mp\sin\theta(q_yl_0)-i
(\sin\theta q_xl_0+\cos\theta q_zl_0)\lambda_2}{\sqrt{2\lambda_2}}\right)
^{m'_{\alpha\beta}}
\nonumber\\
&&\times
L_{n_{\alpha\beta,min}}^{m_{\alpha\beta}}\left(\frac{\cos^2\theta(q_yl_0)^2+
(\cos\theta q_xl_0-\sin\theta q_zl_0)^2\lambda_1^2}{2\lambda_1}\right)
L_{n'_{\alpha\beta,min}}^{m'_{\alpha\beta}}\left(\frac{\sin^2\theta(q_yl_0)^2+
(\sin\theta q_xl_0+\cos\theta q_zl_0)^2\lambda_2^2}{2\lambda_2}\right),
\label{A_explicit}
\end{eqnarray}
where $\pm$ is the sign of $n_\alpha^{}{}^{(}{}'^{)}
-n_\beta^{}{}^{(}{}'^{)}$ for each
bracket, $n_{\alpha\beta,min(max)}^{}{}^{(}{}'^{)}\equiv
Min(Max)\left(n_\alpha^{}{}^{(}{}'^{)},n_\beta^{}{}^{(}{}'^{)}\right)$, and
$m_{\alpha\beta}^{}{}^{(}{}'^{)}\equiv |n_\alpha^{}{}^{(}{}'^{)}
-n_\beta^{}{}^{(}{}'^{)}|$;
$\lambda_{1,2}=l^2_{1,2}/l_0^2=\omega_\perp/\omega_{1,2}$ 
are dimensionless parameters,
$L_n^m(x)$ is the generalized Laguerre polynomial \cite{ryzhik}
and $\tan(2\theta)\equiv
-2\omega_\perp\omega_\|/(\omega_b^2-\omega_\perp^2)$.

As for a double well system with zero well width, we note that  
the noninteracting single electron wavefunction in 
Eq. (\ref{wavefunctions_W_y}) becomes 
$L_y^{-1/2}e^{iky}\psi^{(0)}_{n}(x+l_0^2k)\sqrt{\delta(z)}$ for
$\omega_0\to+\infty$, where $\psi^{(0)}_{n}(x)$ is the same as 
Eq. (\ref{wf_i}) with $l_i$ replaced by the magnetic length, $l_0$.
Separating the $z$ component and integrating $q_z$ first, we obtain
the following isotropic form function:
\begin{eqnarray}
A^{D}_{n_\alpha n_\beta}(\vec{q}_\perp)&=&\sqrt{\frac{n_{\alpha\beta,min}!}
{n_{\alpha\beta,max}!}}\exp\left[-\frac{q^2l_0^2}{4}\right]
\left(\frac{\pm q_yl_0-iq_xl_0}{\sqrt{2}}\right)^m
L_{n_{min}}^m\left(\frac{q^2l_0^2}{2}\right),
\end{eqnarray}
where $q=|\vec{q}_\perp|$, and all other notations are the same as above.

\section{Topological charge density in skyrmion stripe phase}
\label{topo_charge}

According to Ref. \cite{moon95}, the charge density induced by
the topological isospin density 
in the double well system at $\nu=1$ can be obtained from 
the isospin density function, $\vec{m}(\vec{r}_\perp)$:
\begin{eqnarray}
\rho_{topo}(\vec{r}_\perp)=-\frac{1}{\pi}\varepsilon_{\mu\nu}
\vec{m}(\vec{r}_\perp)\cdot\left[\partial_\mu\vec{m}(\vec{r}_\perp)\times
\partial_\nu\vec{m}(\vec{r}_\perp)\right],
\label{topo_charge_density}
\end{eqnarray}
where the magnitude of $\vec{m}(\vec{r}_\perp)$ has 
been normalized to $\frac{1}{2}$. 
In this section we will show that the extra electron charge density,
$\rho_{ex}(\vec{r}_\perp)$, obtained in Eq. (\ref{rho_local}) for a skyrmion
stripe phase is related to the induced charge density fluctuation calculated
by Eq. (\ref{topo_charge_density}).
Using the isospin density operator defined in coordinate space,
Eq. (\ref{newisospin_I}), and the trial wavefunction in
Eq. (\ref{wavefunction_general}), 
we obtain the mean values
of the isospinor in the 2D well plane after renormalization by
the electron average density, $(2\pi l_0^2)^{-1}$:
\begin{eqnarray}
\langle {\cal I}_z(\vec{r}_\perp)\rangle
&=&\frac{1}{2}\left\{ 1-\sum_{q_x}\Theta_2(q_x)
\left[A_{\Uparrow\Uparrow}(q_x,0,0)\cos(q_x(x-Q_yl_0^2/2))
+A_{\Downarrow\Downarrow}(q_x,0,0)\cos(q_x(x+Q_yl_0^2/2))\right]\right\},
\label{m_bar_z}
\end{eqnarray}
and
\begin{eqnarray}
\langle {\cal I}^W_+(\vec{r}_\perp)\rangle &=&
\langle {\cal I}^W_-(\vec{r}_\perp)\rangle^\ast
=\sum_{q_x}\Theta_3(q_x-Q_x)A_{\Uparrow\Downarrow}(q_x,Q_y,0)
e^{i(q_xx+Q_yy-\gamma)}.
\label{m_bar_+}
\end{eqnarray}
Since Eq. (\ref{topo_charge_density}) is valid only to the 
leading order in (weak) isospin density modulation 
\cite{moon95}, we should take
the limits of small isospin spiral and stripe orders
(i.e. $\tilde{q}l_0\ll 1$ and $\tilde{q}Q_yl_0^2\ll 1$ for
$\tilde{q}=q_x$) in above
equations and obtain
\begin{eqnarray}
\langle {\cal I}_z(\vec{r}_\perp)\rangle&\sim&
\frac{1}{2}\left\{ \cos(\psi_0)-4\Delta\sin(\psi_0)
\cos(\tilde{q}Q_yl_0^2/2)\cos(\tilde{q}x)\right\}
=\frac{1}{2}\cos(\psi_{x/l_0^{2}}),
\label{m_bar_z_approx}
\end{eqnarray}
where we have used $\psi_k=\psi_0+4\Delta\cos(k\tilde{q}l_0^2)$ 
(Eq. (\ref{perturb_psi})) for the last equation, and 
\begin{eqnarray}
\langle {\cal I}_+(\vec{r}_\perp)\rangle = 
\langle {\cal I}_-(\vec{r}_\perp)\rangle^\ast
&\sim&\frac{1}{2}A_{\Uparrow\Downarrow}(\vec{Q}_\perp,0)
\sin(\psi_{x/l_0^2})\,e^{i\vec{Q}_\perp\cdot\vec{r}_\perp-i\gamma}.
\label{m_bar_+_approx}
\end{eqnarray}
Note that the long wavelength limit behavior of 
$\langle {\cal I}_\pm(\vec{r}_\perp)\rangle$ is different in 
wide and double well systems: for the former case, 
the two isospin states, $\Uparrow$ and $\Downarrow$, 
have different Landau level
indices, $\vec{n}_1$ and $\vec{n}_2$, and 
therefore $A_{\Uparrow\Downarrow}(\vec{Q}_\perp,0)=
A^W_{\vec{n}_1\vec{n}_2}(\vec{Q}_\perp,0)\to 0$ 
as $|\vec{Q}_\perp|\to 0$ (see Eq. (\ref{A_explicit})). However, 
for the latter case, the two levels of different isospins are of the {\it same}
Landau level index, $N$, and hence $A_{\Uparrow\Downarrow}(\vec{Q}_\perp,0)=
A^D_{NN}(\vec{Q}_\perp,0)\to 1$ as $|\vec{Q}_\perp|\to 0$.
As a result, Eq. (\ref{m_bar_+_approx}) implies that the total magnitude 
of the isospin density in a wide well system 
is not a constant of winding vector, while it is 
a constant for a double well system.
Therefore, in order to apply 
Eq. (\ref{topo_charge_density}) to the wide well system, we have to 
project the isospin vector onto the unit sphere by renormalizing its
magnitude to 1/2 with a space independent constant and redefining
the following isospin density vector, $\vec{m}(\vec{r}_\perp)$: 
\begin{eqnarray}
{m}_x(\vec{r}_\perp)&=&\frac{1}{2}\sin(\psi_{x/l_0^2})\cos(Q_xx+Q_yy-\gamma),
\label{m_x}
\nonumber\\
{m}_y(\vec{r}_\perp)&=&\frac{1}{2}\sin(\psi_{x/l_0^2})\sin(Q_xx+Q_yy-\gamma),
\label{m_y}
\nonumber\\
{m}_z(\vec{r}_\perp)&=&\frac{1}{2}\cos(\psi_{x/l_0^{2}}),
\label{m_z}
\end{eqnarray}
which leads to the following charge density modulation
via Eq. (\ref{topo_charge_density}):
\begin{eqnarray}
\rho_{topo}(\vec{r}_\perp)&=&-\frac{2}{\pi}
\vec{m}(\vec{r}_\perp)\cdot\partial_x\vec{m}(\vec{r}_\perp)\times
\partial_y\vec{m}(\vec{r}_\perp)
\nonumber\\
&=&-\frac{Q_y}{4\pi l_0^2}\sin(\psi_{x/l_0^2})
\frac{\partial\psi_k}{\partial k}\Big\rfloor_{k=x/l_0^2}
\nonumber\\
&\sim&Q_y\Delta\,\tilde{q}\sin(\psi_0)\sin(\tilde{q}x)/\pi.
\label{topo_charge_cal}
\end{eqnarray}
In the same weak isospin density modulation limit,
the extra local charge density shown in Eq. (\ref{rho_local}) becomes 
\begin{eqnarray}
\rho_{ex}(\vec{r}_\perp)&\sim& -\frac{1}{2\pi l_0^2}\sum_{q_x}\Theta_2(q_x)
\left[A_{\Uparrow\Uparrow}(q_x,0,0)\cos(q_x(x-Q_yl_0^2/2))
-A_{\Downarrow\Downarrow}(q_x,0,0)\cos(q_x(x+Q_yl_0^2/2))\right]
\nonumber\\
&\sim&-\frac{2}{2\pi l_0^2}\Delta\sin(\psi_0)
\left[\cos(\tilde{q}(x-Q_yl_0^2/2))-\cos(\tilde{q}(x+Q_yl_0^2/2))\right]
\nonumber\\
&\sim&-Q_y\Delta\,\tilde{q}\sin(\psi_0)\sin(\tilde{q} x)/\pi,
\label{ex_charge_cal}
\end{eqnarray}
which is the same as the charge density shown in 
Eq. (\ref{ex_charge_cal}) above (but with an opposite sign).
Such long wavelength charge density 
modulation is nonzero only when both the spiral order and stripe order
are finite, and when the spiral winding
vector is perpendicular to the stripe normal vector.
Therefore this charge density modulation induced by the topological isospin 
density can be realized as a feature of isospin skyrmion stripe phase.

\section{Relationship between the perturbation study of the 
stripe formation and the collective mode softening at finite wavevector} 
\label{goldstone_mode}

In Section \ref{pert_stripe} we discuss the suitability of the 
perturbation method to study
the stripe formation if it results from an instability of a uniform
phase via a two-step phase transition. In this 
section we will show that this method is equivalent to studying the
stripe formation via the mode softening of a finite wavevector
collective mode {\it inside} an isospin coherent or spiral phase,
which is known as a standard method to investigate the stripe phase
instability in a double well system \cite{yang95,brey00}.
Before we investigate the relationship between these two methods,
it is instructive to mention that in general the collective mode dispersion
can be obtained from the HF energy calculated by using a trial wavefunction 
similar to Eq. (\ref{wavefunction_general})
but based on the symmetry-broken phase. More precisely, we can start from
the following trial wavefunction, which is based on the rotated isospin
basis, $\tilde{c}^\dagger_{1(2),k}$, obtained in 
Eq. (\ref{wavefunction_general}):
\begin{eqnarray}
|\tilde{\Psi}_G\rangle &=& \sum_k\tilde{\tilde{c}}^\dagger_{1,k}|LL\rangle
\nonumber\\
\left[\begin{array}{c}
       \tilde{\tilde{c}}_{1,k}^\dagger \\ \tilde{\tilde{c}}_{2,k}^\dagger
      \end{array}\right]
&=&\left[\begin{array}{cc}
   e^{ik\tilde{Q}_xl_0^2/2}\cos(\tilde{\psi}_0/2)& e^{-ik\tilde{Q}_xl_0^2/2}
    \sin(\tilde{\psi}_0/2) \\
      -e^{ik\tilde{Q}_xl_0^2/2}\sin(\tilde{\psi}_0/2)& 
   e^{-ik\tilde{Q}_xl_0^2/2}\cos(\tilde{\psi_0}/2)
       \end{array}\right]
\cdot
\left[\begin{array}{c}
 \tilde{c}_{1,k-\tilde{Q}_y/2}^\dagger\\ 
 \tilde{c}_{2,k+\tilde{Q}_y/2}^\dagger
      \end{array}\right],
\label{wavefunction_goldstone}
\end{eqnarray}
where we use $\tilde{\tilde{c}}_{1(2),k}^\dagger$, $\tilde{\psi_0}$,
and $\vec{\tilde{Q}}_\perp$ to denote the new state and the relevant
variational parameters for this {\it second} rotation. 
For simplicity, we have taken the phase $\tilde{\gamma}=
\gamma=0$ in the following discussion. As discussed before,
we just need to consider a uniform $\tilde{\psi}_0$ to study the collective
mode dispersion, i.e. when we take $\tilde{\psi}_0$ to be small,
the coefficient in front of leading order (quadratic of $\tilde{\psi_0}$)
term of the HF variational energy calculated by
the new many-body wavefunction, $|\tilde{\Psi}_G\rangle$,
can give the dispersion of a low-lying collective mode. 
This method is basically equivalent to the conventional
time-dependent Hartree Fock theory used in the integer quantum Hall system
\cite{wang02,canted_phase}.

\begin{figure}

 \vbox to 4cm {\vss\hbox to 5.cm
 {\hss\
   {\includegraphics{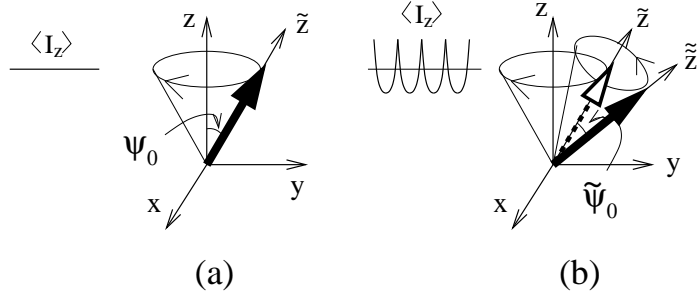}
   }
  \hss}
 }
\caption{(a) Isospin polarization in a coherent or spiral phase: the
new polarization axis is along $\tilde{z}$ axis with an angle $\psi_0$ tilted
from the original $z$ axis. Finite winding wavevector causes 
$\langle {\cal I}_{x,y}\rangle$ oscillating in the real space but keeps 
$\langle {\cal I}_{z}\rangle$ as a constant.
(b) If the isospin coherent (or spiral) phase
is unstable to form a new isospin spiral phase based on the rotated
coordinate, the isospin winding about the new polarization axis, 
$\tilde{z}$, will cause a new periodically oscillating 
$\langle {\cal I}_{z}\rangle$
in the isospin projection onto the original $z$ axis, 
showing a character of stripe phase.
}
\label{isospin_angle}
\end{figure}

\begin{figure}

 \vbox to 6cm {\vss\hbox to 5.cm
 {\hss\
   {\includegraphics{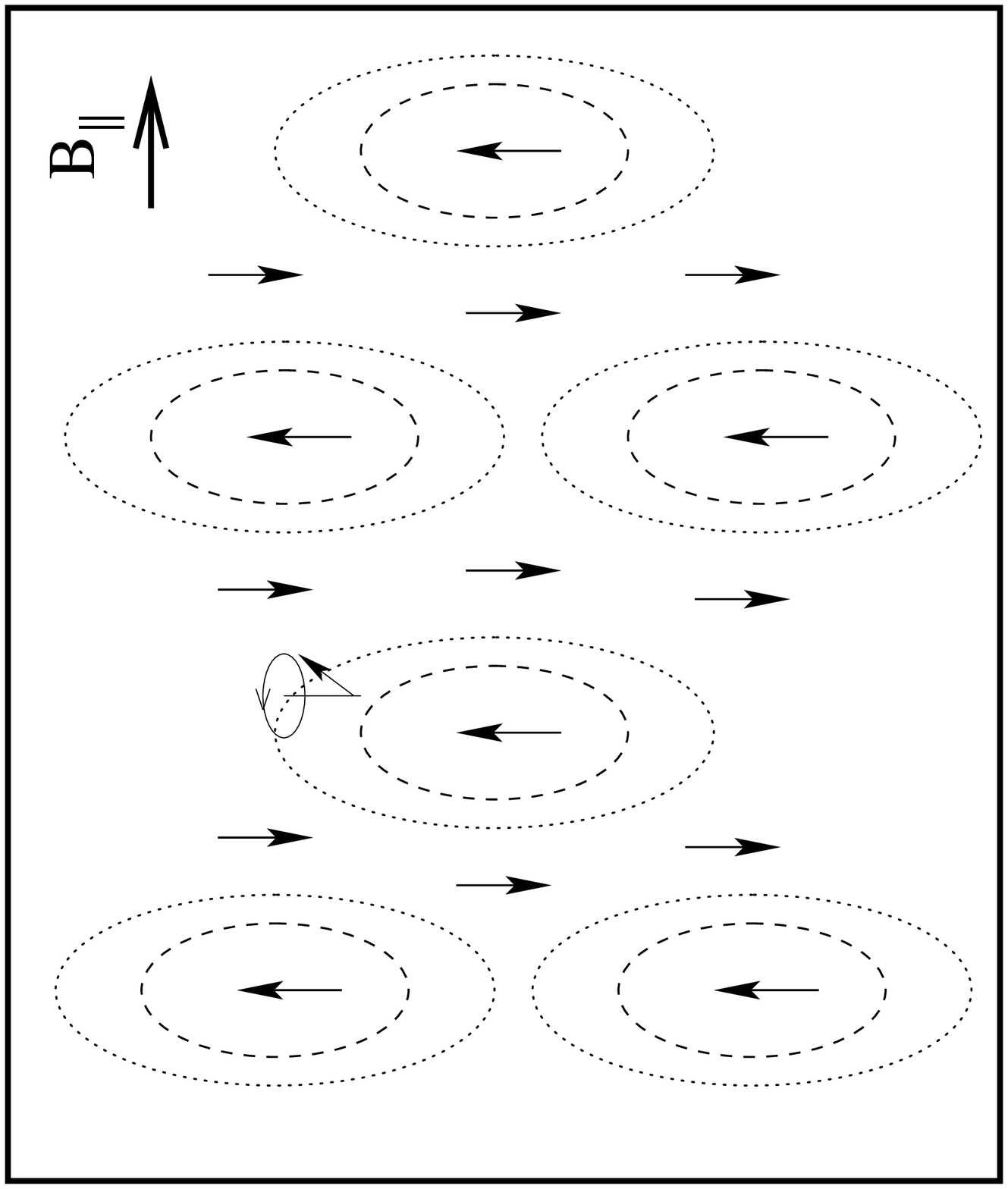}
   }
  \hss}
 }
\caption{Possible domain wall structure in integral quantum Hall system
near level degeneracy region.
Arrows denote the isospin polarization (see Fig. \ref{isospin_angle}(a)).
In the surface of each domain, isospin hybridization may be generated, 
resulting a coherent spiral order.
The in-plane magnetic field is in $+\hat{x}$ direction.
}
\label{domain}
\end{figure}

The study of stripe formation from the collective mode softening 
at finite wavevector inside the coherent phase region
can be realized in the following isospin rotation picture:
When the system has no coherence (i.e. the conventional uniform
QH state), the isospin is polarized along isospin $z$ axis. The trial 
wavefunction of a many-body phase shown in Eq. 
(\ref{wavefunction_general}) can be understood as a result of a rotation
in isospin space, tilting the isospin polarization from 
$+\hat{z}$ to $+\hat{\tilde{z}}$ (see Fig. \ref{isospin_angle}(a)) with a
tilting angle, $\psi_0$. If the spiral winding wavevector is not zero
(i.e. a spiral phase), the isospin polarization vector changes its
$x$ and $y$ components in the old isospin space 
with respect to the guiding center
coordinate, $k$, but keeps its $z$ component, 
$\langle I_z\rangle$, as a constant along the original $z$ axis, i.e.
no stripe order.
When we apply a similar rotation, Eq. (\ref{wavefunction_goldstone}), 
onto the existing coherent
or spiral phase, the isospin polarization vector prefers to have
a second tilting angle, $\tilde{\psi}_0$, about the $\hat{\tilde{z}}$ axis 
(see Fig. \ref{isospin_angle}(b)).  
As shown in Fig. \ref{isospin_angle}(b), if the isospin vector winds 
about the new axis, $\hat{\tilde{z}}$, (i.e. $\vec{\tilde{Q}}_\perp\neq 0$)
the isospin projection onto the original
$\hat{z}$ axis becomes periodically oscillating, showing a 
stripe structure in the original isospin space.
The amplitude of such stripe phase is proportional to the amplitude of 
the second tilting angle, $\tilde{\psi}_0$, if it is small.
Combining with the fact we mentioned above that the collective mode
can be obtained from the HF variational energy by 
taking small $\tilde{\psi}_0$ in the coherent phase, 
we can conclude that the stripe formation, if it results from a two-step
second order phase transition, can be investigated by
the mode softening of the collective mode at a finite wavevector
{\it inside} the coherent (or spiral) phase.
On the other hand, if this collective mode is always gaped at finite
wavevector, we can conclude that there is no
second order phase transition for a stripe formation, and the only 
possible stripe phase is that from the first order transition.

To understand better the relationship between the method 
of studying the stripe formation from the collective mode softening 
at finite wavevector and the method of the perturbation theory
developed in Section \ref{pert_stripe}, we can 
apply Eq. (\ref{wavefunction_goldstone}) to 
Eq. (\ref{wavefunction_general}) (assuming the first tilting angle, 
$\psi_0^\ast\neq 0$, for the first rotation is nonzero and known by 
minimizing the HF energy without stripe order in $\psi_0$),
and let the second tilting angle $\tilde{\psi}_0$ to be small.
Using $\tilde{c}_{1(2),k}^\dagger$ as an intermediate isospin basis,
and considering the conventional Landau gauge,
$\vec{A}_{[y]}(\vec{r}\,)$, with  
the second spiral winding wavevector being along $\hat{x}$, i.e.
$\vec{\tilde{Q}}_\perp=(\tilde{Q}_x,0)$, we obtain
\begin{eqnarray}
\tilde{\tilde{c}}^\dagger_{1,k}&=&
e^{ik\tilde{Q}_xl_0^2/2}\tilde{c}^\dagger_{1,k}
+e^{-ik\tilde{Q}_xl_0^2/2}(\tilde{\psi}_0/2)\tilde{c}^\dagger_{2,k}
\nonumber\\
&=&e^{ik\tilde{Q}_xl_0^2/2}\left[e^{ikQ_xl_0^2/2}\left(\cos(\psi^\ast_0/2)
-\sin(\psi^\ast_0/2)(\tilde{\psi}_0/2)e^{-ik\tilde{Q}_xl_0^2}\right)
{c}^\dagger_{1,k-{Q}_y/2}
\right.\nonumber\\
&&+\left.e^{-ikQ_xl_0^2/2}\left(\sin(\psi^\ast_0/2)+\cos(\psi^\ast_0/2)
(\tilde{\psi}_0/2)e^{-ik\tilde{Q}_xl_0^2}\right)
{c}^\dagger_{2,k+{Q}_y/2}\right].
\label{tt_c}
\end{eqnarray}
Similarly we can obtain the following energetically degenerate state by
using $\vec{\tilde{Q}}_\perp=(-\tilde{Q}_x,0)$:
\begin{eqnarray}
\tilde{\tilde{c}}'{}^\dagger_{1,k}
&=&e^{-ik\tilde{Q}_xl_0^2/2}\left[e^{ikQ_xl_0^2/2}\left(\cos(\psi^\ast_0/2)
-\sin(\psi^\ast_0/2)(\tilde{\psi}_0/2)e^{ik\tilde{Q}_xl_0^2}\right)
{c}^\dagger_{1,k-Q_y/2}
\right.\nonumber\\
&&+\left.e^{-ikQ_xl_0^2/2}\left(\sin(\psi^\ast_0/2)+\cos(\psi^\ast_0/2)
(\tilde{\psi}_0/2)e^{ik\tilde{Q}_xl_0^2}\right)
{c}^\dagger_{2,k+Q_y/2}\right].
\label{tt_c2}
\end{eqnarray}
Then we can define a new operator by adding above two equations together
in the leading order of $\tilde{\psi}_0$:
\begin{eqnarray}
d^\dagger_{1,k}&\equiv&\frac{1}{2}\left[e^{-ik\tilde{Q}_xl_0^2/2}
\tilde{\tilde{c}}^\dagger_{1,k}+e^{ik\tilde{Q}_xl_0^2/2}
\tilde{\tilde{c}}'{}^\dagger_{1,k}\right]
\nonumber\\
&=&e^{ikQ_xl_0^2/2}\left(\cos(\psi^\ast_0/2)
-\sin(\psi^\ast_0/2)(\tilde{\psi}_0/2)\cos(k\tilde{Q}_xl_0^2)\right)
{c}^\dagger_{1,k-Q_y/2}
\nonumber\\
&&+e^{-ikQ_xl_0^2/2}\left(\sin(\psi^\ast_0/2)+\cos(\psi^\ast_0/2)
(\tilde{\psi}_0/2)\cos(k\tilde{Q}_xl_0^2)\right)
{c}^\dagger_{2,k+Q_y/2}
\nonumber\\
&\sim&e^{ikQ_xl_0^2/2}\cos(\psi_k/2){c}^\dagger_{1,k-Q_y/2}
+e^{-ikQ_xl_0^2/2}\sin(\psi_k/2){c}^\dagger_{2,k+Q_y/2},
\end{eqnarray}
where
\begin{eqnarray}
\psi_k &\equiv& \psi^\ast_0+\tilde{\psi}_0\cos(k\tilde{q}l_0^2)
\label{perturb_psi0}
\end{eqnarray}
for $\tilde{\psi}_0\ll 1$.
We note that Eq. (\ref{perturb_psi0}) is exactly the same as Eq. 
(\ref{perturb_psi}) shown in Section \ref{pert_stripe} 
($4\Delta=\tilde{\psi}_0$). Therefore
we have shown that studying the stripe formation instability by using
Eq. (\ref{perturb_psi}) is equivalent to studying the 
finite wavevector mode softening of
the collective mode {\it inside} a coherent or a spiral phase.
Note that another kind of stripe phase with
$I_z$ modulation along $y$ direction can be obtained by the
similar method using the other gauge, $\vec{A}_{[x]}(\vec{r}\,)$, 
in which particle momentum is conserved along
$x$ direction. Therefore the 
perturbative method for the stripe formation developed in Section 
\ref{pert_stripe} is justified and can be applied in a more general
situation.

\section{A general stripe phase function, $\psi_k$}
\label{psi_k}

In general, we can use the following periodic function for $\psi_k$ in
Eqs. (\ref{wavefunction_general}), (\ref{wavefunction_4l_D2}) and
(\ref{wavefunction_4l_W2}):
\begin{eqnarray}
\psi_k=\left\{\begin{array}{cl}
\psi_1 &\mbox{for}\ 0\leq |kl_0^2/a-m|<\xi(1-\eta)/2 \\
-\frac{\varphi}{\xi\eta}(k-\xi/2)+\psi_0
&\mbox{for}\ \xi(1-\eta)/2\leq |kl_0^2/a-m|\leq \xi(1+\eta)/2 \\
\psi_2 &\mbox{for}\ \xi(1+\eta)/2<|kl_0^2/a-m| \leq 1/2 \\
\end{array},
\right.
\end{eqnarray}
where $\psi_0=(\psi_1+\psi_2)/2$ and $\varphi=\psi_1-\psi_2$.
$0<\xi,\eta<1$, $0\leq\psi_1,\psi_2\leq\pi$ and $m$ is an integer.
The meaning of these variational parameters
can be understood more clearly from Fig. \ref{fig_psi}.
Note that when $\eta=0$, we have a rectangular function, while
for $\eta\neq 0$, we have a smooth $\psi_k$ to the linear oder.
To calculate $\Theta_{i}(q_n)$ ($i=1,2,3$) for such $\psi_k$, we can 
first evaluate the following quantity (let $k'=kl_0^2/a$, $q_x'=q_xa$):
\begin{eqnarray}
\Theta&\equiv&\frac{1}{N_\phi}\sum_{k>0} \cos(kq_xl_0^2)e^{i\psi_k}
\nonumber\\
&=&\int_{0}^{\xi(1-\eta)/2}dk'\cos(k'q_x')e^{i\psi_1}
+\int_{\xi(1-\eta)/2}^{\xi(1+\eta)/2}dk'\cos(k'q_x')
e^{-i\varphi(k'-\xi/2)/\xi\eta+i\psi_0}
+\int_{\xi(1+\eta)/2}^{1/2}\cos(k'q_x')e^{i\psi_2}
\nonumber\\
&=&(\cos(\psi_1)+i\sin(\psi_1))\frac{\sin(\xi(1-\eta)q_x'/2)}{q_x'}
+(\cos(\psi_2)+i\sin(\psi_2))\left(\frac{\sin(q_x'/2)}{q_x'}-
\frac{\sin(\xi(1+\eta)q_x'/2)}{q_x'}\right)
\nonumber\\
&&+\frac{2\xi\eta\cos(q_x'\xi/2)\cos(\psi_0)
\left[\xi\eta q_x'\cos(\varphi/2)\sin(\xi\eta q_x'/2)-\varphi
\cos(\xi\eta q_x'/2)\sin(\varphi/2)\right]}
{\xi^2\eta^2 {q_x'}^2-\varphi^2}
\nonumber\\
&&+\frac{2\xi\eta
\sin(q_x'\xi/2)\sin(\psi_0)
\left[-\varphi\cos(\varphi/2)\sin(\xi\eta q_x'/2)+\xi\eta q_x'
\cos(\xi\eta q_x'/2)\sin(\varphi/2)\right]}
{\xi^2\eta^2 {q_x'}^2-\varphi^2}
\nonumber\\
&&-i\frac{2\xi\eta\cos(q_x'\xi/2)\sin(\psi_0)
\left[-\xi\eta q_x'\cos(\varphi/2)\sin(\xi\eta q_x'/2)+\varphi
\cos(\xi\eta q_x'/2)\sin(\varphi/2)\right]}
{\xi^2\eta^2 {q_x'}^2-\varphi^2}
\nonumber\\
&&-i\frac{2\xi\eta
\sin(q_x'\xi/2)\cos(\psi_0)
\left[-\varphi\cos(\varphi/2)\sin(\xi\eta q_x'/2)+\xi\eta q_x'
\cos(\xi\eta q_x'/2)\sin(\varphi/2)\right]}
{\xi^2\eta^2 {q_x'}^2-\varphi^2}.
\end{eqnarray}
Then we have
\begin{eqnarray}
\Theta_2(q_x)&=&\frac{2}{N_\phi}\sum_{k>0} \cos(kq_xl_0^2)\sin^2(\psi_k/2)
=\frac{1}{N_\phi}\sum_{k>0} \cos(kq_xl_0^2)(1-\cos(\psi_k))
=\frac{1}{2}\delta_{q_x,0}-Re[\Theta],
\end{eqnarray}
and
\begin{eqnarray}
\Theta_{3}(q_x)&=&\frac{2}{N_\phi}\sum_{k>0} \cos(kq_xl_0^2)\sin(\psi_k/2)
\cos(\psi_k/2)
=\frac{1}{N_\phi}\sum_{k>0} \cos(kq_xl_0^2)\sin(\psi_k)
=Im[\Theta].
\end{eqnarray}
When replacing $q_x'=q_na=2\pi n$, we have $\sin(q_x'/2)/(q_x'/2)
=\delta_{n,0}$.

\section{Evaluation of $E^{I,o}_{H,F}$}
\label{E_HF_Io}
Here we show the analytical formula of the exchange energies used in Section
\ref{sec_D2} for a double well system at $\nu=4N+2$.
For convenience, we let $\vec{Q}_\perp'\equiv\vec{Q}_\perp l_0$,
$\vec{P}\,'\equiv\vec{P}l_0$, and $d'=d/l_0$, $q'=ql_0$ to be dimensionless.
We have
\begin{eqnarray}
E_F^{I,o}(\vec{0},0)&=&\frac{-1}{\Omega_\perp}
\sum_{\vec{q}}V_{NN,NN}^{I,o}(\vec{q}_\perp)
=\frac{-e^2}{l_0}\int^\infty_0 dq'\,e^{-{q'}^2/2}\left[L^0_{N}
\left(\frac{{q'}^2}{2}\right)\right]^2
\times\frac{1}{2\pi}\int_0^{2\pi}\frac{d\theta}{2}\left(1\pm
\cos(q'(P'_x\cos\theta+P'_y\sin\theta))\,e^{-q'd'}\right)
\nonumber\\
&=&-\frac{e^2}{l_0}\int^\infty_0 dq'\,e^{-{q'}^2/2}\left[L^0_{N}
\left(\frac{{q'}^2}{2}\right)\right]^2\times\frac{1}{2}\left[1\pm
J_0(q'P')\,e^{-q'd'}\right],
\label{E_F_Io_0}
\end{eqnarray}
\begin{eqnarray}
&&E_F^{I,o}(\vec{Q}_\perp,0)=\frac{-1}{\Omega_\perp}
\sum_{\vec{q}}V_{NN,NN}^{I,o}(\vec{q}_\perp)
\cos((Q_xq_y-Q_yq_x)l_0^2)\nonumber\\
&=&-\frac{e^2}{l_0}\int^\infty_0 dq'\,e^{-{q'}^2/2}\left[L^0_{N}
\left(\frac{{q'}^2}{2}\right)\right]^2
\times\frac{1}{2\pi}\int_0^{2\pi}\frac{d\theta}{2}\left(1\pm
\cos(q'(P'_x\cos\theta+P'_y\sin\theta))\,e^{-q'd'}\right)
\cos(q'(Q'_x\sin\theta-Q'_y\cos\theta))
\nonumber\\
&=&-\frac{e^2}{l_0}\int^\infty_0 dq'\,e^{-{q'}^2/2}\left[L^0_{N}
\left(\frac{{q'}^2}{2}\right)\right]^2\times\frac{1}{2}\left[
J_0(q'Q')\right.
\nonumber\\
&&\hspace{1cm}\left.\pm\frac{1}{2}\left(J_0\left(q'\sqrt{(P'_x-Q'_y)^2
+(P'_y+Q'_x)^2}\,\right)
+J_0\left(q'\sqrt{(P'_x+Q'_y)^2+(P'_y-Q'_x)^2}\,\right)\right)\,
e^{-q'd'}\right],
\label{E_HF_Io_Q}
\end{eqnarray}
and
\begin{eqnarray}
&&E_F^{I,o}(\vec{0},q_n)=\frac{-1}{\Omega_\perp}
\sum_{\vec{q}}V_{NN,NN}^{I,o}(\vec{q}_\perp)
\cos(q_nq_yl_0^2)\nonumber\\
&=&-\frac{e^2}{l_0}\int^\infty_0 dq'\,e^{-{q'}^2/2}\left[L^0_{N}
\left(\frac{{q'}^2}{2}\right)\right]^2
\underbrace{\frac{1}{2\pi}\int_0^{2\pi}\frac{d\theta}{2}\left(1\pm
\cos(q'(P'_x\cos\theta+P'_y\sin\theta))\,e^{-q'd'}\right)
\cos(q'_nq'\sin\theta)}_
{\displaystyle \hspace{-3cm}
=\frac{1}{2}\left[J_0(q'q_n')\pm
\frac{e^{-q'd'}}{2}\left(J_0\left(q'\sqrt{{P_x'}^2+(P_y'+q_n')^2}\,\right)
+J_0\left(q'\sqrt{{P_x'}^2+(P_y'-q_n')^2}\,\right)\right)\right]}.
\end{eqnarray}
The most general expression of the exchange energy including both spiral 
and stripe orders ($E^{I,o}_F(\vec{Q}_\perp,q_n)=1\Omega_\perp^{-1}
\sum_{\vec{q}}V_{NN,NN}^{I,o}(\vec{q}_\perp)
\cos(q_nq_yl_0^2)\cos((Q_xq_y-Q_yq_x)l_0^2)$)
can be obtained similarly via the same strategy.


\newpage
\begin{table}
\begin{tabular}{c|cccccl}
{label}&{system}&{$\nu$}&{isospin $\Uparrow$ ($I=1$)}&
{isospin $\Downarrow$ ($I=-1$)} & many-body phases
& broken symmetries \\ \hline
$W1$ &wide well & $2N+1^\ddagger$ 
& $\vec{n}_1=(1,0),
s=\uparrow$&$\vec{n}_2=(0,N),
s=\uparrow$ & coherent &parity \\
& (intersubband) & & & & &\\
$W1'$ &wide well & $2N+1$ 
& $\vec{n}_1=(0,N),s=\uparrow$&$\vec{n}_2=(0,N+1),
s=\uparrow$ & skyrmion stripe & parity \& trans. in\\
& (intrasubband) & & & & & $x$ and $y$ directions\\
$W2$ &wide well & $2N+2^\ddagger$ & 
$\vec{n}_1=(1,0),s=\downarrow(\uparrow)$ 
& $\vec{n}_2=(0,N),s=\uparrow(\downarrow)$ & no & no\\
& (intersubband) & & & & & \\
$W2'$ &wide well & $2N+2$ & $\vec{n}_1=(0,N),s=\downarrow$
& $\vec{n}_2=(0,N+1),s=\uparrow$ & (skyrmion stripe)$^\dagger$ 
& (parity, spin \& trans. \\
& (intrasubband) & & & & & in $x$ and $y$ directions)$^\dagger$ \\
$D1$ & double well & $4N+1$ & $N,l=+1,s=\uparrow$ 
& $N,l=-1,s=\uparrow$ 
& coherent & trans. in $y$ \\
& & & & & spiral & no\\
& & & & & coherent stripe & parity \& trans.\\ 
& & & & & & in $x$ (and $y$) direction\\
& & & & & spiral stripe & parity \& trans. in $y$\\
$D2$ & double well & $4N+2$ & 
$N,\alpha=+1,s=\downarrow$ & $N,\alpha=-1,s=\uparrow$ & coherent$^\S$
& parity \& spin  
\end{tabular}
\caption{
Table of the isospin notations and the many-body phases of 
different systems discussed in this paper. 
$\vec{n}=(n,n')$ is the orbital level 
index of the energy eigenstate of an
electron in a parabolic well subject to 
an in-plane magnetic field (see Eq. (\ref{omega12})). $s=\pm 1/2$ 
and $l=\pm 1$ are spin and layer indices 
respectively. $\alpha=\pm 1$ is for the symmetric(antisymmetric) 
state of the double well system. The many-body phases of $D1$ case 
are described in the layer index basis (see Section \ref{sec_D1}), while
they are described in the symmetry-antisymmetry (noninteracting
eigenstate) basis in the $D2$ case.
The in-plane magnetic field is fixed to be in $+\hat{x}$ direction.
\\
$\ddagger$: Such level crossing exists only when $N>N^\ast$ for 
$N^\ast=\mbox{Max}(\omega_c/\omega_\perp,\omega_\perp/\omega_c)$, 
see Ref. [43].
\\
$\dagger$: We do not really obtain a many-body
phase within the HF approximation, 
but the experimental data of Ref. [15] suggests that
a skyrmion stripe phase may exist in the $W2'$ system 
(see Section \ref{sec_W2}).
\\
$\S$: $=$ "commensurate canted antiferromagnetic phase" in the literature
with spiral order in the layer index basis (see Section \ref{sec_D2}).
}
\label{system_notation}
\end{table}
\begin{table}
\begin{tabular}{c|ccccc}
& & & & \hspace{-4.5cm}In-plane magnetic field, $B_\|$, is along $+\hat{x}$ 
($\longrightarrow$) \\
\hline
systems & $W1'$ & $W2'$ & $D1$ & $D1$ & $D1$ \\
\hline
anisotropic phases & skyrmion stripe & skyrmion stripe & spiral
& spiral stripe & coherent stripe \\
\hline
isospin spiral & & & & \\
winding vector, $\vec{Q}_\perp$  
& {\LARGE$\updownarrow$} & {\LARGE($\updownarrow$)} 
& {\LARGE$\uparrow$} & {\LARGE$\uparrow$} & {\Large$\times$}\\
\hline
isospin stripe & & & & \\ 
normal vector, $\hat{n}$ & {\LARGE$\leftrightarrow$} & 
{\LARGE($\leftrightarrow$)} 
& {\Large$\times$} & {\LARGE$\updownarrow$} & arbitrary  
\end{tabular}
\caption{A table summarizing the directions of the
spiral and the stripe orders in different systems. The in-plane 
magnetic field is fixed in $+\hat{x}$ axis.
For $W2'$ system (wide well at even filling), we speculate that
the skyrmion stripe
is a possible ground state according to the experimetal results
and our analysis in Section \ref{sec_D2}.
Note that in a wide well system,
the spiral order is degenerate at $\vec{Q}^\ast_\perp=(0,\pm Q_y^\ast)$, 
while in a double
well system, it is fixed along $+\hat{y}$ only in the commensurate state.
Winding of the phases in $D1$ system is set by the in-plane magnetic 
field, and does not imply breaking of the translational symmetry
(see discussion in Section \ref{symmetry_CI}).
}
\label{directions}
\end{table}

\begin{thebibliography} {}
\bibitem{subband_note}
Note that the intersubband crossing may not exist if electron
filling factor is so small (or the subband energy sepration, $\omega_1$
is so large) that no higher subband 
levels are occupied in the whole range of $B_\|$ field.
The criterion to have such crossing is
$N>N^\ast=\mbox{Max}(\omega_0/\omega_\perp,\omega_\perp/\omega_0)$.
\bibitem{footnote_t}
We note that the generalized Laguerre polynomial in the effective
tunneling amplitude, $t_{n,P}$, may reverse the sign of tunneling amplitude
for $n>0$ and hence reduce the tunneling effects when layer separation or
$B_\|$ is in a certain range \cite{cote02}. But we will not consider this
situation for simplicity and all of our results should be correct
without losing generality.
\bibitem{footnote_group}
We note that a generalized two-dimensional 
translation operator, $\hat{\cal T}(\vec{R}_\perp)\equiv\exp(-i\vec{R}_\perp
\cdot\hat{\cal M})$, does not form a close
translational group if both $R_x$ and $R_y$ are nonzero.
This is because in general 
$\hat{\cal T}(\vec{R}_\perp)\cdot\hat{\cal T}(\vec{R}_\perp')$ 
cannot be equivalently obtained by the third 
translation, $\hat{\cal T}(\vec{R}_\perp'')$ by an additional phase
factor, which follows from the fact that ${\cal T}_x(R_x)$ and
${\cal T}_y(R_y)$ do not commute with each other.
However, it is easy to see that ${\cal T}_x(R_x)$ and
${\cal T}_y(R_y)$ individually can form 
close translational groups in $x$ and $y$ direction respectively.
Therefore we write them separately to avoid any confusion.
\bibitem{footnote_stripedirection}
This is because the electron Hartree energy in each
Landau level has nodes in $x$ direction, but quiet uniform in
$y$ direction due to the anisotropic distortion of the electron wavefunction in
the strong in-plane magnetic field.  
A simple scenario to understand this is to view an isospin
stripe phase as a two independent (incoherent) stripes in the two
degenerate levels, each of which is partially filled, and therefore the
stripe longitudinal direction must be aligned to be perpendicular to
the in-plane field direction to
optimize the anisotropy energy in each level, 
similar to the stripes at half filling factors extensively studied 
in the literature \cite{pan00,jungwirth99}.
When we turn on the interlevel
coherence between the two degenerate levels, a spiral order is introduced
also and competes with the anisotropy energy to determine the direction
of stripe phase at integer filling factors.
\end{thebibliography}
\end{document}